\documentclass[a4paper,11pt]{article}

\usepackage{jheppub}

\usepackage[T1]{fontenc} 
\usepackage{amsmath,epsf,amssymb,latexsym,amsthm,setspace,bbm,longtable}

\usepackage[matrix,arrow,curve,arc]{xy}
\usepackage{graphics,makecell}
\usepackage{array,multirow,multicol}

\usepackage{etoolbox}
\usepackage{titlesec}
\usepackage{cleveref}

\newsavebox\ltmcbox

\titleformat{\paragraph}
{\normalfont\normalsize\bfseries}{\theparagraph}{1em}{}
\titlespacing*{\paragraph}
{0pt}{3.25ex plus 1ex minus .2ex}{1.5ex plus .2ex}
\makeatletter
\def\@fpheader{\relax}
\makeatother

\newtheorem{thm}{Theorem}[section]

\newtheorem{result}[thm]{Result}

\theoremstyle{definition}

\crefmultiformat{equation}{(#2#1#3)}%
{ and~(#2#1#3)}{, (#2#1#3)}{ and~(#2#1#3)}

\def\P{{\mathbb P}}

\def\ord{{{\text{ord}}}}

\def\ord{{{\text{ord}}}}

\def\C{{\mathbb C}}

\def\cO{{\cal O}}
\newcommand{\Z}{\mathbb{Z}}
\newcommand\Deltah{\widehat{\Delta}}
\newcommand\Deltat{\widetilde{\Delta}}

\newcommand\fh{\widehat{f}}
\newcommand\gh{\widehat{g}}
\newcommand\at{\widetilde{a}}
\newcommand\bt{\widetilde{b}}

\newcommand\dt{\widetilde{d}}

\newcommand\ft{\widetilde{f}}
\newcommand\gt{\widetilde{g}}

\def\ff#1#2{{\textstyle\frac{#1}{#2}}}

\def\e{{\mathfrak{e}}}
\def\f{{\mathfrak{f}}}
\def\g{{\mathfrak{g}}}
\def\h{{\mathfrak{h}}}

\def\nzero{{n${}_0$}}
\def\IVns{{IV${}^{ns}$}}
\def\IVs{{IV${}^{s}$}}
\def\II{{II}}
\def\Izerostar{{I${}_0^\ast$}}
\def\Izerostarns{{I${}_0^{\ast ns} $}}
\def\Izerostars{{I${}_0^{\ast s} $}}
\def\Izerostarss{{I${}_0^{\ast s s} $}}
\def\Instar{{I${}_n^\ast$}}
\def\Izero{{I${}_0$}}
\def\In{{{I${}_n$}}}
\def\Ione{{{I${}_1$}}}
\def\Itwo{{{I${}_2$}}}
\def\Ithree{{I${}_3$}}
\def\Ifour{{{I${}_4$}}}
\def\Ifive{{{I${}_5$}}}
\def\Isix{{{I${}_6$}}}

\def\Ieight{{{I${}_8$}}}

\def\Ionestar{{I${}_1^\ast$}}
\def\Ionestarns{{I${}_1^{\ast ns}$}}
\def\Ionestars{{I${}_1^{\ast s}$}}

\def\Fourstars{{IV${}^{\ast s}$}}
\def\Twostar{{II${}^\ast$}}

\def\IVstarns{{IV${}^{\ast ns}$}}
\def\IVstars{{IV${}^{\ast s}$}}
\def\IIstar{{II${}^\ast$}}
\def\IIIstar{{III${}^\ast$}}
\def\Threestar{{III${}^\ast$}}

\def\Ins<#1>{I${}_{#1}^s$}
\def\Inns<#1>{I${}_{#1}^{ns}$}
\def\InstarNoMon<#1>{I${}_{#1}^{\ast }$}
\def\Instars<#1>{I${}_{#1}^{\ast s}$}
\def\Instarns<#1>{I${}_{#1}^{\ast ns}$}
\def\su{\operatorname{\mathfrak{su}}}
\def\so{\operatorname{\mathfrak{so}}}
\def\sp{\operatorname{\mathfrak{sp}}}
\def\suText<#1>{$\su({#1})$}
\def\soText<#1>{$\so({#1})$}
\def\spText<#1>{$\sp({#1})$}
\def\eText<#1>{$\e_{#1}$}
\def\fText{$\f_4$}
\def\gText{{$\g_2$}}
\def\Deltat{{\tilde\Delta}}
\def\at{{\tilde a}}
\def\bt{{\tilde b}}
\def\dt{{\tilde d}}
\def\gg{{\g_{\text{gauge}}}}
\def\gG{{\g_{\text{global}}}}
\def\ggs{{\g_{\text{GS}}}}
\def\ggst{{\widetilde{\mathfrak{g}}_{\text{GS}}}}
\def\T{{\mathcal{T}}}

\title{Classifying Global Symmetries of 6D SCFTs}
\author[a]{Peter R. Merkx} 
\affiliation[a]{Department of Mathematics, U.C. Santa Barbara, Santa Barbara CA, 93106}
\emailAdd{merkx@math.ucsb.edu}

\abstract{We characterize the global symmetries for the conjecturally complete collection of all six dimensional superconformal field theories (6D SCFTs) which are realizable in F-theory and have no frozen singularities. We provide comprehensive checks of earlier 6D SCFT classification results via an alternative geometric approach yielding new restrictions which eliminate certain theories. We achieve this by directly constraining elliptically fibered Calabi-Yau (CY) threefold Weierstrass models and find this allows bypassing all anomaly cancellation machinery. This approach reduces the problem of classifying which 6D SCFT gauge and global symmetries are realizable in F-theory models before RG-flow to characterizing features of elliptic fibrations associated to these theories obtained by analysis of polynomials determining their local models. We supply an algorithm with implementation producing from a given SCFT base an explicit listing of all compatible gauge enhancements and their associated global symmetry maxima consistent with the geometric constraints we derive while making manifest the corresponding geometric ingredients for these symmetries including any possible Kodaira type realizations of each algebra summand. In mathematical terms, this amounts to determining all potentially viable non-compact CY threefold elliptic fibrations at finite distance in the moduli space with Weil-Petersson metric which meet certain requirements including the transverse pairwise intersection of singular locus components. We provide local analysis exhausting nearly all CY consistent transverse singular fiber collisions, global analysis concerning all viable gluings of these local models into larger configurations, and many novel constraints on singular locus component pair intersections and global fiber arrangements. We also investigate which transitions between 6D SCFTs can result from gauging of global symmetries and find that continuous degrees of freedom can be lost during such transitions.
}

\begin{document} 
\maketitle
\flushbottom

\begin{section}{Introduction}

Six-dimensional superconformal field theories (6D SCFTs) are uniquely well-suited to shed light on the structure of the string landscape. Nearly two decades after the surprising appearance of the first arguments demonstrating their existence~\cite{Witten:1995zh,Seiberg:1996vs,Seiberg:1996qx} resolved earlier proposals suggesting they might exist in principle but must be non-Lagrangian~\cite{Nahm:1977tg},
renewed interest over the last several years~\cite{kumar2009bound,kumar2010mapping,kumar2010global,kumar2011string,newTate,grimm2012structure,morrison2012toric,morrison2012f,bases,heckman2015more,DelZotto:2014hpa} has enabled classification results for these theories~\cite{atomic,classifyingSCFTs} relying heavily on tools from F-theory. Among the features of SCFTs these classifications leave implicit are their global symmetries.

Our primary focus here is to provide a characterization of these symmetries for conjecturally all 6D SCFTs realizable in F-theory without frozen singularities (i.e., those without O$7^+$ planes in the language of type IIB string theory)~\cite{wittenToroidal,deBoer:2001wca,frozen} that have recently been conjecturally classified~\cite{atomic}. Global symmetries play a central role in recent lines of inquiry including investigations concerning renormalization group (RG) flows of 6D SCFTs~\cite{Heckman:2015ola,heckman20166d}, but a systematic treatment has remained lacking. We outline the general structure of 6D SCFT global symmetries, provide summary rules to determine global symmetry maxima for each known 6D SCFT. We also enable explicit listing of the maxima via implementation of an exhaustive search algorithm making manifest the potentially viable Kodaira type realizations of each gauge and global symmetry summand which may occur for a given SCFT base $B\cong \C^2/\Gamma$ determining a family of F-theory models where $\Gamma$ is a discrete $U(2)$ subgroup meeting stringent requirements~\cite{classifyingSCFTs,4DFtheory}. In the process, we show that some of the theories appearing in that classification can be eliminated. We also carry out a check that our methods suffice to otherwise match the previously reported ``Atomic Classification''~\cite{atomic} via manifestly geometric constraints without appeal to Coulomb branch anomaly cancellation machinery. With these tools at our disposal, we then briefly examine the SCFT transitions obtained by promoting global symmetry subalgebras to gauge summands and find that continuous degrees of freedom can be lost during these ``gaugings.''

The rest of this note is organized as follows. We give a general overview of relevant background material and our approach in Section~\ref{s:overview}. In Section~\ref{s:strategy}, we review several features of Weierstrass models in F-theory and previous 6D SCFTs global symmetry classification results for the small class of theories for which these have been treated systematically. We then detail an algorithm determining these symmetries for general 6D SCFTs based upon restrictions we derive in Appendix~\ref{s:theRestrictions} and other previously established constraints~\cite{BertoliniGlobal1,gaugeless}. We turn in Section~\ref{s:newGaugeRestrictions} to a discussion of novel restrictions on 6D SCFT bases and their gauge enhancements providing slight refinements of the classification from~\cite{atomic} determined using our algorithm. In Section~\ref{s:GSRules} we summarize the geometrically realizable global symmetries of 6D SCFTs in terms of the permitted length two subquiver Kodaira type assignments for each valid base. We then discuss the general structure of 6D SCFT global symmetries via summands arising from the ``atomic'' base decomposition constituents in Section~\ref{s:generalStructure}. The transitions between 6D SCFTs that can be obtained by occupying the degrees of freedom permitting global symmetry summands to instead allow further gauge summands (i.e.,\ by ``gauging global symmetries'') are discussed in Section~\ref{s:gaugingGS}. Concluding remarks and an outline of applications and open problems appear in Section~\ref{s:conclusions}. Instructions for using the accompanying computer algebra workbook appear in Appendix~\ref{s:GSWorkbook}. Finally, tables summarizing global symmetries for several key cases helping to complete our analysis appear in Appendix~\ref{s:GSTables}.
\end{section}

\begin{section}{Overview}\label{s:overview} 

Global symmetries of theories with 1D Coulomb branch and those without non-abelian gauge algebra (for cases with $B$ containing a single compact singular locus component) have previously been treated~\cite{BertoliniGlobal1,gaugeless}. Here we extend that approach which involves determining geometrically realizable SCFT global symmetries via the properties of non-compact Calabi-Yau threefold elliptic fibrations of the form $\pi: X\to B$ underlying F-theory 6D SCFT models. Flowing to a conformal fixed point after taking a limit in which all compact components of the singular locus are contracted yields a CFT whose {\it geometrically realizable} global symmetries are constrained by the permissible non-abelian algebras which can be carried on non-compact components of the singular locus of $X$ via a correspondence of these algebras to global symmetries of the SCFT dating to early F-theory descriptions of the
small $E_8\times E_8$ instanton~\cite{FCY1,WitMF,FCY2} also used recently in a number of works~\cite{classifyingSCFTs,DelZotto:2014hpa,atomic}.

The geometrically realizable global symmetry maxima of F-theory 6D SCFT models for the cases treated previously~\cite{BertoliniGlobal1,gaugeless} have only a single compact singular locus component. For those theories, these maxima are subalgebras of the Coulomb branch global symmetry algebras permitted via field-theoretic gauge and mixed anomaly cancellation requirements which govern the gauge enhancement prescriptions of the ``Atomic Classification''~\cite{atomic}.
We find this is true more generally and appears to hold for all 6D SCFTs admitting an F-theory description and having no frozen singularities. 

While it has been argued that all continuous SCFT global symmetries must be gauged upon coupling to gravity~\cite{banks2011symmetries}, precise rules for determining these degrees of freedom for an arbitrary 6D SCFT and provision of gauging consistency conditions before and after such coupling have not been systematically treated in cases with Coulomb branch of dimension larger than one. While we will not discuss this global symmetry gauging mandated by coupling to gravity in any detail, we will briefly discuss gauging of global symmetries taking one SCFT to another. However, our primary focus in this note is simply to constrain the manifest geometrically realizable flavor symmetries of each F-theory 6D SCFT model. Note that the construction we study identifies these degrees of freedom in the UV though such models only give rise to a conformal theory after RG flow. This means that the actual global symmetries of an SCFT associated to each model may differ from those degrees of freedom we shall identify. As shown in earlier work~\cite{BertoliniGlobal1,gaugeless}, these geometrically realizable global symmetries are (in some cases strictly) more constrained than those permitted on the Coulomb branch of the theory. Typically, the latter constrain the actual global symmetries of a CFT since these also act on the Coulomb branch of the theory. However, additional field theoretic constraints can in some cases provide reductions beyond Coulomb branch gauge and mixed anomaly cancellation prescriptions, for example when we have $\su(2)$ gauge algebra~\cite{Ohmori:2015fk}.

The approach we take reduces our central task to a mathematically well-defined problem. This consists of providing a series of constraints on non-compact elliptically fibered Calabi-Yau threefolds we study by means of a singular elliptic fibration $\pi$ determined by Weierstrass equation of the form
\begin{align}\label{eq:weierstr}
y^2 = x^3 + f x + g~
\end{align}
with auxiliary data detailed in Section~\ref{s:Weierstrass} and $f,g$ locally defined polynomials on a complex surface. More precisely, $f,g$ are sections of $\cO(-4K_B),$ $\cO (-6 K_B),$ respectively, with $K_B$ the canonical bundle over the base $B,$ as above. The constraints we obtain involve a careful analysis of local models for these elliptic fibrations. We treat sufficiently many cases that it is convenient to constrain the global Weierstrass models through implementation of exhaustive computer search routines. 

Note that we do not claim the existence of globally consistent F-theory models achieving the flavor symmetry maxima we report. While proofs to that end are often possible and even trivial in sufficiently many cases that one might expect only limited tightenings of these maxima may be obtained, doing so in full generality is delicate and beyond the scope of this work. Among the key difficulties in demonstrating global consistency of the models underlying the maxima we report is the construction $f,g$ in a neighborhood of a compact curve having intersection with multiple transverse type \In{} or \Instar{} fibers. When only a pair of transverse curves is considered, we can often bypass explicit checks using a suitable coordinate system or a sequence of blow-ups. However, when three or more transverse curves are present explicit construction is often highly involved. Further work is needed to show that global symmetry inducing non-compact curves meeting distinct compact singular locus components remain uncoupled in cases where sequences of blow-ups do not immediately suffice to this end. (For example, determining via explicit construction whether $\Sigma$ in the base $2_{\Sigma'}1_{\Sigma}$ having Kodaira types \Ione{} and \Izero{} on $\Sigma'$ and $\Sigma,$ respectively, can support simultaneous transverse intersection with the triple of curve stubs \Isix{},\Ithree{},\Itwo{} permitted as a point singularity collection along $\Sigma$ by ``Persson's List''~\cite{PerssonsList} is rather tedious. Advancing to analogous cases where we modify the type on $\Sigma'$ and replace the type on $\Sigma$ curve by \Instar{} often presents similar but magnified challenges.) Hence, except where the maxima we report dictate only trivial global symmetry is permitted, these algebras should strictly speaking be viewed as upper bounds on the actual global symmetries of each theory in the UV.

Note that restrictions on the geometry of Weierstrass models are seemingly necessary to reach the precise conclusions of the ``Atomic Classification''~\cite{atomic}. It is hence natural to ask whether a parsimonious approach giving a ``purely geometric'' characterization of all known constraints on 6D SCFT F-theory models is possible. In this work, we provide strong evidence towards answering this question in the affirmative.  

Our methods extend earlier work~\cite{BertoliniGlobal1} to the general case, thus resulting in a geometric classification of gauge and flavor symmetries realizable in F-theory models for all 6D SCFTs of the aforementioned classification~\cite{atomic}. We shall proceed without appeal to field-theoretic tools based on Coulomb branch anomaly cancellation requirements involving hypermultiplet count pairing restrictions. This enables us to provide consistency checks on results derived via the latter approach where known. We rely instead on algebro-geometric analysis of elliptically fibered Calabi-Yau threefolds with our efforts focusing on local polynomial expansions of the sections $f,g$ occurring in~\eqref{eq:weierstr}.

To enable explicit listing of global symmetry maxima for each gauge enhancement of any fixed 6D SCFT base, we derive a series of constraints enabling our calculations to proceed via computer algebra system. These fall into three main categories. First, we determine which pairs of curves in $B$ with generic fibers having specified Kodaira types are permitted to intersect without introducing singularities so severe that Calabi-Yau resolution of the fibration would be prevented. Second, we analyze local models for transverse singular curve collisions to determine the minimal ``intersection contributions'' (giving counts towards certain degrees of freedom along each curve) from every relevant permitted intersection. These both entail generalizing previous analysis limited to certain single curve theories~\cite{BertoliniGlobal1,gaugeless}. The final category concerns elimination of certain arrangements multiple singular locus components. 

Together these tools enable us to constrain 6D SCFT gauge and global symmetry algebras independently of and more strongly than Coulomb branch gauge and mixed anomaly cancellation techniques~\cite{atomic,BertoliniGlobal1,gaugeless}. 
Our approach is parsimonious in that we obtain constraints via a manifestly geometric approach based on inspecting elliptic fibrations in correspondence with 6D SCFTs rather than via the hybrid approach underlying~\cite{atomic} which invokes representation theoretic anomaly cancellation tools supplemented by geometric restrictions.

We discuss the previously reported gauge enhancements and bases~\cite{atomic} which our methods eliminate in Section~\ref{s:newGaugeRestrictions}. Gauge enhancement prescriptions for certain constituents of SCFT bases, namely ``links,'' obtained using the accompanying computer algebra routines are compared with outputs of routines provided in conjunction with~\cite{atomic} after minor edits aimed to match gauge prescriptions therein. Extending comparisons for links to those for general 6D SCFT bases yields novel restrictions in certain cases.

While established local analysis including intersection contribution data~\cite{BertoliniGlobal1,gaugeless} plays a key role, our route is complicated by the following issues meriting significant extension of earlier treatments. The theories addressed therein involve models having a singular locus with only one compact curve $\Sigma$ associated to a simple gauge algebra $\g.$ Determining which global symmetry algebras are realizable in such models can be treated via consideration of non-compact curve collections $\{\Sigma_i'\}$ with each $\Sigma_i'$ transverse to $\Sigma$ and carrying non-abelian simple Lie algebra $\g_i'.$ Relatively maximal algebras arising as $\oplus_i\g_i'$ from a permissible configuration are identified as global symmetry maxima via a limit with $\Sigma$ contracted. For such cases, analysis concerning non-compact curves collections with $\oplus_i\g_i'$ potentially maximal suffices while configurations resulting in ``small'' $\oplus_i\g_i'$ are irrelevant. 

To treat more general theories with multiple compact components $\{\Sigma_i\}$ in the singular locus giving rise to gauge algebra $\oplus_i\g_i$, we consider collections of non-compact curves having transverse intersection with some $\Sigma_i.$ While previously determined constraints on maximal algebra yielding configurations transverse to $\Sigma_i$~\cite{BertoliniGlobal1,gaugeless} are helpful, the curves meeting $\Sigma_i$ now include any compact neighbors $\Sigma_{j\neq i}$ which may carry ``small'' algebras corresponding to gauge summands comprising part of the data specifying a theory. Such local configuration hence may not be among the previously studied maximal configurations. This presents the significantly more involved combinatorial problem of finding the maximal configurations meeting each fixed $\Sigma_i$ given not only the Kodaira type along $\Sigma_i$ but also those of any $\Sigma_j$ meeting $\Sigma_i$ with global considerations introducing various subtleties. Further complicating our task is that intersection with a compact transverse curve often requires distinct local analysis an analogous intersection with a non-compact curve of the same Kodaira type and can yield different intersection data.

The minimal orders of vanishing determining Kodaira types become insufficient to realize a permitted type assignment $\{T_i\}$ on each $\{\Sigma_i\}$ in the general case, e.g.~\eqref{eq:minimalABEx}. This leads us to develop local models for many transverse intersections of curve pairs with designated orders of vanishing nearly exhausting all permissible transverse singularity collisions for CY threefold fibrations. The algorithm we supply incorporates this analysis to yield a significant step towards explicit classification of such fibrations including a treatment of codimension-two singularities.

In addition to enabling explicit listing of gauge enhancements and their flavor symmetry maxima, we shall discuss the general structure of these symmetries. We provide two complementary prescriptions involving rules dictating the flavor symmetry maxima that may occur for each 6D SCFT in terms of Kodaira types on curves determining a fixed gauge enhancement. The first consists of rules summarizing these maxima in terms of length two curve chains and constraints imposed additional neighboring curves. This summarizes results obtained with computer routines and organizes them into constraint equations and structured listings. Second, we detail these maxima for longer bases in terms of contributions arising from each of the building blocks in the ``atomic'' decomposition of 6D SCFT bases into ``link'' and ``node'' constituents~\cite{atomic}.
\end{section}

\begin{section}{The strategy}\label{s:strategy}
Our approach generalizes earlier work~\cite{BertoliniGlobal1} to cases with discriminant locus consisting of more than one compact curve. The first ingredient involves imposing the restrictions derived therein for single curve cases with associated non-abelian gauge algebra along with the remaining single curve cases without non-abelian gauge algebra treated in~\cite{gaugeless}. 

We next check a given global symmetry summand inducing configuration for consistency with the number of vanishings of $f,g,$ and $\Delta$ required along each curve in a quiver. There are a handful of additional restrictions we shall impose including the elimination of a few configurations which are barred via earlier analysis (appearing in Appendix E.3 of~\cite{atomic}) and a similar discussion we derive here in Appendix~\ref{s:theRestrictions}.

Before moving to treat the local analysis and other tools we shall require, we begin in this section with a review of our general approach for determining global symmetry maxima within F-theory on purely geometric grounds (i.e.,\ sans anomaly cancellation tools). We also pause to detail an algorithm and our accompanying implementation which allows us to reach conclusions through an exhaustive search of configurations meeting known restrictions outlined here and those derived in Appendix~\ref{s:theRestrictions}.

We shall make use of the maximal configurations for single curve theories studied previously~\cite{BertoliniGlobal1, gaugeless} updated with tightenings in certain cases having non-abelian gauge algebra illustrated in Table~\ref{t:updatedSummary} and in one case with trivial gauge algebra discussed in Section~\ref{s:gaugelessGSDiscussion}. In the remainder of Appendix~\ref{s:theRestrictions}, we turn to a detailed local analysis on the number of vanishings required along each compact curve for all possible pairwise intersections of discriminant locus components which we shall encounter. In the process we also uncover various forbidden curve intersections. These restrictions are central to the approach we describe in this section, but we postpone their discussion until we have outlined the task at hand since their details are somewhat involved. 

\begin{subsection}{Weierstrass models and gauge algebras in F-theory}\label{s:Weierstrass}
The essential geometric ingredient for an F-theory formulation of an SCFT is a Weierstrass model determining a singular elliptic fibration given by $\pi:X\rightarrow B$ with fibers determined by a Weierstrass equation of the form~\eqref{eq:weierstr}
with $B \cong \C^2/\Gamma,$ as above, in the case of 6D F-theory. The discriminant of this equation,
\begin{align}
\Delta:=4f^3+27g^2~,
\end{align}
is a section of $\cO(-12 K_B)$ with its ``discriminant locus,''
\begin{align}
\{\Delta =0\}~,
\end{align}
determining where the fibration is singular. The types of singularities that are permitted without being so severe as to prevent a Calabi-Yau resolution $\widetilde{X}$ are given by the Kodaira classification~\cite{KodairaII,KodairaIII,Neron} with summary appearing in Table~\ref{t:singtypes}. ``Non-minimal'' fiber types indicated in the final row have resolution of singularities containing a curve which can be blown down. Blow-down of such a curve down leads to a new Weierstrass model with orders of vanishing of $(f,g,\Delta)$ along $\Sigma$ reduced by $(4,6,12).$ We hence discard such cases without loss of generality. Similarly, a two-dimensional Weierstrass model is minimal at $P$ defined by $\{\sigma =0\}$ provided $\ord_{\sigma=0}(f)<4,$ or $\ord_{\sigma=0}(g)<6.$ To reiterate, we confine our study in this work to those models lacking non-minimal points. (From the Calabi-Yau condition, we could have blown up such points. Without loss of generality, we take such a model as our starting point.)

\begin{table}[thbp]
\begin{center}
\begin{tabular}{ccc|c|c|c}
ord($f$)	& ord($g$)		& ord($\Delta$)		&type			&singularity			&non-abelian algebra \\
\hline
$\geq0$	&$\geq0$		&0				&I${}_0$			&none				& none \\
$0$		&$0$			&1				&I${}_1$			&none				& none \\
$0$		&$0$			&$n\geq2$		&I${}_n$			&$A_{n-1}$			&$\su(n)$ or $\sp([n/2])$ \\
$\geq1$	&$1$			&2				&II				&none				& none \\
$1$		&$\geq2$		&3				&III				&$A_1$				&$\su(2)$\\
$\geq2$	&$2$			&4				&IV				&$A_2$				& $\su(3)$ or $\su(2)$ \\
$\geq2$	&$\geq3$		&6				&I${}_0^\ast$		&$D_4$				&$\so(8)$ or $\so(7)$ or $\mathfrak{g}_2$ \\
$2$		&$3$			&$n\geq7$		&I${}_{n-6}^\ast$	&$D_{n-2}$			&$\so(2n-4)$ or $\so(2n-5)$\\
$\geq3$	&$4$			&8				&IV${}^\ast$		&$\mathfrak{e}_6$		&$\mathfrak{e}_6$ or $\mathfrak{f}_4$ \\
$3$		&$\geq5$		&9				&III${}^\ast$		&$\mathfrak{e}_7$		&$\mathfrak{e}_7$ \\
$\geq4$	&$5$			&10				&II${}^\ast$		&$\mathfrak{e}_8$		&$\mathfrak{e}_8$ \\
$\geq4$	&$\geq6$		&$\geq12$		&non-minimal		&-					&- 
\end{tabular}
\end{center}
\caption{Singularity types with associated non-abelian algebras.}
\label{t:singtypes}
\end{table}

The precise gauge algebra which occurs in the cases of ambiguity is determined by inspection of the auxiliary polynomials appearing in Table~\ref{t:monodromyCoverTable} where $\Sigma$ is a curve along the singular locus $\{z=0\};$ larger gauge algebras result with more complete factorizations. A significant portion of our analysis concerns which splittings can take place in various intersection arrangements. Several existing results towards this end~\cite{atomic,BertoliniGlobal1,gaugeless} are crucial to our work.

\begin{table}[thbp]
\begin{center}
\begin{tabular}{|c|c|}
\hline
type  & equation of monodromy cover\\ \hline\hline
I${}_n^{s/ns}$, $n\ge3$
&$\psi^2+(9g/2f)|_{z=0}$
\\ \hline 
IV${}^{s/ns}$ 
&$\psi^2-(g/z^2)|_{z=0}$
\\ \hline
I${}_0^{\ast s/ss/ns}$
& $\psi^3+(f/z^2)|_{z=0}\cdot\psi+(g/z^3)|_{z=0}$
\\ \hline
I${}_{2n-5}^{\ast s/ns}$, $n\ge3$
&$\psi^2+\frac14(\Delta/z^{2n+1})(2zf/9g)^3|_{z=0}$
\\ \hline 
I${}_{2n-4}^{\ast s/ns}$, $n\ge3$
&$\psi^2+(\Delta/z^{2n+2})(2zf/9g)^2|_{z=0}$
\\ \hline 
IV${}^{\ast s/ns}$
& $\psi^2-(g/z^4)|_{z=0}$
\\ \hline 
\end{tabular}
\end{center}
\caption{Monodromy cover polynomials determining non-abelian gauge algebras.}
\label{t:monodromyCoverTable}
\end{table}

The association of non-abelian algebras to Kodaira types indicated in Table~\ref{t:singtypes} including non-simply laced cases dates to~\cite{BershadskyEtAl}. The essential idea is that the resolution of singularities $\widetilde{X}$ for a given Kodaira type gives rise to a graph of curves we may naturally associate to a Dynkin diagram determining a Lie algebra $\g.$ This same $\g$ arises as the gauge algebra of a corresponding physical theory associated to an F-theory model compactified on a Calabi-Yau threefold $\widetilde{X}$ with metric furnished via work of Yau~\cite{YauCalabi}. Cases in which non-simply laced algebras may occur further involve a cover of $\widetilde{X}$ with properties determined by the auxiliary ``monodromy cover'' polynomials appearing in Table~\ref{t:monodromyCoverTable}. These together with the local polynomial expansions for certain Kodaira types as detailed previously (in Appendices A,B of~\cite{BertoliniGlobal1}) are the main objects involved in our discussion.

\end{subsection}

\begin{subsection}{Global symmetries from F-theory geometry}\label{s:1Dgauged}
The field-theoretic and geometrically realizable global symmetries of 6D SCFTs with an F-theory model having discriminant locus with a single compact component were studied in previous work~\cite{BertoliniGlobal1,gaugeless} to treat 1D Coulomb branch cases and certain trivially gauged theories. This leaves us to focus on theories with models having more than one compact component after providing a few tightenings of earlier results detailed in Sections~\ref{s:singleCurveGS}. 

Two ingredients are essential in our derivations. We make heavy use of earlier results giving expansions of $f,g,$ and $\Delta$ that determine general forms for expansions giving local models of Kodaira type \Instar{} and \In{} curves~\cite{newTate,BertoliniGlobal1}. We also rely on Tables~\ref{t:sumloccontr},\ref{t:maxn} taken (up to minor corrections) from~\cite{BertoliniGlobal1} which give forbidden curve intersections and intersection contributions obtained from certain local models. Though we require significant generalizations derived in Appendix~\ref{s:theRestrictions}, these local intersection models and contribution data play a key role. We also will use the maximal configurations derived in those works and several geometric restrictions on curve pair intersections derived in~\cite{atomic}.

\begin{paragraph}{Setup and notation}\label{s:setup}
Let us begin by reviewing aspects of our setup, notation and terminology largely based on~\cite{BertoliniGlobal1,atomic}. 

Global symmetries arise in F-theory via non-compact components of the discriminant locus. Coupling of the associated gauge
group on each of these curves becomes zero after a rescaling, hence leading to a global symmetry~\cite{BertoliniGlobal1}. Our main focus here involves extending earlier discussions~\cite{BertoliniGlobal1,gaugeless} concerning symmetries constructed in this way for cases with discriminant locus containing a single compact curve (determining an SCFT with gauge algebra consisting of at most a single simple Lie algebra summand) to the much broader collection of theories appearing in~\cite{atomic}. The latter typically have bases with multiple compact components of the discriminant locus. We shall see that this extension is simple in principle, but presents significant combinatorial challenges.

Let us pause to discuss two limitations of our approach also noted elsewhere~\cite{BertoliniGlobal1,gaugeless}. First, the Tate's algorithm prescription~\cite{newTate,matter1} may not capture the most general possible forms for Kodaira types I${}_{7\leq n\leq 9}$. Since we shall rely on the Tate forms in these cases, it is possible that a limited class of configurations may be missed by our approach. Second, where local analysis permits non-compact curves to allow any monodromy designation, we report global symmetry maxima with those possibilities giving the largest algebras consistent with our analysis though additional restrictions could in principle be obtained in some cases. However, the maxima report always appear to be subalgebras of the Coulomb branch global symmetries which we generally expect to find as subalgebras of the actual global symmetry algebra for a given theory. This suggests that only limited further reductions from these maxima may be obtained. Note that multiple global symmetry maxima persist in many cases even when monodromy designations are unambiguous both in cases for fixed geometric realization of a given gauge algebra (e.g.\ $\su(2)$ on a $-1$ curve realized by Kodaira type \IVns{} with maxima appearing in Table~\ref{t:updatedSummary}) and in cases where we compare all geometries for fixed base realizing a given gauge assignment (e.g.\ $\su(3)$ on a $-1$ curve with $\su(10)$ and $\su(3)^{\oplus 3}$ maxima, as read from Table 6.1 of~\cite{BertoliniGlobal1}).

We now pause to review a few details of the ``Atomic Classification''~\cite{atomic}. Theories are classified therein by detailing all possible connected trees of compact curves $\Sigma_i \subset \{\Delta=0\}\subset\C^2$ having $\Sigma_i\cdot \Sigma_i=m_i$ over which the fibration is singular. This ``atomic'' decomposition into permitted subgraphs given by the values $m_i$, rules for their gluing and the gauge summands from each $\Sigma_i$ dependent on nearby attachments leaves certain ambiguities in the Kodaira types which can realize a given gauge summand, e.g.,\ types \Izero{}, \Ione{}, and II each yield trivial summand. One of our secondary objectives here is to resolve these ambiguities and provide a geometric check of this classification.

Contraction of all $\Sigma_i$ yields the orbifold base $B\cong \C^2/\Gamma$ with $\Gamma$ a discrete $U(2)$ subgroup determined by the values $m_i$ or alternatively the values $\widetilde{m_i}$ obtained after iterated blow-down of all $-1$ curves to yield an ``endpoint.'' These were classified in~\cite{classifyingSCFTs} and restructured in~\cite{4DFtheory}. Distinct curves $\Sigma_j$ must have transverse intersections in at most a single point. The curves $\{\Sigma_i\}$ must be contractible at finite distance in the Calabi-Yau moduli space with Weil-Petersson metric dating in this context to~\cite{tian1987smoothness,todorov1989weil} which leads to a pair of conclusions via~\cite{hayakawa1995degeneration,wang1997incompleteness}: i) $\Sigma_j\cong \P^1$ with negative self-intersection (that is, $\Sigma_i\cdot \Sigma_i<0$), and ii) the graph consisting of the $\Sigma_j$ must have positive definite adjacency matrix given by
\begin{equation}\label{eq:adjMatrix}
A_{ij} = -\Sigma_i \cdot \Sigma_j \ .
\end{equation}

To simplify notation, we will often omit the minus signs giving curve self-intersections with the understanding that all self-intersections are negative (e.g., in place of the chain $-3,-2$, writing instead $3,2$). Any two digit self-intersections will be given with parentheses where ambiguous. For example, when writing $1,12,1$ without commas, we shall write $1(12)1.$ 

A key ingredient for our work is the general result of~\cite{atomic}: with few exceptions, every 6D SCFT base in F-theory is of the form
\begin{align}\label{eq:atomicStructure}
S_0 \overset{S_1} g_1 L_1 \overset{I^{\oplus r}}{g_2}L_2g_3L_3\cdots
                           g_kL_kg_{k+1}\cdots \overset{I^{\oplus s}}{g_{m-1}}L_{m-1}\overset{I^{\oplus t}}{g_m}S_m,
\end{align}
where $g_i \in \{4,6,7,8, 9,(10),(11),(12)\}$ are ``DE-type'' nodes (referring to the gauge algebras supported on these curves), $I^{\oplus l}$ are subgraphs of the form $122....2$ consisting of $l$ curves called ``instanton links'', and $S_i,L_i$ are ``side links'' and linear ``interior links'', respectively; attachment to DE-type nodes occurs via the exterior $-1$ curves when possible. Truncations of this general form are also permitted. Briefly, allowed bases are linear chains of curves with branching possible only near the ends. We refer to~\cite{atomic} for the details of two exceptions to the above structure. The first allows a limited class of bases with a single 4-valent curve that are linear away from this curve. The second allows up to four instanton branches for certain bases with precisely five nodes.

We will now review and extend the setup, notation, and terminology introduced in~\cite{BertoliniGlobal1}. We let the compact irreducible effective divisors of~\eqref{eq:adjMatrix}, namely $\Sigma_i\subseteq \{\Delta=0\}$ lie at $\{z_i=0\}$ and designate their self-intersection numbers via $m_i= -\Sigma_i\cdot\Sigma_i$. Let $P_{i,k}$ denote the intersections of $\Sigma_i,\Sigma_k$ for $i\neq k$ when non-empty. 

We shall consider non-compact collections of curves $\{\Sigma_{i,j}'\}_{j\in J_i}$ transversely intersecting $\Sigma_i$ at $P_{i,j}'$ with $\Sigma_{i,j}'\cap \Sigma_k=\emptyset$ for $k\neq i.$  Let $(a_i,b_i,d_i) =(a,b,d)_{\Sigma_i}$ indicate orders of vanishing of $f$, $g$ and $\Delta$ along $\Sigma_i$ and similarly define $(a_{i,j}',b_{i,j}',d_{i,j}')$ those for $\Sigma_{i,j}'.$ Should we embed a neighborhood of the curves $\Sigma_{i}$ in some larger space, it is conceivable that one could derive more stringent requirements on global symmetries arising from the $\Sigma_{i,j}'.$ Nonetheless, configurations in such contexts must still obey the constraints we shall derive which are strictly local in $i$ up to propagation of these local constraints along the compact curves of the quiver (in the sense detailed at the start of Appendix~\ref{s:theRestrictions}) due to purely local analysis of intersections. Consequently, no extra freedom is introduced for example by allowing the $\Sigma_{i,j}'$ to have intersection with multiple $\Sigma_i.$  

For $\Sigma$ any of the curves $\Sigma_i$ or $\Sigma_{i,j}'$ with orders $(a,b,d)= (a,b,d)_{\Sigma}$ we let
\begin{align}
&\ft_\Sigma\equiv {f \over z^a}~,  &\gt_\Sigma &\equiv {g \over z^b}~,    &\Deltat_\Sigma&\equiv{\Delta \over z^d},~
\end{align} 
abbreviating these quantities as $\ft,\gt,\Deltat$ where unambiguous and noting that these are sections of the line bundles
\begin{align}
\label{eq:fgDeltares}
\ft& \leftrightarrow \cO(-4K_B-a\Sigma)~,		&\gt&\leftrightarrow \cO(-6K_B-b\Sigma)~, 	&\Deltat&\leftrightarrow \cO(-12 K_B-d\Sigma)~.
\end{align}
We will refer to $\Deltat$ as the {\em residual discriminant.} When $\Sigma$ is any of the compact divisors $\Sigma_k$, setting $m=m_k$ we have 
\begin{align}
\label{eq:genusrat}
K_B\cdot\Sigma =m-2~,
\end{align}
since $\Sigma\cong \P^1$ with genus $g=0$; we define {\it residual vanishings on $\Sigma$} as the quantities
\begin{align}
\label{eq:resvan}
& \at_\Sigma = (-4K_B -a \Sigma)\cdot \Sigma = -4 (m-2) + ma  ~,\nonumber\\
& \bt_\Sigma = (-6K_B -b \Sigma)\cdot \Sigma = -6(m-2)+mb  ~,\nonumber\\
& \dt_\Sigma = (-12K_B -d \Sigma)\cdot \Sigma = -12(m-2)+md  ~,
\end{align}
which count the number of zeros with multiplicity of the restrictions to $\Sigma$ of $\ft$, $\gt$ and $\Deltat,$ respectively. To improve the na\"ive constraints on $\Sigma_i,\Sigma_{i,j}$ reading
\begin{align}\label{eq:naiveDegCon}
    & (-4K_B-\sum_i a_i \Sigma_i - \sum_{j} a_{k,j} \Sigma_{k,j}')\cdot \Sigma_k \geq 0 \ ,             \nonumber    \\
	&(-6K_B-\sum_i b_i \Sigma_i - \sum_{j} b_{k,j} \Sigma_{k,j}')\cdot \Sigma_k \geq 0 \ ,             \nonumber       \\    
	&(-12K_B-\sum_i d_i \Sigma_i - \sum_{j} d_{k,j} \Sigma_{k,j}')\cdot \Sigma_k \geq 0 \ ,
\end{align}
we begin by defining for an intersection of two curves $\Sigma,\Sigma'$ at $P$ the {\it intersection contributions} from $\Sigma'$ to $\Sigma$ towards the residual vanishings given by the quantities 
\begin{align}
\label{eq:defordP}
\at_P &\equiv \ord_P \ft\big|_{z=0}~,  &\bt_P &\equiv \ord_P \gt\big|_{z=0}~,  &\dt_P &\equiv \ord_P \Deltat\big|_{z=0}~,
\end{align}
abbreviating these as $(\at_P,\bt_P,\dt_P)_{\Sigma}.$ Note that strict inequalities in 
\begin{align}\label{eq:aprioriContributions}
 \at_P &\geq a_{\Sigma'}~,         &\bt_P &\geq b_{\Sigma'}~,     &\bt_P &\geq b_{\Sigma'}~,\nonumber \\
\ord_P f &\geq a_\Sigma + a_{\Sigma'}~, 		&\ord_P g &\geq b_\Sigma + b_{\Sigma'}~,	&\ord_P \Delta &\geq d_\Sigma + d_{\Sigma'} ~,
\end{align}
often follow from local analysis and that non-minimality at the intersection requires one of
\begin{align}
\label{eq:4612}
\ord_Pf&<4~, 		&\ord_Pg&<6~.
\end{align}
Constraints tightening~\eqref{eq:naiveDegCon} using intersection contributions following from local analysis then read
\begin{align}\label{eq:intConConstraints}
   \at_\Sigma &\geq \sum_{j} \at_{P_{k,j}'}+\sum_{i\neq k}\at_{P_{k,i}}=\sum_{i}\at_{P_{i,\Sigma}}~, \nonumber \\ 
   \bt_\Sigma &\geq \sum_{j} \bt_{P_{k,j}'}+\sum_{i\neq k}\bt_{P_{k,i}}=\sum_{i}\at_{P_{i,\Sigma}}~, \nonumber\\
  \dt_\Sigma &\geq \sum_{j} \dt_{P_{k,j}'}+\sum_{i\neq k}\at_{P_{k,i}}=\sum_{i}\at_{P_{i,\Sigma}}~,
  \end{align}
  where $P_{i,\Sigma}$ are relabellings of any intersection points between $\Sigma$ and other components of the discriminant locus, namely $\{\Sigma_j\}_{j\neq i}$ and $\{\Sigma_{k,j}'\}_{j\in J}.$ Further restrictions come from consistency checks for gluing these local models into globally well-defined configurations.
  
We shall employ the terminology of~\cite{BertoliniGlobal1} referring to curves such that $\Sigma$ at $\{z=0\}$ has discriminant of the form
\begin{align}
{\Delta \over z^d} &= \left( 4 \ft^3  + 27z^p \gt^2 \right)~,
\end{align}
for some $p>0$ and $z \nmid \ft$ as having {\it odd} type and indicate the orders of vanishing as $(a,b+B,d)_\Sigma$, where $B=0,1,\cdots$. 
For such curves, the second term in the right hand side vanishes identically upon restriction to $\Sigma$ and hence $\dt_P = 3 \at_P$.
When instead $(a+A,b,d)_\Sigma$, $A=0,1,\cdots$, the residual discriminant has the form
\begin{align}
{\Delta \over z^d} &= \left( 4 z^p \ft^3  + 27 \gt^2 \right)~
\end{align}
for some $p>0,$ and $z \nmid \gt$ making $\dt_P = 2\bt_P,$ we refer to such curves as {\it even} type.
The remaining cases with Kodaira types I${}_n$ and I${}_n^\ast$ are termed {\it hybrid} types, these having both $\ft\big|_{z=0}$ and $\gt\big|_{z=0}$ involved in contributions to vanishings of the residual discriminant.

Note that the analysis of~\cite{BertoliniGlobal1} for single curve theories addressed the general cases having any permitted $A\geq 0$ and $B\geq 0$ via the observation that maximal global symmetry algebras arise for a given Kodaira type when $A=B=0.$ Theories having multiple compact components of $\Delta$ however often require some $\Sigma_i$ have $A>0$ or $B>0$ to realize a maximal configuration making our analysis somewhat more demanding. For example, consider the gauge enhancement with type assignment given by
\begin{align}\label{eq:needAB}
 \underset{(\text{\IVs{}},\su(3))}{3} \ \ \underset{(A,B,0)}{\underset{(\text{\Izero{}},-)}{1}}\ \  [\g_{GS}]\ .
\end{align}
Observe that the maximal global symmetry algebra $[\g_{GS}]$ which can arise is $\e_6$ realized by a type \IVstars{} fiber meeting the $-1$ curve. This requires $A\geq1$ in~\eqref{eq:needAB} by~\eqref{eq:naiveDegCon}. Hence, to search for consistent geometric assignments yielding maximal algebras, it becomes relevant to consider non-zero $A,B.$ In fact, non-zero $A,B$ values are often required by a gauge enhancement before any global symmetry considerations as noted above, e.g.\ the configuration of~\eqref{eq:minimalABEx}. Together, it is hence natural to study nearly all $A,B$ local intersection models in development of a brute force algorithmic approach to finding maximal configurations.
\end{paragraph}

\end{subsection}

\begin{subsection}{Theories with one compact singular locus component}\label{s:singleCurveGS}

We now review details of global symmetry algebra maxima realizable in F-theory for single curve theories. Our approach to treat arbitrary 6D SCFT bases found in F-theory constructions via our algorithm requires also the data consisting of the maximal transverse configurations for single curve theories as detailed in~\cite{BertoliniGlobal1, gaugeless}. Many of these configurations do not lead maximal algebras for single curve theories. They do, however, constrain the transverse configurations we encounter in trying to do determine these maxima for more general bases.

\begin{subsubsection}{Non-trivially gauged theories}
Global symmetries realizable in F-theory for those cases where a single compact curve in the base carries non-abelian gauge algebra first appeared in~\cite{BertoliniGlobal1}. We use these restrictions with a few new tightenings for type III and IV curves that we indicate with a `$\dagger$' symbol in Tables~\ref{t:updatedSummary}. In cases with non-abelian gauge algebra, the Coulomb branch global symmetry predictions from field theory are remarkably close to the constraints we find from F-theory geometry, the latter being more restrictive in some cases.\footnote{As noted in~\cite{BertoliniGlobal1}, results of~\cite{Ohmori:2015fk} constrain field theory global symmetries beyond Coulomb branch gauge anomaly cancellation requirements in some cases including one with $\su(2)$ gauge algebra in which the reduction resolves a mismatch with geometrically realizable global symmetries to bring agreement with predictions from F-theory. Whether yet unknown further constraints on field theory might allow precise matching in all cases remains an intriguing question.} 

\begin{table}[htbp]
\begin{center} 
\begin{tabular}{c|c|c|c}
type	along $\Sigma$ & algebra on $\Sigma$	&	$-\Sigma^2$	& max. global symmetry algebra(s) \\\hline
\multirow{4}{*}{III}	& \multirow{4}{*}{$\su(2)$}			&2	&$\so(7)$\\\cline{3-4}
&	&	\multirow{3}{*}{1}&	$\so(7)\oplus\so(7)\oplus\su(2)$ \\
&	&				&	$\so(7)\oplus\sp(3)\ \ (\dagger)$ \\
&	&				&	$\sp(5)\ \ (\dagger)$ \\
\hline
\multirow{9}{*}{IV} & \multirow{2}{*}{$\su(2)$} & 2	&$\g_2 \ \ (\dagger)$ \\\cline{3-4}
			&		& \multirow{3}{*}{1}		& 	$\g_2\oplus\g_2\oplus\su(3) \ \ (\dagger)$ \\
			   & &                                     & $\g_2 \oplus \sp(2)\ \ (\dagger) $ \\					  
			    & &                                     &  $\sp(3)\ \ (\dagger)$ \\ 	
			\cline{2-4}
    & \multirow{7}{*}{$\su(3)$} 
& \multirow{1}{*}{3} & - \\ \cline{3-4}
&
& \multirow{2}{*}{2}	&$\su(3)\oplus\su(3)$\\
    & & &$\sp(2)$\\\cline{3-4}
    &	&	\multirow{4}{*}{1}&	$\su(3)^{\oplus4}$ \\
    &	&				&	$\su(3)^{\oplus2}\oplus\sp(2)$\\
&	&				&	$\su(3)\oplus\sp(3)$ \\
&	&				&	$\sp(4)$ \\\hline
\end{tabular}
\end{center}
\caption{Global symmetries of gauged F-theory models on Kodaira type III and IV curves.}
\label{t:updatedSummary}
\end{table}

\end{subsubsection}

\begin{subsubsection}{Gaugeless theories}\label{s:gaugelessGSDiscussion}
The global symmetries and maximal transverse configurations which may arise in F-theory models lacking non-abelian gauge algebra (i.e.,\ where the discriminant locus contains only a single trivially gauged compact curve) first appeared in~\cite{gaugeless} as Tables 3,5,7, thus completing a characterization of the geometrically realizable global symmetries for single curve theories. We shall use the fact that a further tightening is possible for an \Ione{} curve; the flavor symmetry maximum coming from an \Instar{} transverse fiber with $n=4$ can be constrained slightly by observing that such a fiber must have monodromy, yielding a reduction for one of the maxima from $\so(16)$ to $\so(15).$ We shall also use the maximal configurations and tabulations of intersection contributions (appearing in~\cite{gaugeless} as Tables 4,6) in these gaugeless cases. Generalizations of this data to cases with $A,B>0,$ are treated in the Appendix~\ref{s:theRestrictions}. Arguments yielding these results appearing in~\cite{gaugeless} use the same approach we take here and in~\cite{BertoliniGlobal1}. We extend this analysis in part for treatment of arbitrary bases since gaugeless compact components of the discriminant locus often can appear in longer bases only if $A,B>0.$ We detail the relevant intersection contributions and forbidden intersections for such curves. This extension plays a key role in our algorithm determining global symmetry maxima consistent with the other restrictions derived in Appendix~\ref{s:theRestrictions}.

A few comments in the case of a single type \Izero{} curve $\Sigma$ may be helpful. The maximal configurations from~\cite{PerssonsList} are those permitted as collections of singular points along $\Sigma,$ but we impose stronger requirement that these arise from a transverse curve configuration. As with many other cases we study here, these maximal configurations may place distinct restrictions on the singularity type of the compact curves they intersect. As an example, among the maximal gauged curve configurations above for a type \Izero{} are $[$\Threestar{},III$]$ and $[$\Fourstars{},\IVs{}$],$ from which the algebras $\e_7 \oplus \su(2)$ and $\e_6 \oplus \su(3)$ arise. These require $B>0$ and $A>0,$ respectively where $(a,b,d)_\Sigma=(A,B,0).$ Since $A,B$ cannot simultaneously be non-zero in type \Izero{}, a given fixed compact component of the discriminant locus with designated orders of vanishing permits only one of these transverse curve collections. In this sense, they can only arise in distinct geometries. This phenomenon is one motivation for tracking the geometric data captured by the orders of $f,g,\Delta$ in our work. In our example, we have a larger algebra in which these are both subalgebras, namely $\e_8.$ We might hastily conclude the geometrically realizable global symmetry algebra for all models with a single \Izero{} curve is then always $\e_8.$ However, when $B>0,$ intersection with any $e_8$ bearing curve is non-minimal. Our approach intends to enable a broader determination of whether the distinct global symmetries we find may arise from {\it distinct SCFTs} with their data specified by the geometry of the fibration at a level of precision beyond specification of the gauge algebra. For this reason and to confirm preliminary existence checks for geometric realizations of each configuration, we further track the orders of $f,g,\Delta$ along each of the compact and non-compact components of the discriminant locus.
\end{subsubsection}

\end{subsection}

\begin{subsection}{Algorithm summary}\label{s:algorithm}
In this section we describe an algorithm we have implemented via series of computer algebra system routines. These routines are intended to allow adaptation for other purposes, though the primary focus in their development is the computation of 6D SCFT global symmetries. There are three main groupings of methods. The first computes 6D SCFT gauge enhancements from geometric considerations. The second handles semisimple Lie algebra inclusion rules. The final grouping determines curve configurations leading to geometrically realizable global symmetry maxima. Several subroutines dual a purpose role in the first and third groupings. The reason for this is that many of the restrictions on which Kodaira types may be paired in curve collisions often hold even when one of the curves is non-compact.

A summary of the algorithm determining gauge enhancements and global symmetries for each enhancement on a quiver given by the values $m_i$ follows.~\footnote{Bases permitting infinitely many enhancements complicate our summary and we shall separate the statements for such quivers. The core algorithm applies identically for such ``rogue bases.'' Since our implementation includes an argument effectively giving an upper bound on the algebra rank permitted for any gauge summand to yield a finite time algorithm through bypassing the infinity of enhancements permitted on a rogue base, awkward summary statements extending those we provide follow in the obvious way with this caveat introduced by such a user determined rank bound.} Certain subroutines are more elaborate than indicated to allow efficiency boosts and result formatting including `sewing' results by combining shorter quivers together to treat longer quivers, storage of partial results during computation, writing data to file for later use and presentation in text, and enabling parallel computation. Since these are non-essential to the underlying algorithm, we will not further discuss these aspects here, instead providing the precise work-flow via inclusion of our implementation with arXiv submission of this note.

Given a quiver $Q$ specified via the values $m_i$, the first leg of the algorithm finds all compatible gauge algebra assignments on $Q$ while tracking the Kodaira types $T_i$ yielding gauge algebras $\oplus_i\g_i.$ For each of these type assignments $\mathcal{T}\sim\{T_i\}$, the second leg determines all geometrically realizable global symmetry maxima for $\T.$ 

This process begins by assigning all possible orders of vanishing $(a,b,d)_{\Sigma_i}$ on each $\Sigma_i$ in $Q$ compatible with naive non-minimality constraints~\eqref{eq:naiveDegCon} up to user specified maximum values $A_{\text{max},T},B_{\text{max},T}$ to be allowed for each Kodaira type $T.$ Each order assignment $\{(a,b,d)_{\Sigma_i}\}$ is then paired with every na\"ively permitted monodromy assignment appearing in Table~\ref{t:singtypes}. Intersection contributions are computed for each $T_i$ in each assignment $\T$ to check~\eqref{eq:intConConstraints} with failing assignments discarded. Any remaining $\T$ are checked against restrictions on pairwise intersections of compact curves. Each length three subquiver is then checked against restrictions on maximal configurations involving such triplets. If $Q$ is a branching quiver, a final check against restrictions on transverse trio configurations mostly derived in~\cite{BertoliniGlobal1,gaugeless} is applied. The resulting list of permitted $\T$ determines the gauge enhancements we allow for the given quiver.

The second leg of the algorithm constrains any geometrically realizable global symmetries by finding transverse configurations $\{\Sigma_{i,j}'\}$ permitted for each assignment $\T.$ This beings by assigning for each fixed $\T$ every transverse collection $\{\Sigma_{i,j}'\}$ of non-trivially gauged non-compact curves, again in two phases with the second involving decoration by a monodromy assignment with constraints on configurations provided by~\eqref{eq:naiveDegCon},\eqref{eq:intConConstraints}, respectively. Each collection is then checked against restrictions on pair intersections, transverse duets and triplet degenerations, and certain larger maximal configurations slightly generalizing those determined in~\cite{BertoliniGlobal1}. The remaining transverse configurations $\{\T'\sim\{\Sigma_{i,j}'\}\}_{\T'\in J_\T}$ determine all possible geometrically realizable global symmetry summands for fixed $\T.$

Each configuration $\T'$ for a fixed $\T$ is then explicitly associated to a global symmetry algebra summand $\g_{\text{GS},\T'}\cong\oplus_{i,j}\g_{i,j}$. Any relatively maximal algebras among $\{\g_{\text{GS},\T'}\}_{\T'\in J_\T}$ are determined via an implementation of semisimple Lie algebra inclusion rules and analysis of maximal semisimple Lie subalgebras from~\cite{dynkin}. These are the global symmetry algebras we permit for the enhancement $\T.$ Any $\T'$ with strictly smaller associated algebras are discarded and the resulting list returned to give the explicit geometric realizations of any global symmetry maxima for $\T.$ In making the maxima comparisons, we put all $\T$ having differing $A_i,B_i$ assignments on the same footing provided the Kodaira types are matching.

A repackaged version of the program provided in conjunction with~\cite{atomic} is included along with methods allowing direct comparisons of enhancements described in the literature and those we compute via other methods enabling tracking of geometric data not previously available. We generally find agreement for the enhancements permitted on links with those of~\cite{atomic} and consequentially also for most 6D SCFTs with a handful of exceptions eliminated via geometric constraints on F-theory bases as detailed in Section~\ref{s:newGaugeRestrictions}. The input $Q$ for our implementation is not confined to links; only the finitely many 4-valent bases are barred. 
\end{subsection}

\begin{subsection}{Flavor summand locality}
We now pause to provide an example illustrating one result of our analysis. Briefly, global symmetry contributions are local in quiver position for fixed Kodaira type but not for fixed gauge algebra. Each assignment $\T_{\alpha}$ for $\alpha\sim2215$ with $\gg\cong\su(2)\oplus\e_6$ appears in Table~\ref{t:GS2215noSimMax}, where n${}_0$ denotes the trivial algebra, flavor summands appear below the curves on which they arise, and the right-most column contains the flavor symmetry maxima for a given type assignment. While we can realize a $\g_2$ global symmetry summand from the left-most $-2$ curve and an $\su(3)$ summand from the $-1$ curve, these options are mutually exclusive, i.e.,\ all global symmetry algebras for an $\su(2)\oplus\e_6$ gauge theory realizable in F-theory via this quiver are strictly smaller than $\su(3)\oplus \g_2.$  

{
    \small\setlength{\tabcolsep}{6pt}
     \begin{table}[htbp] \footnotesize
    \begin{center}
      \begin{tabular}{ccccc}2 & 2 & 1 & 5 & GS Total: \\
(III,\suText<2>{}) & (\II{},\nzero{}) & (\Izero{},\nzero{}) & (\IVstars{},\eText<6>{}) &  \\
   $ A_{1} $  &  $ 0 $  &  $ A_{1} $  &  $ 0 $  &  $ A_{1}^2 $ \\
  (\IVns{},\suText<2>{}) & (\II{},\nzero{}) & (\Izero{},\nzero{}) & (\IVstars{},\eText<6>{}) &  \\
   $ \g_{2} $  &  $ 0 $  &  $ A_{1} $  &  $ 0 $  &  $ A_{1} \oplus \g_{2} $ \\
  (\Itwo{},\suText<2>{}) & (\Ione{},\nzero{}) & (\Izero{},\nzero{}) & (\IVstars{},\eText<6>{}) &  \\
   $ A_{2} $  &  $ 0 $  &  $ A_{2} $  &  $ 0 $  &  $ A_{2}^2 $ \\
  (\suText<2>{}) & (\nzero{}) & (\nzero{}) & (\eText<6>{}) &  \\
   $ A_{2} $  &  $ 0 $  &  $ A_{2} $  &  $ 0 $  &  $ A_{2}^2 $ \\
   $ \g_{2} $  &  $ 0 $  &  $ A_{1} $  &  $ 0 $  &  $ A_{1} \oplus \g_{2} $ \\
 (\II{},\nzero{}) & (III,\suText<2>{}) & (\Izero{},\nzero{}) & (\IVstars{},\eText<6>{}) &  \\
   $ 0 $  &  $ A_{1} $  &  $ 0 $  &  $ 0 $  &  $ A_{1} $ \\
  (\II{},\nzero{}) & (\IVns{},\suText<2>{}) & (\Izero{},\nzero{}) & (\IVstars{},\eText<6>{}) &  \\
   $ 0 $  &  $ \g_{2} $  &  $ 0 $  &  $ 0 $  &  $ \g_{2} $ \\
  (\Ione{},\nzero{}) & (\Itwo{},\suText<2>{}) & (\Izero{},\nzero{}) & (\IVstars{},\eText<6>{}) &  \\
   $ 0 $  &  $ A_{2} $  &  $ 0 $  &  $ 0 $  &  $ A_{2} $ \\
  (\nzero{}) & (\suText<2>{}) & (\nzero{}) & (\eText<6>{}) &  \\
   $ 0 $  &  $ \g_{2} $  &  $ 0 $  &  $ 0 $  &  $ \g_{2} $ 
   \end{tabular} 
       \caption{All global symmetry maxima for $2215$ with gauge algebra $\su(2)\oplus\e_6$ along with each possible Kodaira type assignment to the quiver realizing this gauge algebra.} 
       \label{t:GS2215noSimMax}
          \end{center}
       \end{table}
}        

For a quiver $\alpha,$ we thus fix a Kodaira type assignment $\T_\alpha$ before addressing which global symmetries are geometrically realizable. We can then compare results among all $\T_\alpha$ having shared gauge algebra, as above. Somewhat surprisingly, while various constraints are ``non-local'' in curve position including the permitted orders $(a,b,d)_{\Sigma_i}$ (as illustrated in Table~\ref{t:nonlocalAB}), the $\gG$ maxima for fixed $\T_\alpha$ instead appear to always include every relatively maximal product of any permitted curve contributions. In contrast, this ceases to hold when varying $\T_\alpha$ for $\gg$ fixed as illustrated by the above example.
\end{subsection}

\begin{subsection}{Distinguished Calabi-Yau threefolds from global symmetry maxima}
In this section we introduce a distinguished class of elliptically fibered CY threefolds determined by global symmetry maxima of 6D SCFTs. We examine the role of these symmetries in the field-theory to geometry ``dictionary'' and show that a nearly bijective correspondence results when including these symmetries among the data specifying an SCFT. 

Briefly, the idea is to consider which singular elliptically fibered CY threefolds $\pi: X\to B$ give F-theory models for a 6D SCFT having data $(\gg,\gG,\Gamma),$ where $\gg$ and $\gG$ are gauge and global symmetry Lie algebras, respectively, and $\Gamma$ discrete $U(2)$ subgroup gauge fields. As discussed in~\cite{atomic}, $\Gamma$ determines a unique quiver $\{m_i\}$ that is the minimal blowup of the endpoint associated to $\Gamma$ permitting gauge enhancement given by $\g_{\text{gauge}}.$ Such quivers do not exhaust those for F-theory models with matching gauge content, i.e.,\ dropping $\gG$ from the SCFT data leaves the geometry severely underspecified. Models compatible with $(\gg,\Gamma)$ often allow many choices for $\{m_i\},\T$ with geometrically realizable global symmetry algebras $\{\gG\}$ so different that their common merger at a conformal fixed point upon renormalization would be highly surprising. When instead tentatively regarding $\gG$ as an essential ingredient in specifying a CFT, matching models become so constrained that we find a nearly bijective map from 6D SCFTs to corresponding CY threefolds. The distinguished collection of all threefolds determined via this correspondence from the 6D SCFT landscape appears to be a natural candidate for further study.

We now turn to an example before further discussing this correspondence more generally.
Consider the collection of theories having $\gg\cong\e_6.$ The compatible bases include: 1,2,21,3,31,131,4,41,141,5,51,151,512,1512. The number of curves in the base is not fixed by $\gg,$ nor by also specifying $\Gamma.$ We can say more upon fixing $\gG.$ We consider the case that $\gG\cong\su(3)$ and note that any compatible base has at least two curves since each single curve theory with $\gg\cong\e_6$ gauge algebra has $\gG$ trivial via Table 6.2 of~\cite{BertoliniGlobal1}. We can eliminate the bases $131,141,151,1512$ since each has all $\gg\cong\e_6$ compatible enhancements with $\gG$ being too large. 

We should now specify $\Gamma$ to distinguish between the bases $51,512,$ and others. Consider the $\gg\cong \e_6$ compatible $\T$ on $51,512.$ These appear in Tables~\ref{t:GS51All} and \ref{t:GS512e6}. Upon fixing either base, $\T$ is determined by a choice of $\gG.$ Now we observe that the field theory data $\Gamma$ distinguishes between the remaining $\gG\cong\su(3)$ compatible bases and hence in conjunction with $\g_{\text{global}}$ specifies the geometry uniquely up to quiver and Kodaira type assignment. 
{\small\setlength{\tabcolsep}{6pt}
     \begin{table}[htbp] \footnotesize
    \begin{center}
      \begin{tabular}{ccc}5 & 1 & GS Total: \\
   (\IVstars{},\eText<6>{}) & (\Izero{},\nzero{}) &  \\
    $ 0 $  &  $ A_{2} $  &  $ A_{2} $ \\
   (\IIIstar{},\eText<7>{}) & (\Izero{},\nzero{}) &  \\
    $ 0 $  &  $ A_{1} $  &  $ A_{1} $ \\
   (\IVstarns{},\fText{}) & (\Izero{},\nzero{}) &  \\
    $ 0 $  &  $ A_{2} $  &  $ A_{2} $ \\
   (\IVstarns{},\fText{}) & (\II{},\nzero{}) &  \\
    $ 0 $  &  $ \g_{2} $  &  $ \g_{2} $ \\
   (\IVstarns{},\fText{}) & (\Ione{},\nzero{}) &  \\
    $ 0 $  &  $ A_{2} $  &  $ A_{2} $ \\
   (\fText{}) & (\nzero{}) &  \\
    $ 0 $  &  $ \g_{2} $  &  $ \g_{2} $ \\
             \end{tabular} 
        \caption{All gauge and global symmetry options for 51 with each possible Kodaira type specification realizing a given gauge theory.} 
        \label{t:GS51All}
           \end{center}
        \end{table}
 }
{
    \small\setlength{\tabcolsep}{6pt}
     \begin{table}[htbp] \footnotesize
    \begin{center}
      \begin{tabular} {cccc}5 & 1 & 2 & GS Total: \\
  (\IVstars{},\eText<6>{}) & (\Izero{},\nzero{}) & (\Izero{},\nzero{}) &  \\
   $ 0 $  &  $ A_{2} $  &  $ 0 $  &  $ A_{2} $ \\
  (\IVstars{},\eText<6>{}) & (\Izero{},\nzero{}) & (\II{},\nzero{}) &  \\
   $ 0 $  &  $ A_{1} $  &  $ A_{1} $  &  $ A_{1}^2 $ \\
  (\IVstars{},\eText<6>{}) & (\Izero{},\nzero{}) & (\Ione{},\nzero{}) &  \\
   $ 0 $  &  $ A_{2} $  &  $ A_{1} $  &  $ A_{1} \oplus A_{2} $ \\
  (\eText<6>{}) & (\nzero{}) & (\nzero{}) &  \\
   $ 0 $  &  $ A_{2} $  &  $ A_{1} $  &  $ A_{1} \oplus A_{2} $ 
            \end{tabular} 
       \caption{All global symmetry maxima for 512 with gauge algebra $\e_6$ and each Kodaira type assignment realizing this algebra.} 
       \label{t:GS512e6}
          \end{center}
       \end{table}
} 

This example suggests that it is natural to consider a field-theory/geometry ``dictionary'' relating elliptic fibrations compatible with a choice of field theory data $(\Gamma, \g_{\text{gauge}},\g_{\text{global}}).$ Here $\g_{\text{global}}$ constrains which blowups of the aforementioned minimal $(\gg,\Gamma)$ compatible base should be considered, for example by requiring blowup of a base with a single $-4$ curve to $51$ when $\gg\cong\f_4$ and $\gG$ is non-trivial, or with another $\Gamma,$ from $-3$ to $512.$ Such a dictionary formulation thus gives a natural route from field-theory to the bulk of geometries in correspondence with 6D SCFTs, with the role of $\gG$ essential in accessing the better part of this geometric landscape. The fibration with minimally blown-up base can be viewed as a degenerate case in which we have omitted any $\gG$ specification.

Let us return to our $\gg\cong\e_6$ example, instead now fixing $\Gamma_{-3}$ corresponding to the endpoint $-3.$ Since $\g_{\text{gauge}}$ has a single summand and all curves $\Sigma$ having $m_\Sigma\geq3$ carry non-trivial $\g_\Sigma,$ there can only be one $m\geq3$ curve in any compatible base. All curves must have $1\leq m \leq 6$ since $m_\Sigma \geq7$ requires $\Sigma$ minimally support $\g_\Sigma\cong\e_{\geq7}.$ For the base $-3,$ we have $\gG=0,$ and this is the only such base. The remaining bases with shared endpoint which can match $\gg$ have
\begin{align*}
 \alpha \in \{41,151,512,1612,1\overset{1}{6}1\}\ .
\end{align*}
For $41,$ we have a unique $\gG\cong \su(3).$ From the data above for $51,$ we can deduce that when $\alpha$ given by $151,$ the unique $\gG\cong \su(3)^{\oplus 2}.$ For the only trivalent option, this becomes $\su(3)^{\oplus 3}$ (noting Table~\ref{t:GS61e6}). For $1612,$ options for $\gG$ match those from $512$ after appending an $\su(3)$ summand coming from the outer $-1$ curve. To simplify the correspondence, let us consider only the geometries leading to the maximal $\gG$ on each quiver. This gives for $512,$ $\gG\cong \su(2)\oplus \su(3)$ and for $1612,$ $\gG\cong \su(2)\oplus \su(3)^{\oplus 2}.$ To summarize, $\alpha$ and $\T$ are determined uniquely in each case by $\gG.$ 

Observe that ``special'' CY threefold fibrations are singled out by this ``dictionary,'' namely those which have singular curve collections carrying one of the $\gG$ maxima for a 6D SCFT. Denoting this collection of varieties $\mathcal{M}_{\gG},$ our example summary amounts to concluding that inverting the map 
\begin{align*}
\mathcal{M} _\gG |_{(\gg \cong\e_6,\Gamma_{-3})} \mapsto (\e_6,\gG,\Gamma_{-3})
\end{align*}
from bases with enhancements specified up to Kodaira type to their relatively maximal global symmetries gives an injective map from $\gG$ to $\mathcal{M}_{\gG}.$ Generalizing this statement to all 6D SCFTs is non-trivial due to subtleties including the presence of multiple relative maxima for $\gG.$ Note that there is an analog of this distinguished class of threefolds in the moduli space of compact Calabi-Yau threefolds upon consideration of a compact base giving an F-theory model coupled to gravity wherein the global symmetries we describe are promoted to gauge symmetries via assignment of values $m_i$ to non-compact $\gG$ carrying curves when possible while our transversality requirements are relaxed.

The distinguished threefolds $\mathcal{M}_{\gG}$ in contact with F-theory models are remarkably sparse among CY threefold elliptic fibrations. Consider a fibration with base containing a single compact curve $\Sigma$ and $T_\Sigma\sim\text{\Izero{}}.$ The unique degenerate fiber collection along $\Sigma$ corresponding to the unique flavor symmetry maximum for such models, namely $\gG\cong\e_8$ arising from a transverse curve $\Sigma'$ with $T_{\Sigma'}\sim\text{\Twostar{}},$ corresponds to the a distinguished geometry among the many others appearing as entries of ``Persson's list'' from~\cite{PerssonsList}. Enhancing $\Sigma$ to reach $T_\Sigma\sim\text{\Inns<n>{}}$ for $n$ odd makes the number of transverse configurations grow exponentially in $n$ while only $\frac{n+3}{2}$ geometries are in correspondence with the $\gG$ maxima of Tables (6.1-2) of~\cite{BertoliniGlobal1}. This sparsity is not limited to single curve bases. For example, there are infinitely many minimally enhanced bases with outer links permitting an $\e_8\oplus \e_8$ global symmetry. For each, we have a variety with that flavor symmetry arising in the singular limit which is distinguished from the remaining varieties inducing any of the $6757$ proper $\gG$ subalgebra isomorphism classes.
{
    \small\setlength{\tabcolsep}{6pt}
     \begin{table}[tbhp] \footnotesize
    \begin{center}
      \begin{tabular} {ccc}6 & 1 & GS Total: \\
  (\IVstars{},\eText<6>{}) & (\Izero{},\nzero{}) &  \\
   $ 0 $  &  $ A_{2} $  &  $ A_{2} $ \\
  (\IIIstar{},\eText<7>{}) & (\Izero{},\nzero{}) &  \\
   $ 0 $  &  $ A_{1} $  &  $ A_{1} $ 
   \end{tabular}
        \caption{All gauge and global symmetry options for 61 with each possible Kodaira type specification realizing a given gauge theory.} 
        \label{t:GS61e6}
           \end{center} 
        \end{table} 
 }

\end{subsection}

\end{section}

\begin{section}{Gauge algebras}\label{s:newGaugeRestrictions}
In this section we discuss gauge algebra assignments for each base permitted via~\cite{atomic} though forbidden via the geometric constraint algorithm outlined in Section~\ref{s:strategy}. We also outline our method for comprehensive comparison of gauge enhancements permitted via~\cite{atomic} versus our algorithm. The latter are strictly more constrained with a minority of cases excluded.
\begin{subsection}{Link enhancements}
We now inspect the link gauge assignments compatible with some explicit Kodaira type specification meeting our geometric constraint algorithm. We compare our prescriptions for links with those determined in~\cite{atomic} and then extend to a comparison for general 6D SCFT bases via the node attachment restrictions of~\cite{atomic}.
\begin{subsubsection}{Consequences of $\so(13)$ global symmetry constraints}\label{s:so13}
An $\so(13)$ gauge summand can only be carried on a curve of negative self-intersection $m=2$ or $m=4.$ The F-theory global symmetry from~\cite{BertoliniGlobal1} for such a curve, $\sp(5),$ is independent of $m,$ while the Coulomb branch global symmetry is given by $\sp(9+\Sigma^2).$ These agree for $m=4,$ but the discrepancy for $m=2$ leads to gauge enhancements for a family of bases which are more constrained than those characterized in~\cite{atomic}. In particular, gauge enhancements for quivers which are truncations of $21414\cdots$ containing the link $21$ beginning $\so(13),\sp(6\leq l\leq 7),\cdots$ are eliminated.
\end{subsubsection}
\begin{subsubsection}{Further link enhancement restrictions}
We have carried out a comprehensive comparison between the link gauge enhancements permitted by our approach versus those prescribed in~\cite{atomic}. The few discrepancies between these prescriptions are discussed in this section. Our comparisons are drawn after adjustments to the gauge enhancement prescription algorithm implementation accompanying~\cite{atomic} aimed to make these fully consistent with the gauge algebra constraints of~\cite{atomic} and identification of $\su(2)$ and $\sp(1)$.\footnote{These edits resolve mismatches with prescriptions of~\cite{atomic} for the bases $13,$ $213,$ and $2\overset{1}{3}1$ (affecting results for longer quivers) resolved by adjustments bringing listings for these into agreement with underlying gauging rules. Details of edits appear in the workbook accompanying the arXiv submission of this note in the subroutines for those quivers.} We find agreement for all link enhancements except those appearing in Table~\ref{t:linkDiscrepancies} or contained in the family detailed in Section~\ref{s:so13}. We now briefly discuss geometric elimination of the former.
     \begin{table}[!h] \small
    \begin{center}
      \begin{tabular} {ccccccccccccccc}
    Enhancement & Permitted via~\cite{atomic} & Geometrically realizable \\
    $\underset{\sp(1)}{2} \ \ \overset{\underset{\sp(1)}{2}}{\underset{\so(7)}{3}} \ \ 1 \ \ \underset{\so(9)}{3}$ &  $\checkmark$ & X \\ \\
    $\underset{\text{\gText{}}}{3} \ \  \underset{}{1} \ \  \overset{\substack{\underset{\text{\gText{}}}{3} \\ \underset{\ }{1}}}{\underset{\text{\fText{}}}{5}} \ \  \underset{}{1} \ \  \underset{\text{\gText{}}}{3}$&  $\checkmark$ & X \\ \\
    $\vdots$ &     $\vdots$ &     $\vdots$ \\ \\
    $\underset{\sp(1)}{2}\ \  \underset{\text{\gText{}}}{3} \ \  \underset{}{1} \ \  \overset{\substack{\underset{\text{\gText{}}}{3} \\ \underset{\ }{1}}}{\underset{\text{\fText{}}}{5}} \ \  \underset{}{1} \ \  \underset{\text{\gText{}}}{3} \ \ \underset{\sp(1)}{2} \ \ 2$&  $\checkmark$ & X  
   \end{tabular}
        \caption{Discrepancies with gauge enhancements prescribed via~\cite{atomic}.} 
        \label{t:linkDiscrepancies}
           \end{center} 
        \end{table}
\begin{paragraph}{No $\g_2$ trifecta branching from $\f_4$}\label{s:noG2Trifecta}
Certain enhancements permitted via~\cite{atomic} contain the configuration
\begin{align}\underset{\text{\gText{}}}{3} \ \  \underset{}{1} \ \  \overset{\substack{\underset{\text{\gText{}}}{3} \\ \underset{\ }{1}}}{\underset{\text{\fText{}}}{5_\Sigma}} \ \  \underset{}{1} \ \  \underset{\text{\gText{}}}{3} 
\end{align}
\noindent on a sub-quiver. This can be excluded via the following geometric considerations.

For a gaugeless curve to support an \fText{},\gText{} neighbor pair, it must be a type II curve. This leaves only
\begin{align}
    \underset{\text{\gText{}}}{3} \ \ \  \underset{\text{II}}{1} \ \ \  \overset{\substack{\underset{\text{\gText{}}}{3} \\ \\  \underset{\text{II} }{1}\\ \\} }{\underset{\text{\fText{}}}{5}} \ \ \ \underset{\text{II}}{1} \ \ \ \underset{\text{\gText{}}}{3}  
\end{align}
as an assignment of Kodaira types potentially realizing this enhancement. However,  this too is ruled out since $\dt_\Sigma=\deg(\Deltat_\Sigma)=4$ along a $-5$ curve $\Sigma$ with \fText{} algebra while each type II contributes two vanishings of $\Deltat_\Sigma.$

This in turn eliminates each enhancement from~\cite{atomic} containing the above gauge assignment, these being the five truncations of
\begin{align}
    \underset{\sp(1)}{2}\ \  \underset{\text{\gText{}}}{3} \ \  \underset{}{1} \ \  \overset{\substack{\underset{\text{\gText{}}}{3} \\ \underset{\ }{1}}}{\underset{\text{\fText{}}}{5}} \ \  \underset{}{1} \ \  \underset{\text{\gText{}}}{3} \ \ \underset{\sp(1)}{2} \ \ 2
\end{align}
\noindent obtained by removing any choice of $-2$ curves.
\end{paragraph}

\begin{paragraph}{Eliminating a $2\overset{2}{3}13$ enhancement}
Enhancing the quiver $2\overset{2}{3}13$ to yield the enhancement 
\begin{align}
\underset{\sp(1)}{2} \ \ \  \overset{\underset{\sp(1)}{2}}{\underset{\so(7)}{3_{\Sigma'}}} \ \ \ 1_\Sigma \ \ \ \underset{\so(9)}{3}
\end{align}
appears to be permitted by~\cite{atomic} gauging requirements and is among the companion workbook enhancement listings. However, geometric considerations eliminate this enhancement.

We proceed by inspecting which Kodaira types might be permitted to realize the relevant gauge algebra assignments. The $-2$ intersections with an $\so(7)$ algebra carrying $-3$ curve requires these have Kodaira type III. This implies that of the $6$ residuals in $\Deltat_{\Sigma'}$ available along the $-3$ curve, we have used three for each $-2$ curve intersection. This leaves no residual vanishings for the $-1$ curve to carry a Kodaira type other than \Izero{}. However, a type \Izero{} assignment is not permissible since $6+7$ vanishings of $\Deltat_\Sigma$ are required along this curve to support intersections with the neighboring curves which must have Kodaira types \Izerostarss{} and \Instarns<1>{} (as we can read from their algebra content) making their $\dt_\Sigma$ contributions to the $-1$ curve at least $6$ and $7$, respectively.
\end{paragraph}

\end{subsubsection}

\begin{subsubsection}{Summary of link comparisons}
After compensating for the aforementioned issues, we find agreement with the gauge enhancement structure on all links with that of~\cite{atomic}. In other words, there is some Kodaira type assignment and chosen orders of $f,g,\Delta$ along each curve of the link that meets all geometric constraints known to us and realizes each link enhancement dictated via the auxiliary computer algebra workbook listings of~\cite{atomic} after the minor aforementioned edits with the above eliminated enhancement exceptions.
\end{subsubsection}
\end{subsection}

\begin{subsection}{Comprehensive enhancement comparison}
We now compare enhancements for bases with nodes we obtain with previous results. After accounting for a few technical exceptions, the enhancements we permit match those of~\cite{atomic}. We begin with a summary of comparisons for single node attachments to a link and then turn to discuss attaching a pair of nodes to an interior link.
\subsubsection{$\e_6$, $\e_7$ and $\e_8$ attachments}
Enhancements of bases formed by left attachment of an $\e_7$ or $\e_8$ node to a link of the form $L\sim 1223\cdots$ yield matches as do instanton links (taking the form $122\cdots$) with $\e_8$ node attachment. Explicit comparisons for the latter are made awkward by differences in the handling of infinite link enhancement families but can be treated by confining listings of~\cite{atomic} to those with empty gauge summand on the two leftmost curves and the rightmost matching a corresponding term from our listings.

We also find agreement for left $\e_6$ and $\e_7$ attachments to links of the form $123\cdots,$ $1223\cdots,$ and $122\cdots.$ Comparisons for the latter can be made with a procedure analogous to $\e_8$ instanton link case, here instead via restriction to link enhancements obeying the gauging condition of~\cite{atomic} making the leftmost $-2$ curve empty, $\sp(1),$ or $\su(3)$ (with the latter only for $\e_6$ attachment).

Excluding the links $1\overset{2}{2}3$ and $122315131$ discussed in Section~\ref{s:novelAttachmentRestrictions}, explicit comparisons yield agreement for the remaining branching links meeting $\e_{\geq6}$ curves, thus confirming all link attachment prescriptions of~\cite{atomic} for a single compact curve carrying an $\e_{\geq6}$ algebra with exceptions noted above. 
\subsubsection{Attachments to an interior link}
Gauging rules of~\cite{atomic} for interior links with node attachments match those following from our approach in all cases except for exclusion of a forbidden node pair to $122315131$ discussed shortly. Matching for cases with attachment of a $-6$ or $-4$ curve to the link $1315131$ and all allowed attachments to $131513221$ or $13151321$ can be confirmed by comparison of these gauging rules with our listings for each quiver of this form. Agreement for enhancements of node attachments to the links $12231,\ 12321,$ and $1231$ follows from link enhancement agreement and inspection that our prescriptions respect those attachment gauging rules as does that for $131$, though checks for the latter are involved as specification of enhancements permitted by~\cite{atomic} require supplementing attachment rules with convexity conditions.

These checks can be extended to confirm that all bases with interior links and no side-links having up to two nodes except bases of the form $(1)4141\cdots$ yield matches.\footnote{This accounts for two exceptions we have noted elsewhere. The first concerns $12231513221,$ which does not permit $(\e_6,\e_6)$ attachments as a consequence of blowdowns induced on neighboring nodes, namely $(5,5).$ The second concerns attachment of a $-4$ curve to $122315131$ curve with $\so(N)$ with $8\leq N\leq 12$, which appears to be permitted.} In fact, after matches confirmed in the following subsections, we can conclude that with the aforementioned exceptions for noble branching link discrepancies, $\so(13)$ caveats, certain pairings for $122315131,$ and bases with node attachment to $1\overset{2}{2}3,$ we find agreement for all gauge enhancements. This follows from checks on multi-node bases revealing no further eliminations. Whether stronger geometric restrictions on short quivers can be derived to more significantly reduce the 6D SCFT landscape via our algorithm remains an intriguing question for future work.

\subsubsection{Enhancements on quivers of the form $(2)(1)414\dots$}
Excluding the restriction discussed in Section~\ref{s:so13} above, our method yields matching enhancements for quivers of the form $(2)(1)414\cdots$ with those prescribed via~\cite{atomic}. Carrying out this check is delicate as each such quiver permits infinitely many enhancements. However, these obey a simple set of rules determined in~\cite{atomic} which match the restrictions on length three sub-quiver gauge algebra assignments dictated by geometric global symmetry restrictions excluding the aforementioned $\so(13)$ caveat. 

To explicitly confirm matching away from those special cases detailed in Section~\ref{s:so13} using our algorithm, we begin by fixing a quiver in this family and choosing an upper bound on gauge summand rank. Listing all compatible enhancements and discarding those having summands matching the rank bound allows confirmation via inspection that remaining enhancements obey the corresponding gauging conditions of~\cite{atomic}. Checks through large rank and quiver length reveals the claimed matching.

\begin{subsubsection}{Novel links and link attachment restrictions}\label{s:novelAttachmentRestrictions}
The links $1\overset{1}{5}13215$ and $31\overset{1}{5}1315$ appear to be allowed, though absent from the listings of~\cite{atomic}. These blow-down consistently and permit multiple valid gauge assignments including options with $\f_4$ on the rightmost $-5$ curve. It thus appears that each is a properly a noble link rather than its truncation (of the outer $-5$ curve) being an alkali link permitting only right $\e_{6}$ or $\e_7$ attachment.

Among the links listings of~\cite{atomic} is $3\overset{2}{2}1_\Sigma$ with indicated attachments for (only) $\e_6,\e_7,$ though neither appears to be permitted. The $\e_6$ algebra is not possible, since this requires attachment to a curve with $m\leq 6$ which then cannot satisfy the adjacency matrix condition. For $\e_7$ algebra, we have $m\leq 7$ similarly barred, leaving $m=8$ as the only potentially consistent option. However, this is also barred as after one blow-down we reach
\begin{align}
3\overset{2}{1}7,
\end{align} 
which is inconsistent with the normal crossings condition.
We conclude this link is not permitted any attachments (making it a noble link), though an $\e_6$ global symmetry can arise from $\Sigma.$
\end{subsubsection}

\begin{subsubsection}{Link and attachments summary}
Novel links and link attachment prescriptions (comparing with~\cite{atomic}) appear in Table~\ref{t:linkFixes}.
{ \small\setlength{\tabcolsep}{5pt} \begin{table}[!h] \begin{center}\begin{tabular}{cccccc}
$3\overset{2}{2}1$ & $1\overset{1}{5}13215$ & $31\overset{1}{5}1315$ & $12231513221$ & $122315131$\\
 no attachments & $\checkmark$ & $\checkmark$ & $\underset{k_l+k_r>0}{(\e_{\geq 6+k_l};\e_{\geq 6+k_r})}$ & $(\e_{\geq 6};\ \f_4,\e_{\geq 6} \text{ or } \underset{8\leq N\leq 12}{\so(N)})$
\end{tabular}
\caption{Summary of novel links and link attachment prescriptions versus those~\cite{atomic}. Here `$\checkmark$' indicates a link that appears to be allowed though not listed in~\cite{atomic}.}
\label{t:linkFixes}
\end{center} \end{table} } 

\end{subsubsection}

\end{subsection}

\begin{subsection}{No trios of branching side-links, implementation scope}
Our implementation of the algorithm outlined in Section~\ref{s:algorithm} does not treat 4-valent bases. The only bases not treated directly are the single instanton link decorations 
\begin{align}
S_L \underset{I^{\oplus n}}{\overset{S_U}{\g}} S_R
\end{align}
of single node bases which are treated directly, these being of the form
\begin{align}
S_L \overset{S_U}{\g} S_R \ .
\end{align}
Our implementation also does not treat branching from any vertical branches, but this does not impose any limitations. While {\it a priori} the classification of~\cite{atomic} allows for a pair of branching side-links $S_L,S_U$ meeting a node which then joins the backbone giving
\begin{align}
S_L \overset{S_U}{\g} L_{R} \cdots \ ,
\end{align}
this situation is never encountered in practice. To check this, we confirm that $-12$ is the only possible node allowing a pairs of branching side links and an interior link $L_R$. In this case, we can only have $L_R$ be interior link $-1$ and no attachments to this link are possible. In fact, there is only one such base and it can be rewritten in the form 
\begin{align}
S_L \overset{1}{(12)} S_R \ ,
\end{align}
where $S_L$ is given by $1\overset{1}{5}13221$ and $S_R$ is the reverse of this link.
\end{subsection}
\end{section} 

\begin{section}{Global symmetry classification summary via local contributions}\label{s:GSRules}
Since flavor symmetries for bases with a single compact singular locus component were characterized in~\cite{BertoliniGlobal1,gaugeless}, we are left to contend with bases containing at least two compact curves. In this section, we detail analogous results capturing the general case obtained via the algorithm of Section~\ref{s:algorithm} by dictating the flavor summand maxima arising from each segment of a base with specified enhancement. 

Though curves $\Sigma_i$ in a base may have any of the values $1\leq m_{\Sigma_i} \leq 8$ or $m_{\Sigma_i}=12,$ length two chains are highly constrained with the only options being $\alpha \in \{1k \text{ for }k>1,\ 22, 23\}.$ We shall use this fact to characterize flavor summands arising from each curve of every viable short subquiver decorated with a Kodaira type assignment. This yields a classification of 6D SCFT flavor symmetries via decomposition of an arbitrary base into short chains for which we prescribe summands with a combination of configuration listings and short constraint equations. In many cases, these tighten the permissible symmetries beyond those permitted by the free multiplet counts for the cases detailed in~\cite{atomic}.

A fact we shall use frequently is that any curve carrying gauge algebra $\f_4,\e_6,\e_7,$ or $\e_8$ does not support any non-abelian global symmetry summand. As elsewhere in this note, bracketed terms indicate global symmetry summands. Note that for a curve $\Sigma$ with $7\leq m_{\Sigma}\geq 8,$ only the assignment \IIIstar{}$\leftrightarrow\e_7$ is permitted; for $6\leq m_{\Sigma}\geq 8,$ type \IVns{}$\leftrightarrow\f_4$ is barred while $\e_{6\leq k\leq7}$ types are permitted; for $m_{\Sigma}\leq 4,$ each of $\f_4$ and $\e_{6\leq k\leq 7}$ assignments are valid; finally, $\e_8$ assignment requires $m_{\Sigma}=12.$ We shall use these facts to treat multiple quivers simultaneously and refer to them as ``the permitted gauging rules.''

It will be convenient to introduce notation $[\g_{\alpha}]$ for flavor symmetry summands arising from a subquiver $\alpha\subset\beta$ (e.g.\ $23\subset 232$) upon fixing a type assignment on $\beta$ given by $\T_\beta,$ taking $\beta=\alpha$ when no containing quiver is apparent from context. For example, since the type assignment $\T\sim\T_{23}\sim\text{III,\Izerostarss{}}$ to $23$ allows an $\su(2)$ flavor summand from the $-3$ curve via~\eqref{eq:GS23}, we shall write $[\g_{23,\T}]\cong\su(2)]$ or simply $[\g_{23}]\cong \su(2)$ when context makes $\T$ unambiguous. We will similarly refer to flavor summands arising from a given curve $\Sigma$ by $[\g_{\Sigma}]$ when context makes the containing type assignment clear, e.g.\ in $23_{\Sigma}$ with $T_{\Sigma}\sim\text{\Izerostarss{}},$ we have $[\g_{\Sigma}]\cong\su(2).$ 
\begin{subsection}{$23$}
The only permissible Kodaira type assignments for the bare quiver $23$ are
\begin{align}\label{eq:GS23}
\begin{array}{cccccc}
     & 2 & & 3_\Sigma \\
     &\underset{\su(2)}{\text{III/\IVns{}}} &\qquad \quad  &\underset{\g_2}{\text{\Izerostarns{}}} & \\ [1.4em]
     &\underset{\su(2)}{\text{III}} & &\underset{[\su(2)]}{\underset{\so(7)}{\text{\Izerostarss{}}}} & \ ,
\end{array}
\end{align}
noting that the indicated $[\g_{\Sigma}]$ may be further constrained by the presence of an additional neighboring curve $\Sigma'$ which must have $1\leq m_{\Sigma'}\leq2.$ Resulting reductions of $[\g_{\Sigma}]$ are dictated by the Kodaira type $T_{\Sigma'}$ on $\Sigma'.$ When $m_{\Sigma'}=2,$ a unique assignment
\begin{align}\label{eq:GS232}
\begin{array}{ccccccc}
     & 2 & \qquad \quad & 3 & \qquad \quad & 2\\
     &\underset{\su(2)}{\text{III}} & &{\underset{\so(7)}{\text{\Izerostarss{}}}} &&\underset{\su(2)}{\text{III}} & \ 
     \end{array}
\end{align}
results and $[\g_{232}]$ is trivial. When $m_{\Sigma'}=1,$ $[\g_{23}]$ is trivial provided $T_{\Sigma'}$ is non-trivially gauged or type II. When $T_{\Sigma'}\in\{\text{\Izero{}},\text{\Ione{}}\}$, $[\g_{\Sigma}]\cong\su(2)$ persists.

\end{subsection}

\begin{subsection}{$(12)1,81,71,61,51$}\label{s:51}
Since the unique gauge assignment for $(12)1$ has no flavor summand arising from either curve and a unique Kodaira type assignment given by
\begin{align}
\begin{array}{cccccc}
  & 12 & \qquad \quad & 1 &\\
& \underset{\e_8}{\text{\IIstar{}}} & & \underset{-}{\text{\Izero{}}} & \ ,
\end{array}
\end{align}
neighboring curves are irrelevant to prescribe contributions from this curve pair. 

The permitted configurations for the bare bases $m1$ with $5\leq m\leq 8$ are
\begin{align}\label{eq:GS51}\arraycolsep=9pt%
\begin{array}{cccc}
           5\leq  m\leq 8 & 1 \\
     \underset{\e_7}{\text{\IIIstar{}}} & {\text{\Izero{}}} & [\su(2)] \\     
     \underset{\f_4}{\text{\IVstarns{}}}/\underset{\e_6}{\text{\IVstars{}}} & {\text{\Izero{}}} & [\su(3)] \\      
     \underset{\f_4}{\text{\IVstarns{}}} & \text{\Ione{}} & [\su(3)] \\
     \underset{\f_4}{\text{\IVstarns{}}} & \text{II} & [\g_2] & \ .
\end{array}
\end{align}
Note that the permitted gauging rules constrain the allowed values of $m$ in various cases. This condition will be implicit in the further listings with indeterminate $m$ in this section. Since flavor summands which can arise from the curve pairs $m1$ in longer quivers depend on Kodaira types of right neighboring curves, we next detail the effect of these attachments while observing the irrelevance of any left attachment.
 
Via~\cite{atomic}, the only forms for links which can attach to an $\e_6$ node are
\begin{align}
 12\overset{\tiny \vdots}{3}\cdots\ , \quad 1223\cdots\ , \quad  \text{  or  } \quad 12(2)(2) \cdots \ .
\end{align} 
We proceed through the type assignments for these links compatible with the presence of node with $7\leq m\leq 8$ and detail flavor summands arising from the subquiver $m1$ for each. All remaining links allow only $m\leq6$ attachment, these being of the form $13\cdots.$ Briefly, a $-m$ curve carrying $\f_4$ or $\e_6$ gauge have an $\su(N\leq 3)$ maximum for $[\g_{\Sigma}]$ in $m1_\Sigma\cdots$ with $N$ dependent on other curves attached $\Sigma;$ for $m1\cdots$ with $m\geq 7,$ this is reduced to $\su(2).$ Note that $T_{\Sigma}\sim\text{\Izero{}}$ is required for $m\geq 7$ as intersection with \IIIstar{} is otherwise non-minimal. 

\begin{itemize}
\item  $12\overset{\tiny \vdots}{3}\cdots:$ Attaching a $-7$ or $-8$ curve $\Sigma$ to a link of this form requires the $-2$ curve have $\su(2)$ gauge summand from Kodaira type III. When $7\leq m\leq 8,$ $[\g_{m1}]$ is trivial. The only difference when attaching a curve with $5\leq m\leq 6$ is that we can realize this $\su(2)$ gauge summand via \IVns{}. All configurations are among
\begingroup
\renewcommand*{\arraystretch}{1.2}
\begin{align}\arraycolsep=9pt%
\begin{array}{ccccccc}
           5\leq  m\leq 8 & 1 & 2 & 3 \\
     \text{\IIIstar{}/\IVstars{}/\IVstarns{}} & \text{\Izero{}} & \text{III} & \underset{[\su(2)]}{\text{\Izerostarss{}}} &  \\ 
     \text{\IIIstar{}/\IVstars{}/\IVstarns{}} & \text{\Izero{}} & \text{III} & \text{\Izerostarns{}} \\
      \text{\IVstars{}/\IVstarns{}} & \text{\Izero{}} & \text{\IVns{}} & \text{\Izerostarns{}} \\
      \text{\IVstarns{}} & \text{II} & \text{\IVns{}} & \text{\Izerostarns{}} & . 
\end{array}
\end{align}
\endgroup

\item $1223\cdots:$ Link attachments of this form to a curve $\Sigma$ with $5 \leq m_{\Sigma}\leq 8$ require $[\g_{m1}]$ for $\Sigma$ enhanced to $\e_7$ as is required when $m_{\Sigma}\geq 7.$ Otherwise an $\su(2)$ flavor summand can occur. All potentially viable configurations are
\begingroup
\renewcommand*{\arraystretch}{1.2}
\begin{align}\arraycolsep=9pt%
\begin{array}{ccccccc}
           5\leq  m\leq 8 & 1 & 2 & 2 & 3 \\
     \text{\IIIstar{}} & \text{\Izero{}} & \text{II} & \text{\IVns{}} & \text{\Izerostarns{}} \\
     \text{\IVstars{}/\IVstarns{}} &\underset{[\su(2)]}{\text{\Izero{}}} & \text{II} & \text{\IVns{}} &  \text{\Izerostarns{}} & . 
\end{array}
\end{align}
\endgroup

\item $12_\Sigma(2)(2) \cdots:$ When $\Sigma$ has non-trivial gauge summand, $[\g_{m1}]$ is trivial as required by~\cite{PerssonsList,gaugeless}. For gaugeless $\Sigma,$ instead $[\g_{m1}]$ are constrained by $T_{\Sigma}$ according to
\begin{align}\arraycolsep=9pt%
\begin{array}{cccc}
           5\leq  m\leq 8 & 1 & 2 \\
     \underset{\e_7}{\text{\IIIstar{}}} & \underset{[\su(2)]}{\text{\Izero{}}} & \text{\Izero{}/\Ione{}} \\ 
     \underset{\e_7}{\text{\IIIstar{}}} & \underset{}{\text{\Izero{}}} & T_{\Sigma}\notin \{\text{\Izero{},\Ione{}}\} \\      
     \underset{\f_4}{\text{\IVstarns{}}}/\underset{\e_6}{\text{\IVstars{}}} & \underset{[\su(3)]}{\text{\Izero{}}} & \text{\Izero{}/\Ione{}} \\      
     \underset{\f_4}{\text{\IVstarns{}}}/\underset{\e_6}{\text{\IVstars{}}} & \underset{[\su(2)]}{\text{\Izero{}}} & \text{II} \\      
     \underset{\f_4}{\text{\IVstarns{}}}/\underset{\e_6}{\text{\IVstars{}}} & \underset{}{\text{\Izero{}}} & T_{\Sigma}\notin \{\text{\Izero{},\Ione{},II}\} \\       
     \underset{\f_4}{\text{\IVstarns{}}} & \underset{[\su(2)]}{\text{II}} & \text{II} \\  
     \underset{\f_4}{\text{\IVstarns{}}} & \underset{[\su(2)]}{\text{\Ione{}}} & \text{\Ione{}} & .
\end{array}
\end{align}
All configurations for the bases $m12$ with $m\geq5$ appear in Table~\ref{t:GS512}. Note that attachment of additional $-2$ curves does not ensure further constraints except when their types require $T_\Sigma$ be raised beyond \Ione{}.

\item $1_\Sigma3\cdots:$ Attaching links of this form to $\Sigma'$ requires $m_{\Sigma'}\leq 6$ and $[\g_{m1}]$ being trivial. This follows since the unique assignment to $613$ is \Fourstars{}\Izero{}\IVs{}, which leaves no further residual vanishings of $\Deltat_\Sigma$ along $\Sigma.$ The only possible assignments for $m13$ with $5\leq m\leq 6$ are
\begin{align}\label{eq:GS513}\arraycolsep=12pt%
\begin{array}{ccccccccc}
     & 5\leq m\leq 6 & 1 & 3\\
     &\underset{\e_6/\f_4}{\text{\IVstars{}/\IVstarns{}}} & \text{\Izero{}} & \underset{\su(3)}{\text{\IVs{}}}  \\ [1.3em]
     &\underset{\f_4}{\text{\IVstarns{}}} & \text{II} & \underset{[\su(2)]}{\underset{\g_2}{\text{\Izerostarns{}}}} & . 
\end{array}
\end{align}
\end{itemize}
Note that the above treatment also captures each possible linking to a curve with $m\leq4$ enhanced to one $\f_4,\e_{6\leq k\leq 7}$ provided we substitute $m=5$ in the above discussion with the appropriate value of $m$ and consider only links consistent with the number of blowdowns permitted by $m$ as detailed in~\cite{atomic} Appendix D. (For $m=4,$ links of each form discussed are permitted.)
\end{subsection}

\begin{subsection}{$41$}\label{s:41rules}

Our discussion here more involved since $-4$ node permits gauged $-1$ neighbors. We first discuss the highly constrained configurations for certain bases with subquiver $4_\Sigma 1_{\Sigma'}.$ As there are infinitely many enhancements of the base $41,$ determining $[\g_{41}]$ presents certain difficulties. However, there is little freedom for the Kodaira types of curves permitted intersect a $-4$ curve while maintaining non-minimality. The only infinite family of enhancements have the form $\so(N),\sp(N')$ and flavor summands for these enhancements can be characterized by a few simple conditions with the rest easily handled by inspection.

Non-trivial $[\g_{\Sigma}]$ may arise depending on the neighboring Kodaira type $T_{\Sigma'}$ provided $T_{\Sigma}\leftrightarrow\so(N).$ The only compatible gauged types are \Inns<n>{} curves carrying $\sp(N')$ algebras. A constraint on $[\g_{\Sigma}]$ accounting for any neighboring curves can be derived via simple tallying conditions for contributions towards the residual vanishings $\dt_\Sigma.$ Simple conditions capturing this constraint treat all but finitely many cases which we resolve first by working through the permitted attaching links and discussion of the bare base $41.$ We confine ourselves to $T_{\Sigma}\leftrightarrow\so(N)$ since $\f_4$ and $\e_{\geq 6}$ assignments are captured in Section~\ref{s:51}.

The only permitted right attachments to a $-4$ curve are among
\begin{align}\label{eq:141ConvexityCondition}
1\overset{2}{3}2 \ , \quad 123\cdots \ , \quad 1223 \ , \quad 13\cdots \ , \quad 14\cdots \ ,  \quad \text{ and } \quad 12(2) \ .
\end{align}
We proceed through these possibilities, in each turning to consider simultaneous left attachments after discussing the bare cases, i.e.,\ those without a left attachment to $\Sigma.$
\begin{itemize}
\item $41\overset{2}{3}2:$ This base has a single type assignment given by 
\begin{align*}
 \underset{\text{(\Izerostars{},\soText<8>{})}}{4} \ \ \ \underset{\text{(\Izero{},\nzero{})}}{1} \ \ \ \overset{\underset{\text{(III,\suText<2>{})}}{2}}{\underset{\text{(\Izerostarss{},\soText<7>{})}}{3}} \ \ \  \underset{\text{(III,\suText<2>{})}}{2} 
\end{align*}
with trivial flavor symmetry from all curves, thus making left attachments irrelevant.

\item $412_{\Sigma_2}3\dots:$ Here $[\g_{41}]$ is trivial except for gaugeless $\Sigma',$ as we can read from~\eqref{eq:GS232} and Table~\ref{t:updatedSummary}. This is also trivial when $\g_{\Sigma}$ is $\f_4$ or $\e_{\geq6},$ and again when $\g_{\Sigma}\cong\so(8)$ and $T_{\Sigma'}\sim\text{II};$ we can confirm the latter using the maximal configurations for type II and \Izerostars{} curves from~\cite{gaugeless} and~\cite{BertoliniGlobal1}, respectively. 

The remaining gaugeless types, \Izero{} and \Ione{} arise in few configurations. For $\g_{\Sigma}\cong\so(8),$ these have a $-1$ curve summand $[\g_{\Sigma'}]\cong\su(3)$ when $T_{\Sigma_2}\sim\text{III}$ and $[\g_{\Sigma'}]$ otherwise trivial when $T_{\Sigma_2}\sim\text{\IVns{}}$. In the remaining cases $T_{\Sigma}\leftrightarrow\so(N)$ for $9\leq N \leq 12$ with $[\g_{\Sigma'}]$ is trivial for $11\leq N\leq 12.$ When $9\leq N\leq 10$, $[\g_{\Sigma'}]$ is trivial for $T_{\Sigma_2}\sim\text{\IVns{}}$ otherwise has maximum for $T_{\Sigma_2}\sim\text{III}$ given by $\su(2).$ Finally, the $-4$ curve summand $[\g_{\Sigma}]\cong \sp(N-8)$ except for the unique assignment to the quiver $4123$ featuring a type \Ione{} curve along $\Sigma'$ when is reduced to $[\g_{\Sigma}]\cong \sp(1).$

\item $413\dots:$ Since the quiver $413$ permits a (large but) finite number of enhancements, all configurations with flavor symmetry maxima indicated can be listed directly with the accompanying workbook.\footnote{To provide a listing including all enhancements, the parameter restricting $N$ in $\so(N)$ gauge summands, namely `maxNForInstar' should be set to allow $N\geq 24$ (i.e.,\ by setting the value as at least 8 to allow I${}_{n\leq 8}^\ast$) and the global symmetry \In{} fiber maximum chosen accordingly via consultation of Table (6.1) of~\cite{BertoliniGlobal1}.} The key features of every such configuration can be captured as follows.
 
When $\g_{\Sigma}$ is any of $\f_4,\e_6,$ or $\e_7$ algebra, $[\g_{41}]$ is trivial. Remaining cases involve an $\so(N)$ algebra on $\Sigma$ with $8\leq N \leq 24,$ with $[\g_{\Sigma}]$ obeying the following constraints. For $N=9,$ it is trivial unless $T_{\Sigma'}\sim\text{\Izero{}}$ when $[\g_{\Sigma}]\cong\su(2).$ For $N\geq10$ and
$T_{\Sigma'}\sim\text{\Izero{}}$ we have $[\g_{\Sigma}]\cong\sp(2)$ and $[\g_{\Sigma'}]$ is trivial. Otherwise, we have (in agreement with convexity conditions) $T_{\Sigma'}\leftrightarrow\sp(M)$ for $M\geq1$ and $[\g_{\Sigma}]\cong \sp(M')$ with $M+M'\leq \lfloor N/2 \rfloor-1.$ Attachment of further $-1$ curves $\Sigma_{L},\Sigma_{U}$ meeting $\Sigma$ give $[\g_{\Sigma}]$ obeying $M+M'+M_L+M_U\leq \lfloor N/2 \rfloor-1$ where $\g_{\Sigma_{L}},\g_{\Sigma_{U}}$ are given by $\sp(M_L),\sp(M_U),$ respectively. Tracking the Kodaira types along the $-1$ curves gives a stronger constraint depending on the types realizing neighboring gauge summands (particularly when $N$ is even) via~\eqref{eq:intConConstraints} used with the raised intersection contributions appearing in Table~\ref{t:InstarLocalContributionsToIn}. To simplify the statement here, we consider the configuration
\begin{align}\label{eq:141GSconfig}
\begin{array}{ccccc}
& 1_{\Sigma_1} & \\
1_{\Sigma_2} & 4_{\Sigma} & 1_{\Sigma_3} \\
& [\Sigma_{4}] & 
\end{array}
\end{align}
with types $T_{\Sigma_i}\sim\text{\Inns<n_i>{}}$ and $T_{\Sigma}\sim\text{\Instar{}}\leftrightarrow\so(N).$ Configurations are then constrained by $\dt_{\Sigma}=4n$ via
\begin{align}\label{eq:141GSConstraint}
4n \geq (\sum_i n_i) + \delta_{N,\text{even}} (\sum_i \delta_{n_i,\text{odd}})\ .
\end{align}
This extends to restrictions on sub-configurations obtained by omission of terms corresponding to any removed outer curves (e.g.\ to also constrain $141$).

Here $[\g_{\Sigma'}]$ can also be non-trivial. When $\g_{\Sigma}\cong\so(N_L)$ with $N_L\leq 9,$ this summand is trivial with a single exception for $T_{\Sigma'}\sim\text{\Inns<3>{}}$ and $N_L=9$ allows $[\g_{\Sigma'}]\cong\su(2).$ If the $-3$ curve $\Sigma_R$ has $\g_{\Sigma_R}\cong\g_2,$ then $[\g_{\Sigma'}]$ is trivial except again when $T_{\Sigma'}\sim\text{\Inns<3>{}}$ and $N_L=9$ allows $[\g_{\Sigma'}]\cong\su(2).$ The remaining enhancements appear as
\begin{align}
(\text{I${}_{n_L}^*$},\so(N_L)) \ \ \overset{[\su(N_T)]}{(\text{\Inns<n>{}},\sp(M))} \ \ (\text{I${}_{n_R}^*$},\so(N_R))
\end{align}
 with $10 \leq N_L\leq 24$ and $7\leq N_R\leq 12;$ note that $\lfloor n/2\rfloor = M$ and that $7+ n_L+\delta_{N_L,even}= N_L,$ where the Kronecker symbol is nonzero for even values of $N_L.$ Values of $N_R$ and $n_R$ are related similarly. These cases have $[\g_{\Sigma'}]$ trivial or of the form $\su(N_T)$ with $N_T$ constrained via a generalization of conditions from~\cite{atomic} obtained by type tracking which reads
 \begin{align}\label{eq:SoSpSoCondition414}
 4M + 2\delta_{n, 2M+1}  \geq N_L+N_R+2N_T-14 + (2\delta_{n,2M+1}-1)(\delta_{N_L,\text{even}}+\delta_{N_R,\text{even}}).
 \end{align}

\item $414 \dots:$ Quivers of this form support infinitely many enhancements, but enhancement of either $-4$ node to $\f_4$ or $\e_{\geq 6}$ is forbidden. Constraints here mirror those for $413.$ Again $[\g_{\Sigma'}]$ obeys with~\eqref{eq:SoSpSoCondition414}. The summand $[\g_{\Sigma'}]$ is essentially captured by gauging condition of~\cite{atomic} (which can be viewed as counting roots $\Deltat_\Sigma$ needed for \Inns<n_i>{} junctions requiring $n_i$ vanishings). In the presence of a left neighboring $-1$ curve $\Sigma_L$ (giving $\cdots 1_{\Sigma_L}41\cdots$) noting assignments to $141$ are of the form $\sp(M_1),\so(N),\sp(M_2)$ while $[\g_\Sigma]\cong\sp(M_3)$ with $M_3$ constrained by
\begin{align} 
N\geq M_1+M_2+M_3+8,
\end{align} 
with the corresponding term dropped for $\Sigma_L$ absent.
Similarly, the flavor summands which may appear from the $-4$ curve at the T-junction in $1\overset{1}{4}1$ are constrained by
\begin{align} \label{eq:GS141}
N\geq M_0+M_1+M_2+M_3+8,
\end{align} 
where $\sp(M_0)$ gives the gauge algebra along the upper $-1$ curve. 

The gauging restriction on $M$ of~\cite{atomic} for $\so(N_L),\sp(M),\so(N_R)$ enhancements reading
\begin{align*}
4M &\geq N_L+N_R-16
\end{align*}
naturally generalizes to constrain $[\g_\Sigma']\cong\su(M')$ in $41_{\Sigma'}4$ via~\eqref{eq:SoSpSoCondition414}. Note that Kodaira type specification beyond the precision needed to determine the gauge content is essential.

\item $412_{\Sigma_2}\dots:$ The discussion for $413$ captures all details here except in cases with $\Sigma_2$ gaugeless or carrying $\su(n).$ The $[\g_{41}]$ restriction for $\so,\sp,\so$ enhancements is again given by~\eqref{eq:SoSpSoCondition414}. We shall treat the remaining cases in three groups. The first features gaugeless $\Sigma',\Sigma_2.$ The second feature type III or IV. Both are quickly characterized explicitly using the accompanying workbook by selecting small bounds for \In{} and \Instar{} fibers. The final group contains infinitely many enhancements with $\so,\sp,\su$ gauge summands arising from \InstarNoMon<n_L>{},\Inns<n_M>{},\Ins<n_R>{} assignments. 

The latter are again governed by extension of~\cite{atomic} conditions on $412$ gauging. We shall account for a triple of curves meeting $\Sigma'$ with fibers of the types
\begin{align}
[\g_L]\ \  \underset{\text{\InstarNoMon<n_L>{}}}{\overset{\so(N_L)}{4}} \ \  \overset{[\g_M']}{\underset{\text{\Inns<n_M>{}}}{\overset{\sp(\lfloor n_M/2\rfloor )}{1}}} \ \ \underset{\text{\Ins<n_R>{}}}{\overset{\su(n_R)}{2}} \ \ [\g_R] \ .
\end{align}
Here $\g_L\cong \sp(N_L')$ where $N_L$ obeys the same restriction leading to condition~\eqref{eq:GS141} with the appropriate terms dropped in the absence neighboring curves, namely 
\begin{align}
N_L\geq  N_L' + \lfloor n_M/2\rfloor +8 \ .
\end{align} 
The summand $\g_R \cong \su(N_R')$ must satisfy
\begin{align}
2n_R \geq n_M + N_R'.
\end{align}
Finally, $[\g_{M'}]$ maxima may be realized by a type \InstarNoMon<m'>{} fiber giving $[\g_{M'}]\cong \so(M')$ with one of $M'=2m'+8$ or $M'=2m'+7$ and $M'$ required to satisfy constraints on $\dt$ along the $-1$ curve reading
\begin{align}\label{eq:mprime}
n_M\geq n_R + m' + n_L
 +\delta_{n_M,\text{odd}}(\delta_{N_L,\text{even}}+\delta_{M',\text{even} }),
\end{align}
or by a second type of potentially relatively maximal flavor summand with one of the forms $\su(M')$ or $\oplus_{M'_k}\su(M'_k)$ obeying $\dt_{\Sigma'}$ constraints (generalizing the maximal configurations for an I${}_n^{ns}$ fiber appearing in~\cite{BertoliniGlobal1} to include multiple $\su$ curves and a single $\so$ curve) which reads 
\begin{align}\label{eq:mprime2}
\sum_{k}M'_k \leq 4+n_M -n_L- \delta_{n_M,\text{odd}}(1+\delta_{N_L,\text{even}})-n_R \ .
\end{align}
One can check this implies that in all cases where $m'$ is permitted to be non-negative and $n_M>2$, the $\so(M')$ summand is always the unique maximal global contribution from the $-1$ curve and when~\eqref{eq:mprime} requires $m'<0$, instead the second type of summand gives the maximal contribution and does so when $k=1.$ When $n_M\leq 2,$ a \IVs{} intersection is also permitted, which makes the analysis more involved and leads to multiple relative maxima for $[\g_{\Sigma'}]$ under certain conditions, in particular when $n_M=2,$ $N_L=9$ and $n_R=1$ where these are $\so(8)$ and $\su(2)\oplus \su(3)$ (the latter arising from \Itwo{},\IVs{}). 
\end{itemize}

\end{subsection}

\begin{subsection}{$21$}
Our treatment here is simplified by the fact that left attachments to $2_{\Sigma} 1_{\Sigma'}$ can only begin with a $-2$ or $-3$ curve. Attachments to $\Sigma'$ can result in many enhancements, in particular when the curve attached is a $-4,-3$ or $-2$ curve. We shall proceed by determining the $[\g_{21}]$ in cases without left attachment before returning to treat their presence. Observe that we have already treated $21m$ via Table~\ref{t:GS512} when $5\leq m\leq8$ or $m=12.$ We begin by reviewing these cases.
\begin{itemize}

\item $21(12)\cdots:$ Here $[\g_{21}]$ is trivial since $\Sigma'$ must have type \Izero{}. Left attachments are hence irrelevant.

\item $218 \cdots$ or $217\cdots:$  An $\su(2)$ flavor summand can arise from each of $\Sigma,\Sigma'$ simultaneously provided $T_{\Sigma'}\sim\text{\Ione{}}$ and $T_{\Sigma'}\sim\text{\Izero{}}.$ As we can read from Table~\ref{t:GS512} (providing identical data via replacement in the $\e_7$ cases of the $-6$ curve with a $-8$ or $-7$ curve), for $[\g_{21}]\cong\su(2)$ with $\T_{21}\sim\text{\Izero{},\Izero{}}$ or $\T_{21}\sim\text{II,\Izero{}}.$ Otherwise $[\g_{21}]$ is trivial.

\item $21m_{\Sigma_R}\cdots$ with $5\leq m\leq 6:$ All configurations for these quivers appear in Table~\ref{t:GS512}. In all cases the rank of $[\g_{21}]$ is bounded by 5. Enhancement of $\Sigma_R$ to $\e_7$ is covered by the above treatment for $217.$ When $\Sigma_R\leftrightarrow \e_6,$ non-trivial $[\g_{\Sigma}]$ may arise depending on $\T_{21}$ beyond the precision for gauge specification and $[\g_{\Sigma}]$ may be trivial, $\su(2),$ or $\su(3).$ When $m=5$ and $\g_{\Sigma_R}\leftrightarrow\f_4$ specification of $[\g_{21}]$ is sufficiently involved that we defer to Table~\ref{t:GS512} noting the bound $[\g_{\Sigma}]\subset\sp(4).$ 

\item $214_{\Sigma_R}\cdots:$ Left attachments here can only yield $(3)2214$ or $3214.$ The highly constrained configurations on these quivers are treated in our Section~\ref{s:sideLinksWithNodes} discussion for side-links. Right attaching links and further nodes do not affect the structure of $[\g_{21}].$ 
The presence of $\Sigma_R$ places strong restrictions on permitted enhancements since $\g_{\Sigma_R}\cong\so(N_R)$ requires $T_{\Sigma'}$ be gaugeless or type \Inns<N_M>{}. The discussion of $[\g_{\Sigma}]$ and $[\g_{\Sigma'}]$ complements that for $412$ and only involves conditions introduced there together a restriction on $[\g_{\Sigma}]$ given by~\eqref{eq:SuSuConstraint}. 

\item The bare quiver $21$: Infinitely many enhancements of the form $\su,\su$ and $\su,\sp$ realized by types \Ins<n_L>{},\Ins<n_R>{} and \Ins<n_L>{},\Inns<n_R>{}, respectively, are permitted here. The main ingredient determining $[\g_{21}]$ is the discussion for \In{} fibers appearing~\cite{BertoliniGlobal1}. There are finitely many remaining enhancements, though numerous. These are readily listed explicitly via the accompanying workbook (via rank bounds on \In{} and \Instar{} similar to those we detailed to $413.$). We hence abbreviate their discussion here accordingly. 

Let us first detail $[\g_{\Sigma}]$ for the aforementioned infinite families. In $\su,\su$ enhancements with \Ins<n_L>{},\Ins<n_R>{} realization, $[\g_{\Sigma}]$ is of the form $\su(n_M)$ where
\begin{align}\label{eq:SuSuConstraint}
2n_L \geq n_R + n_M.
\end{align}
This constraint also governs the contributions from the $-2$ curve in $\su,\sp$ enhancements realized by \Ins<n_L>{},\Inns<n_R>{}.

Prescribing $[\g_{\Sigma'}]$ is slightly more involved. Most relevant details appear in our treatment for $412.$ Consider the cases with such $\su,\su$ enhancements. When $n_R\geq3,$ intersection with an $\so(M)$ fiber is barred via non-minimality requirements. Resulting maxima are given by $\su(n_{M'})$ terms with $8+n_R+\delta_{6,n_R} \geq n_L + n_{M'}.$ When $n_R \leq2$ the $\su(M')$ contributions are governed by the same conditions as the cases with $\su,\sp$ enhancements. The latter allow two possible forms for $[\g_{\Sigma}]$ consisting of $\so(M')$ and $\su(M')$ summands. The former is maximal when permitted with $M'>8.$ These appear from a type \InstarNoMon<m'>{} fiber having $M'=6+2m'$ or $M'=5+2m'$ (depending on monodromy) and must obey
\begin{align*}
4+n_R \geq n_L +m'+ \delta_{n_R,\text{odd}}(1+\delta_{M',\text{even}}).
\end{align*}
The $\su(M')$ summands which may appear from a type \Ins<M'>{} fiber are slightly more constrained than in the $\su,\su$ enhancement cases. These obey
\begin{align*}
12 + n_R \geq n_L +M' + 4+2\delta_{n_R,\text{odd}}.
\end{align*}
This leaves us to contend with finitely many remaining $\T_{21}$ involving at least one of the types II, III, IV, or \InstarNoMon<n\leq 4>{}. When the latter occurs as $T_{\Sigma}$ to give an enhancement of the form $\so,\sp$ with gauge summands realized by types \InstarNoMon<n_L>{},\Inns<n_R>{}, $[\g_{\Sigma'}]$ is governed by the maximal configurations discussed in~\cite{BertoliniGlobal1} concerning type \Inns<n>{} curves; these consist of $\so(M'>8)$ flavor summands when permitted and otherwise involve only $\su(N\leq4)$ terms. We omit further discussion for the finitely many remaining $\T_{21}$ since complete listings are quickly obtained via the accompanying workbook.

\end{itemize}

Let us now also briefly discuss consequences of left attachment to $21.$ All such attachments are via a $-2$ or $-3$ curve. The former leads to a short instanton link with further right attachments discussed in Section~\ref{s:instantonLinksWithNodes}. Instead attaching a $-3$ curve is highly constraining as it gives a $32$ subquiver; further right node attachments is discussed in~\ref{s:sideLinksWithNodes}. This leaves discussions for the quiver $321$ and right attachments giving $3215,$ and $3213.$ The latter are detailed in Table~\ref{t:GS3213} while $321$ is discussed in Section~\ref{s:quiversEnding21} with all configurations captured via Table~\ref{t:51321GSRules}.

\end{subsection}

\begin{subsection}{$31$}
Left attachment of $\Sigma_L$ to $3_{\Sigma} 1_{\Sigma'}$ requires $1\leq m_{\Sigma_L}\leq 2.$ Right attachment of $\Sigma_R$ is permitted for $2\leq m_{\Sigma_R} \leq 6.$ There are many enhancements of quiver involving such attachments and these often feature a rich global symmetry algebra structure making this perhaps the most non-trivial case. We begin by discussing $[\g_{\Sigma}]$ in each of these contexts.
\begin{subsubsection}{The $-3$ curve contributions}
\begin{itemize}
\item $\cdots 231\cdots:$
Left attachment with $m_{\Sigma_L}=2$ is highly constraining, as we can read from Table~\eqref{eq:GS23}. Here $[\g_{\Sigma}]$ is at most $\su(2),$ with non-triviality requiring $\g_{\Sigma}\cong\so(7)$ rather than $\g_2,$ the $T_{\Sigma'}\in\{\text{\Izero{},\Ione{}}\}$, and $T_{\Sigma_L}\sim\text{III}$ rather than \IVns{}.

\item $\cdots 315\cdots$ and $\cdots 316\cdots:$ 
Right attachment with $5\leq m_{\Sigma_L}\leq 6$ also places tight restrictions on the gauge and flavor summands from $\Sigma.$ We can read from Table~\eqref{eq:GS513} that an $\su(2)$ flavor summand $[\g_{\Sigma}]$ may arise, but only if $m_{\Sigma_R}=5$ and $\Sigma_R$ is unenhanced (with $\f_4$ algebra) while $\g_{\Sigma}\cong\g_2.$ Any gauged left attachment again makes $[\g_{\Sigma}]$ trivial.

\item The bare base $31:$ This base permits a wide variety of enhancements and flavor symmetries detailed fully in Table~\ref{t:GS31}. The structure of $[\g_{\Sigma}]$ permits a condensed description. These summands are of the form $\sp(N)$ for $N\leq 5$ except when $\g_{\Sigma}$ is given by $\so(8)$ in which case $[\g_{\Sigma}]\cong\sp(1)^{\oplus n}$ for $n\leq 3,$ where $n$ is determined by $T_{\Sigma'}.$ The value $n=3$ only occurs when $T_{\Sigma'}\sim\text{\Izero{}};$ $n=2$ can arise when $T_{\Sigma'}\sim\text{I${}_{1\leq k\leq 2}$};$ we have $n=1$ for $T_{\Sigma'}\sim\text{\Inns<3>{}}.$ For $\g_\Sigma\in\{\su(3),\f_4,\e_6,\e_7\},$ instead $[\g_{\Sigma}]$ is trivial. For $\g_\Sigma\cong\g_2,$ we have $[\g_\Sigma]\subset\sp(1)$ with non-triviality precisely when $\Sigma'$ is gaugeless. The remaining cases have $\g_\Sigma\cong\so(M)$ with $[\g_\Sigma]\cong\so(N)$ where $N$ is determined by $\g_{\Sigma'}\cong\sp(N');$ for $M=7,$ we have $N+N' \leq 2;$ otherwise $9\leq M\leq 12$ and $N$ is constrained by $N+N' \leq M-7.$ 

\item $\cdots 314_{\Sigma_R} \cdots:$
The $\Sigma_R$ attachment still allows enhancement $\Sigma'$ which makes a full treatment somewhat involved. However, quivers containing $314$ permit finitely many enhancements. Hence, all configurations are readily detailed via the accompanying workbook. Sub-enhancements along $31$ are the subset of those for $31$ allowing $\Sigma_R$ and consequently $[\g_{\Sigma}]$ can be read from Table~\ref{t:GS31} by confining to compatible enhancements conveniently detailed in~\cite{atomic}. These $[\g_{\Sigma}]$ are bounded by $\sp(1)$ for $\g_{\Sigma}\cong\so(9),$ by $\sp(2)$ for those with $\g_{\Sigma}\cong\so(10\leq N\leq 11),$ and by $\sp(3)$ for $\g_{\Sigma}\cong\so(12),$ as follows by confining to the subset obeying convexity requirements on $\Sigma'$ with minimally enhanced $\Sigma_R$ neighbor having $\g_{\Sigma_R}\cong\so(8)$.

\item $\cdots 313_{\Sigma_R} \cdots:$ Details here are identical to the previous base except for freedoms introduced when we reduce $\g_{\Sigma_R}$ below $\so(8).$ This reduction allows $[\g_{\Sigma}]\subset\sp(3)$ when $\g_{\Sigma}\cong \so(10\leq N\leq 11)$ and $[\g_{\Sigma}]\subset\sp(4)$ for $\g_{\Sigma}\cong\so(12)$ with proper inclusions mandated by the condition for $\g_\Sigma\cong\so(N)$ having $[\g_{\Sigma}]\cong\sp(M)$ and $\g_{\Sigma'}\cong\sp(M')$ neighbors requiring $M+M'\leq N-7.$ Details for all (of the finitely many) configurations are readily available via the accompanying workbook. Note that left attachment of $\Sigma_L\leftrightarrow\sp(M_L)$ with $m_{\Sigma_L}=1$ reduces $[\g_{\Sigma}]$ according to $N-7 \geq M+M'+M_L$ with further reductions possible for even $N$ dependent on the types realizing $\sp(M'),\sp(M_L)$ through introduction of Kronecker symbols for \Inns<n>{} realizations having $n$ odd via~\ref{eq:131GSConstraint}.

\item $312: $ The noble link $3_\Sigma 1_{\Sigma'}2$ permits a large but finite number of enhancements. We can treat the bare quiver by listing all possible configurations via the accompanying workbook. Since this is a ``noble link'' and enhancements of containing links are strongly limiting, we shall content ourselves with explicit configuration listings for these links without loss of depth in our discussion for longer bases. Attachments on the right must begin with a $-3$ curve while those on the left attachments must begin with a $-2$ curve $\Sigma_L.$ 
Contributions from $[\g_\Sigma]$ are easily determined from $T_{\Sigma_L}$ using subquiver $23$ constraints of~\eqref{eq:GS23} while noting that reduction to trivial $[\g_\Sigma]$ results for $T_{\Sigma'}\sim\text{II}$ or $T_{\Sigma'}$ gauged. 

\item $\cdots 131 \cdots:$ Many of the large but finite number of enhancements on $1_{\Sigma_L} 3_\Sigma 1_{\Sigma'}$ feature multiple flavor symmetry maxima. Outer attachments $\Sigma_a$ and $\Sigma_b$ require $m_{\Sigma_i}\geq 3$ and $m_{\Sigma_a}=m_{\Sigma_b}=3$ is forbidden (to blow-down consistently). As a result, features of $[\g_{31}]$ for enhancements of $131$ occurring in longer bases are captured by discussion in Section~\ref{s:generalStructure} while most $131$ configurations are irrelevant except in discussion of the bare quivers $131$ and $3131$ for which we content ourselves with the availability of explicit workbook listings as in the previous case. 

The largest family of enhancements for this base is of the form
\begin{align}\label{eq:131GSconfig}
\begin{array}{ccccc}
& [1_{\Sigma_1}] & \\
1_{\Sigma_2} & 3_{\Sigma} & 1_{\Sigma_3}
\end{array}
\end{align}
with types $T_{\Sigma_i}\sim\text{\Inns<n_i>{}}$ and $T_{\Sigma}\sim\text{\Instar{}}\leftrightarrow\so(N).$ The analog of~\eqref{eq:141GSConstraint} for these cases is relies on (6.73),(6.77) of~\cite{BertoliniGlobal1} to obtain sharpening beyond bounds imposed by $\dt_{\Sigma}=6+3n,$ this being replaced instead by
\begin{align*}
\dt_\Sigma'\equiv \dt_\Sigma -(2+\delta_{n,\text{odd}})=4+3n-\delta_{n,\text{odd}}.
\end{align*}
to yield
\begin{align}\label{eq:131GSConstraint}
4+3n-\delta_{n,\text{odd}}\geq (\sum_i n_i) + \delta_{N,\text{even}} (\sum_i \delta_{n_i,\text{odd}}) \ .
\end{align}
Analogous restrictions hold on sub-configurations obtained by omission of terms corresponding to removed outer curves (e.g.\ to also constrain $31$).

\end{itemize}
\end{subsubsection}

\begin{subsubsection}{The $-1$ curve contributions}

\end{subsubsection}
We now discuss flavor summands from $31_{\Sigma'}$ arising via $\Sigma'.$ We proceed through these cases for right attachments $\Sigma_R$ which constrain these, starting with the largest values for $m_{\Sigma_R}.$ 

\begin{itemize}
\item $315\cdots$ and $316\cdots:$ Here $[\g_{\Sigma'}]$ is trivial for each enhancement. 

\item $314_{\Sigma_R}\cdots:$ Non-trivial $[\g_{\Sigma'}]$ requires $\g_{\Sigma_R}\cong\so(N_R).$ Compatible assignments appear as $\so(N_L) \sp(M) \so(N_R)$ with $[\g_{\Sigma'}]\cong\su(N_T)$ having $N_T$ constrained via $T_{\Sigma},T_{\Sigma_R}$ (beyond the precision to specify $\g_\Sigma,\g_{\Sigma_R}$) via~\eqref{eq:SoSpSoCondition414}.

\item $312, 3123, 131, 31:$ Each of these cases permits finitely many enhancements and we can proceed by listing all possible configurations via the accompanying workbook. Since there are very few arrangements for $23,$ the configurations for $3123$ are rather limited. Strong restrictions from context in longer bases limits the relevance of a discussion for the bare quivers $312$ and $131,$ thus we leave details to explicit listings without loss of general structural insight.

\item $313_{\Sigma_R}\cdots:$ As for $314,$ enhancements of the form $\so(N_L) \sp(M) \so(N_R),$ all have $[\g_{\Sigma'}]$ obeying~\eqref{eq:SoSpSoCondition414}. The remaining enhancements have trivial $[\g_{\Sigma'}]$ with the exceptions 
\begin{align}\arraycolsep=9pt%
\begin{array}{cccc}
           \g_\Sigma & \g_{\Sigma_R} & T_{\Sigma'} & [\g_{\Sigma'}] \\
          \su(3) & \su(3) & \text{\Izero{}} &\su(2)\oplus \su(2)\\
          & & & \su(3) \\
            \g_2 & \g_2 & \text{III/\Itwo{}} & \su(2)\\
             & & \text{\IVns{}/\Inns<3>{}} & \su(3) \\
            \g_2 & \so(7) & \text{III/\Itwo{}} & \su(2)\\
             & & \text{\Inns<3>{}} & \su(2) \\             
            \g_2 & \so(8) & \text{\Itwo{}/\Inns<3>{}} & \su(2)\\             
            \g_2 & \so(9) & \text{\Inns<3>{}} & \su(2)\\                         
\end{array}
\end{align}
and those obtained by reversing roles of $\Sigma,\Sigma'.$ 
\end{itemize}

\end{subsection}

\begin{subsection}{$22$}\label{s:GS22}
We first discuss the infinite family of classical enhancements and conclude with the few remaining cases featuring non-$\su(N)$ algebras. The only essential restriction for the classical enhancements of bases of the form $\cdots 222 \cdots$ is a condition which extends naturally to constrain flavor symmetries for such configurations. The extended condition reads
\begin{align}\label{eq:222convexity}
  2 N_M \geq N_L+ N_T + N_R 
\end{align}
for a subquiver of the form 
\begin{align*}
\overset{\su(N_L)}{2} \ \ \underset{[\su(N_T)]}{   \overset{\su(N_M)}{2}} \ \ \overset{\su(N_R)}{2}\ ,
\end{align*}
where the flavor symmetry contribution from the middle curve is indicated with brackets. Note that in the absence of a neighboring curve, the condition is that obtained by dropping the corresponding term $N_j.$ In the case of a T-junction, an additional term appears and gives flavor contribution from the $-2$ curve at the junction of
\begin{align*}
\overset{\su(N_L)}{2} \ \ \overset{\overset{\su(N_U)}{2}}{\underset{[\su(N_T)]}{   \overset{\su(N_M)}{2}}} \ \ \overset{\su(N_R)}{2}\ 
\end{align*}
obeying
\begin{align}\label{eq:22GSCondition}
  2 N_M \geq N_L+ N_T + N_R +N_U \ . 
\end{align}

The remaining enhancements of the above quivers feature fiber types other than \In{}. These appear below along with global symmetry summands indicated.
\begingroup
\renewcommand*{\arraystretch}{1.9}
\begin{align}\arraycolsep=18pt%
\begin{array}{ccccccc}
2\ \overset{2}{2}\ 2 \ : & \text{III} \ \ \  \underset{[\su(2)]}{\overset{\text{III}}{\text{\Izerostarss{}}}} \ \ \  \text{III}\\
& \text{III} \ \ \  \underset{[\su(2)]}{\overset{\text{III}}{\text{\Izerostarns{}}}} \ \ \  \text{III}/\text{\IVns{}}\\
& \text{II} \ \ \  {\overset{\text{II}}{\text{\IVns{}}}} \ \ \  \underset{[\su(3)]}{\text{\IVs{}}}\\
& \text{\IVns{}} \ \ \  {\overset{\text{\IVns{}}}{\text{\Izerostarns{}}}} \ \ \  \text{\IVns{}}\\
& \text{II} \ \ \  {\overset{\text{II}}{\text{\IVns{}}}} \ \ \  \text{II}/\text{III}/\text{\IVns{}}
\end{array}
\end{align}
\endgroup

\begingroup
\renewcommand*{\arraystretch}{1.6}
\begin{align}\label{eq:GS22Nonclassical}\arraycolsep=18pt%
\begin{array}{ccccccc}
2\ 2\ : & \text{II} \ \ \  \text{II} \\
& \text{II} \ \ \  \underset{[\su(2)]}{\text{III}}\\
& \text{II} \ \ \  \underset{[\g_2]}{\text{\IVns{}}}\\
& \text{III}/\text{\IVns{}} \ \ \  \text{\IVns{}}\\
&\underset{[\su(2)]}{\text{III}} \ \ \  \underset{[\su(2)]}{\text{III}}\\
& \text{II} \ \ \  \text{II} \\
& \text{III}/\text{\IVns{}} \ \ \ \underset{[\su(3)]}{\text{\IVs{}}} \\
& \underset{[\su(3)]}{\text{\IVs{}}} \ \ \ \underset{[\su(3)]}{\text{\IVs{}}} \\
& \text{III}/\text{\IVns{}}  \ \ \  \underset{[\sp(3)]}{\text{\Izerostarns{}}}\\
& \text{III}  \ \ \  \underset{[\sp(3)\oplus \sp(1)]}{\text{\Izerostarss{}}}
\end{array}
\end{align}
\endgroup

\begingroup
\renewcommand*{\arraystretch}{1.4}
\begin{align}\arraycolsep=18pt%
\begin{array}{ccccccc}
2\ 2\ 2 \ : & \text{II} \ \ \  \text{II}/\text{III} \ \ \ \text{II} \\
& \text{II} \ \ \  \underset{[\su(3)]}{\text{\IVns{}}} \ \ \ \text{II} \\
& \text{II} \ \ \  \text{\IVns{}} \ \ \ \text{III}/\text{\IVns{}} \\
& \text{II}  \ \ \ \text{III}\ \ \  \underset{[\su(2)]}{\text{III}} \\
\end{array}
\end{align}
\endgroup

\begingroup
\renewcommand*{\arraystretch}{1.4}
\begin{align}\arraycolsep=18pt%
\begin{array}{ccccccc}
2\ 2\ 2 \ \text{(continued)}: 
& \underset{[\su(2)]}{\text{III}} \ \ \  \text{III} \ \ \ \underset{[\su(2)]}{\text{III}} \\
& \text{II} \ \ \  \text{\IVns{}} \ \ \ \underset{[\su(3)]}{\text{\IVs{}}} \\
& \text{III}/\text{\IVns{}} \ \ \  \text{\IVs{}} \ \ \ \text{III}/\text{\IVns{}} \\
& \text{III}/\text{\IVns{}} \ \ \  \text{\IVs{}} \ \ \ \underset{[\su(3)]}{\text{\IVs{}}} \\
& \underset{[\su(3)]}{\text{\IVs{}}} \ \ \  \text{\IVs{}} \ \ \ \underset{[\su(3)]}{\text{\IVs{}}} \\
& \text{II} \ \ \  \text{\IVns{}} \ \ \ \underset{[\sp(3)]}{\text{\Izerostarns{}}} \\
& \text{III}/\text{\IVns{}} \ \ \  \underset{[\sp(2)]}{\text{\Izerostarns{}}} \ \ \ \text{III}/\text{\IVns{}} \\
& \text{III} \ \ \  \underset{[\sp(2)\oplus\sp(1)]}{\text{\Izerostarss{}}} \ \ \ \text{III} \\
\end{array}
\end{align}
\endgroup
\end{subsection}

\end{section}

\begin{section}{Flavor symmetry structure via ``atomic'' decomposition}\label{s:generalStructure}
Here we discuss the structure of 6D SCFT global symmetries with help from the ``atomic'' base decomposition~\eqref{eq:atomicStructure} of~\cite{atomic}. Briefly, flavor symmetries only arise from curves near the tail ends of a base with the largest rank summands coming from side-links and short subquivers containing the few permissible $-4$ curves when either is present. The only exceptions are also the only bases for which decomposition summand and total rank bounds are absent, these having the form $(1)2\cdots2$ or $(1)414\cdots4(1).$ The latter are readily treated separately via the methods Section~\ref{s:GSRules} using simple constraint equations.

This decomposition allows us to characterize flavor summands for each theory by way of contributions from each term in~\eqref{eq:atomicStructure} upon fixing an enhancement, i.e.,\ via summands $[\h_{i,j}]$ in
\begin{align}\label{generalGS}
\underset{[\h_{0,r}]}{S_0} \ \ \overset{\overset{[\h_{1,u}]}{S_1}}{\underset{[\h_{1,l}]}{ g_1}}\ \  \underset{[\h_{1,r}]}{L_1}\ \ \cdots \ \ \overset{\overset{[\h_{m,u}]}{I^{\oplus t}}}{\underset{[\h_{m,l}]}{g_m}}\ \ \underset{[\h_{m,r}]}{S_m}
\end{align}
giving total flavor symmetry $\g_{\text{GS}}\cong \oplus_{i,j}\h_{i,j},$ with $\h_{i,j}$ depending on the enhancement. To achieve this, we simply detail contributions from each link with node attachment(s) which can appear in a base. Let us begin with an abbreviated summary of this approach and its results.

For linear bases, global symmetry summands arise only near the tail ends of multi-node bases with the exception of those for which the only nodes are $-4$ curves. This follows from our discussions in Sections~\ref{s:interiorLinkGS},\ref{s:interiorLinkContributions} together with the permitted chains of nodes classified in Appendices B,C of~\cite{atomic} obeying the algebra inclusions of (5.48) from~\cite{atomic} requiring that gauge summands $\g_i$ from the nodes $g_i$ obey
\begin{align}\label{eq:algebraInclusions}
\g_1\subset \g_2 \subset \cdots \subset \g_m \supset \cdots \supset \g_k \ .
\end{align}
This containment property implies that subquivers with $-4$ curves only occur towards the periphery of a base. The classification of node chains places severe limits the number of such nodes. (In the presence of any other node type, at most eight can occur; in nearly all node chain families, far fewer are permitted.) This means that determining flavor symmetries arising for a typical base is achievable in practice via classification of contributions from: (i) interior-link node pairings (Section~\ref{s:interiorLinkContributions}), (ii) side-links attachments (Section~\ref{s:sideLinksWithNodes}), and (iii) instanton-link attachments (Section~\ref{s:instantonLinksWithNodes}).

Flavor summands for theories with a linear base have all rank $r_i$ contributions from curves $\Sigma_i$ such that $r_i>1$ arising in subquivers where the only nodes (if any) are $-4$ curves on the base periphery (via~\eqref{eq:algebraInclusions}) when any $\e_{\geq6}$ gauge summands are present. The total contributions arising from an interior link $L_i$ joining nodes having gauge summands $\g_L$ and $\g_R$ are severely constrained. In fact, we have
\begin{align}\label{eq:linkGS}
\begin{array}{ccccccc}
\g_L & \g_R & \qquad & [\h_{i,r}]\subset  & \\
 \e_{\geq7}& \e_{\geq7} & & 0\\
{\f_4} /\e_6 & \f_4/\e_{\geq6} & & \su(2)\\
  \so(N_L) & \f_4/\e_{\geq6} & & \su(2)\\
 \so(N_L) & \so(N_R) & &  \su(2)^{\oplus 3} \text{ or } \sp(2) & \ \ \text{for $L_i\not\sim 1$} \ ,
\end{array}
\end{align}
with precise specification depending on the enhancement as detailed shortly and summarized in~\cref{eq:linkGS,eq:interiorLinkGSsummary,eq:enhancedLinkGSsummary22,eq:enhancedLinkGSsummary33,eq:enhancedLinkGSsummary33part2}. Flavor summands from $g_i$ are trivial for $g_i\neq4$ and $\g_i\not\cong\so(N);$ for $g_i=4$ we have $[\h_{i,l}]\subset\sp(4)$ unless $L_i=L_{i-1}=1$ where $\sp(M)\cong[\h_{i,l}]$, $M>4$ summands can appear, as detailed Section~\ref{s:interiorLinkGS}. The tight constraints on the number of D-nodes in node chains featuring any E-nodes as classified in Appendices B,C of~\cite{atomic} together with our interior chain, side-link and interior-link node attachment contribution treatments readily yield bounds on the flavor symmetries arising for all enhancements of each base not permitting infinitely many enhancements (i.e.,\ excluding $(2)(1)414\cdots$ and $(1)222\cdots$ and their branching variants). In most cases, these constraints suffice to determine flavor symmetries precisely with little effort up to treating any side-link summands.

Note that explicit listing via the accompanying workbook is also computationally feasible in practice except with very large numbers of nodes for which supplementing outer subquiver results using interior link contributions summands detailed here suffices as a practical route. Enhancement listings even for long quivers are typically computationally inexpensive. Side-link flavor summand maxima detailed here (or readily computed in isolation for a designated node attachment) allow extension to the general case. (In fact, all enhancements and these maxima for every side-link with node attachment can be quickly procured with the accompanying workbook.) Side-link enhancement matching on overlaps with nodes allows easy lookups enabling a practical method for arbitrary base treatment (even when exceedingly large). Assisted by the long and short base classifications of~\cite{atomic} Appendix C, global symmetries for all 6D SCFTs classified therein are thus readily determined. The number of side-links which can attach to certain bases is significant. Hence, for expository purposes we shall stop short of providing explicit listings for each short and long base, instead contenting ourselves with having provided a route that makes explicit listings computationally feasible. 
\begin{subsection}{Interior links with node attachments}\label{s:interiorLinkContributions}
As shown in~\cite{atomic} and reviewed here in Section~\ref{s:setup}, the structure of any 6D SCFT base consists of a linear quiver with decoration possible only near the ends. Hence, treatment for linear bases consisting of an interior link with a pair of outer node attachments is a key ingredient in a general characterization of SCFT flavor symmetries. In the following subsections we determine global symmetry contributions from these subquivers. 
\begin{subsubsection}{Interior of linear base flavor contribution summary}\label{s:interiorLinkGS}
We shall proceed by treating each interior link with a pair of outer node attachments. The left-right symmetrization of rules below applies, i.e.,\ rules for node pairs follow upon reversing the link orientation. We tacitly exclude the single $-1$ curve link as the only permitted node attachments are $-4$ curves yield the $414\cdots$ quivers already treated in Section~\ref{s:GSRules}. Our statements here hold for cases with a node bounding each side of an interior link, which shall also remain implicit. We shall refer to the flavor symmetry contribution from the left and right portion of each quiver by $[\g_L]$ and $[\g_R],$ respectively.
\begin{itemize}
\item Attachment of a node $n$ to the interior link 
\begin{align*}
\alpha \in \{12231513221, 1223151321, 122315131,12231 \}
\end{align*}
 forming $n\alpha\sim n_\Sigma 1_{\Sigma'}\cdots$ for $n\geq 7$ or $\g_\Sigma\cong\e_{\geq 7}$ requires $[\g_L]$ is trivial, while an unenhanced $n=6$ node has $[\g_L]\cong[\g_{\Sigma'}]\cong\su(2).$ 
 \item Right attachment of a curve with $m\geq 6$ to 
 \begin{align*}
 \alpha \in \{1223151321, 122315131, 13151321,1315131, 12321, 12231,1321, 131 \}
 \end{align*}
 leaves $[\g_R]$ trivial.
 \item Left attachment of $\Sigma$ with $m_\Sigma=4$ to an interior link $\alpha$ when $\g_{\Sigma}$ is among $\f_4,\e_{\geq 6}$ yields trivial $[\g_L]$ with two exceptions: (i) an $\f_4$ attachment to $\alpha$ of the form $13_{\Sigma} 1\cdots$ with assignment
 \begin{align*}
 \overset{\f_4}{4} \ \ \overset{\II{}}{1} \ \ \overset{\g_2}{3} \ \  \overset{\II{}}{1} \cdots
 \end{align*}
 allowing $[\g_\Sigma]\cong\su(2),$ (ii) $\alpha\sim1,$ i.e.,\ for $41$ when $[\g_{4_\Sigma1}]\subset\g_2$ with precise specifications in~\eqref{eq:GS51}. 
 \item For $Q\sim\cdots 14141\cdots$, $[\g_{Q}]$ obeys~\eqref{eq:141GSConstraint},\eqref{eq:SoSpSoCondition414} and weaker simplified constraints~\cref{eq:141ConvexityCondition,eq:SoSpSoCondition414} as discussed in Section~\ref{s:GSRules}.
 \item Left attachment of a $-4$ curve $\Sigma$ with $\so(N\geq 9)$ gauge assignment to an interior link of the form $1_{\Sigma_L}3_{\Sigma_M}1_{\Sigma_R}\cdots$ requires $[\g_{\Sigma_L}]=[\g_{\Sigma_M}]=0$ with the exceptions: (i) non-trivial $[\g_L]$ as indicated for the sub-enhancements
 \begin{align*}
 \overset{\so(9)}{4} \ \ \underset{[\su(2)]} {\overset{(\su(2), \text{\Inns<3>{}})}{1}}\ \ \overset{\so(7)}{3} \cdots
\qquad \text{and} \qquad
 \overset{\so(9)}{4} \ \ \underset{[\su(2)]} {\overset{(\su(2), \text{\Inns<3>{}})}{1}}\ \ \overset{\g_2}{3} \cdots\ ,
 \end{align*} 
noting that all $T_{\Sigma_M}\not\sim\text{\Inns<3>{}}$ which realize $\g_{\Sigma_M}\cong\su(2)$ have $[\g_{\Sigma_L}]$ trivial (e.g.,\ for $T_{\Sigma_M}\sim\text{\Itwo{}}$), and (ii) classical enhancements of $41314$ which we cover separately. These are simply governed by~\eqref{eq:141GSConstraint},\eqref{eq:SoSpSoCondition414},\eqref{eq:131GSConstraint}. Contributions from $\Sigma$ when $T_{\Sigma_L}\sim\text{\Izero{}}$ are $[\g_{\Sigma}]\cong\sp(M),$ where $M = N-8.$ 
For $N=8,$ we have $[\g_{\Sigma_M}]$ trivial unless $\Sigma_L,\Sigma_R$ are gaugeless and $[\g_{\Sigma_M}]\cong \su(2)$ may occur for $\g_{\Sigma}\cong\g_2.$
For $\g_{\Sigma_L}\cong\sp(M'),$ maxima $[\g_{\Sigma}]\cong\sp(M)$ obey
 \begin{align}\label{eq:131DAttach}
 M+M' \leq  N-8.
 \end{align}
For $T_{\Sigma_L}\sim\text{\Inns<n'>{}}$ with $n'$ odd and $N$ even,~\eqref{eq:141GSConstraint}) gives the stronger constraint
 \begin{align}\label{eq:131DAttach2}
 M+M' \leq  N-9.
 \end{align}
\end{itemize}
We shall employ the notation of~\cite{atomic} (Section 5.2.1) for interior links via tracking induced blow-down counts to condense our summaries. For example, in place of an interior $-1$ link, we write $\overset{1,1}\oplus.$  Recall this notation for the remaining interior links reads
\begingroup
\renewcommand*{\arraystretch}{1.1}
\begin{align}\arraycolsep=14pt%
\begin{array}{ccc}
\overset{2,2}{\oplus} \leftrightarrow 131 
   &\overset{3,3}{\oplus} \leftrightarrow 12321
       &\overset{4,4}{\oplus} \leftrightarrow 123151321\\
\overset{2,3}{\oplus}\leftrightarrow 1321 
   &\overset{3,3}{\bigcirc} \leftrightarrow 1315131
       &\overset{4,5}{\oplus}\leftrightarrow 1231513221\\
\overset{2,4}{\oplus}\leftrightarrow 13221 
   &\overset{3,4}{\oplus}\leftrightarrow 13151321
       &\overset{5,5}{\oplus}\leftrightarrow 12231513221 \ ,
\end{array}
\end{align}
\endgroup
allowing us to refine~\eqref{eq:linkGS} via~\eqref{eq:interiorLinkGSsummary}.

\begingroup
\renewcommand*{\arraystretch}{1.2}
\begin{align}\label{eq:interiorLinkGSsummary}\arraycolsep=18pt%
\begin{array}{ccccccc}
\text{quiver} & [\h_{i,r}] & [\h_{i,t}]\\
E_{\geq7}\overset{k,l}{\oplus}E_{\geq 7} & 0 & 0\\
E_{6}\overset{k\leq4,l}{\oplus}E_{\geq 6} & 0 & 0\\
D^+\overset{2,4}{\oplus}E_{\geq 7} & 0 & 0\\
E_{6}\overset{5,l}{\oplus}E_{\geq 6} & \su(2) & 0\\
D^+\overset{2,4}{\oplus}E_{6} & \su(2) & 0\\
E_{\geq 6}^+\overset{3,3}{\oplus}E_{\geq 6}^+ & 0 & 0\\
\so(8\leq N\leq 10)\overset{3,3}{\oplus}E_{\geq 6}^+ & \su(2) & \sp(N-8)\\
\so(11\leq N\leq 12)\overset{3,3}{\oplus}E_{\geq 6}^+ & 0 & \sp(N-8)
\end{array}
\end{align}
\endgroup 
Here we have substituted node gauge summands while suppressing their self-intersections while
$D$ indicates any permitted $\so(N\geq8)$ node, $D^+$ an $\so(N\geq8)$ or $\f_4$ node, and $E_n^+$ an $E_n$ or $\f_4$ node. We now extend the above shorthand to treat link enhancements by labeling forced $T_i$ (or $\g_i$ when unambiguous) while unlabeled positions remain {\it a priori} unconstrained. We denote minimally (non-minimally) enhanced links via an `$\sim$' (`$\not\sim$') symbol. For example,
\begingroup
\renewcommand*{\arraystretch}{1.2}
\begin{align}\arraycolsep=10pt%
\begin{array}{ccc}\label{eq:enhancedPairingsEx}
\underset{\sim}{\overset{2,2}{\oplus} } & \leftrightarrow & \ \ \underset{\text{\Izero{}}}{1} \ \ \underset{\su(3)}{3} \ \ \underset{\text{\Izero{}}}{1} \\
\underset{\not\sim}{\overset{1,1}{\oplus} } & \leftrightarrow &  \underset{T\neq \text{\Izero{}}}{1}   \\
{}_{\text{\Inns<n>{}}\g_2-}\underset{\f_4}{\overset{3,3}{\bigcirc}}&\leftrightarrow&\underset{\text{\Inns<n>{}}}{1} \ \ \underset{\g_2}{3} \ \ \underset{}{1} \ \ \underset{\f_4}{5} \ \ \underset{}{1} \ \ \underset{}{3} \ \ \underset{}{1} \ .
\end{array}
\end{align}
\endgroup
We have the following abbreviated summary.
\begingroup
\renewcommand*{\arraystretch}{1.2}
\begin{align}\label{eq:enhancedLinkGSsummary33}\arraycolsep=22pt%
\begin{array}{ccccccc}
\overset{3,3}{\bigcirc} \leftrightarrow 1315131\ \text{   E-to-E pairings:} & [\h_{i,r}] & [\h_{i,t}]\\
E_6^+\underset{\e_6}{\overset{3,3}{\bigcirc}}E_6& 0 & 0\\
\so(8)/\f_4\ \underset{\sim}{\overset{3,3}{\bigcirc}}\  \so(8)/\f_4 & 0 & 0\\
\so(8)/\f_4\ {}_{\g_2-}\underset{}{\overset{3,3}{\bigcirc}}E_6 & \su(2) & 0\\
\so(8)/\f_4\ {}_{\g_2-}\underset{}{\overset{3,3}{\bigcirc}} {}_{-\g_2}\ \so(8)/\f_4 & \su(2)^{\oplus2} & 0\\
\end{array}
\end{align}
\endgroup
\begingroup
\renewcommand*{\arraystretch}{1.2}
\begin{align}\label{eq:enhancedLinkGSsummary33part2}\arraycolsep=18pt%
\begin{array}{ccccccc}
\overset{3,3}{\bigcirc} \leftrightarrow 1315131 \text{  D-to-E pairings:} & [\h_{i,r}] & [\h_{i,t}]\\
\so(N\leq10)\underset{\sim}{\overset{3,3}{\bigcirc}}E_6& 0 & \sp(N-8)\\
\so(9)\ {}_{\text{\Inns<3>{}}\g_2-}\underset{}{\overset{3,3}{\bigcirc}}E_6 & \su(2) & 0\\
\so(9\leq N\leq12)\ {}_{\text{\Inns<2>{}}\g_2-}\underset{}{\overset{3,3}{\bigcirc}}E_6 & 0 & \sp(N-9)\\
\so(10\leq N\leq 13)\ {}_{\text{\Inns<3>{}}\g_2-}\underset{}{\overset{3,3}{\bigcirc}}E_6 & 0 & \sp(N-9-\delta_{N,\text{even}})\\
\end{array}
\end{align}
\endgroup
The remaining D-to-D node pairings via $\overset{3,3}{\bigcirc}$ are more easily characterized using decomposition about the $-5$ curve as detailed above. 
\begingroup
\renewcommand*{\arraystretch}{1.3}
\begin{align}\label{eq:enhancedLinkGSsummary22}\arraycolsep=18pt%
\begin{array}{ccccccc}
\overset{2,2}{\oplus} \leftrightarrow 131\text{  pairing:} & [\h_{i,r}] & [\h_{i,t}\oplus\h_{i+1,t}]\\
E_6^+\ \underset{}{\overset{2,2}{\oplus}}\ E_6 & 0 & 0\\
f_4\ \underset{\sim}{\overset{2,2}{\oplus}}\ f_4 & 0 & 0\\
\so(N_L\leq 10)\ \underset{\sim}{\overset{2,2}{\oplus}}\ E_6^+ & 0 & \sp(N_L-8)\\
\so(N_L\leq 10)\ \underset{\sim}{\overset{2,2}{\oplus}}\ \so(N_R\leq 10) & 0 & \sp(N_L-8)\oplus\sp(N_R-8)\\
\so(8)/f_4\ \underset{\g_2}{\overset{2,2}{\oplus}}\ \so(8)/f_4 & \su(2) & 0\\
\end{array}
\end{align}
\endgroup
The few $\overset{2,2}{\oplus}$ D-to-D node pairings not covered above are characterized by~\eqref{eq:141GSConstraint},\eqref{eq:SoSpSoCondition414},\eqref{eq:131GSConstraint} with complete listings available via the auxiliary workbook. 
\end{subsubsection}
\end{subsection}

\begin{subsection}{Side-links with a node attachment}\label{s:sideLinksWithNodes}
We shall break our discussion of these cases into two parts. The first concerns determining flavor symmetry contributions arising on the inner side-link curves and the second those appearing from outermost curves. We shall tacitly exclude instanton-links from being considered as side-links, instead addressing these in Section~\ref{s:instantonLinksWithNodes}. We also separate our discussion of two particular cases: $413$ (discussed in Section~\ref{s:GSRules}) and $4131$. Each permits finitely many enhancements and is readily treated via direct listings supplied by the accompanying workbook. We hence treat the remaining cases with statements requiring modification when applied to instanton-links and side-links $13$ and $131$ attached to $-4$ curves.

\begin{subsubsection}{Contributions from non-outermost curves of side-links}
For each curve $\Sigma$ in a non-instanton side-link with node attachment, $[\g_\Sigma]$ is highly limited except when $\Sigma$ is ``outermost curve,'' i.e.,\ one having a single compact neighbor. Even these typically have $[\g_\Sigma]$ trivial unless $m_\Sigma=1.$ Non-outermost curve contributions typically obey $[\g_\Sigma]\subset\su(2)$ and are always among $\su(2)^{\oplus k\leq 3}, \su(N\leq 4),$ or $\sp(2)$ when non-trivial (upon the aforementioned tacit exclusion of $13$ and $131$ from our discussion).

We now detail all non-trivial non-outermost curve flavor symmetry contributions. Note that having excluded instanton-links from our discussion rules out non-trivial $[\g_{\Sigma'}]$ for example from $m_\Sigma1_{\Sigma'}$ when $\g_{\Sigma}\cong\e_{\geq7}.$ 
\begin{itemize}
\item $E^+$ linking curve contributions: In $m_\Sigma1_{\Sigma'}r_{\Sigma_R}\cdots$ with $\g\cong\e_{k\geq 6},$ $[\g_{\Sigma'}]$ is trivial unless $k=6$ and $\Sigma_R$ is gaugeless when $[\g_{\Sigma'}]\cong \su(2)$ can arise. When instead $\g_\Sigma\cong \f_4,$ we have $[\g_{\Sigma'}]\cong \su(N\leq 3)$ with $N=3$ requiring $T_{\Sigma_R}\sim\text{I${}_{l\leq 1}$};$ For $T_{\Sigma_R}\sim \text{II},$ $N\leq 2.$ For $\Sigma_R$ gauged, $[\g_{\Sigma'}]$ is trivial. In particular, this allows non-trivial $[\g_L]$ in $E_{k\geq6}^+\alpha$ only when $\alpha\sim1223\cdots$. 

\item Interior $\g_2$ curve contributions: The subquiver $13_{\Sigma'}1$ occurs in many links and permits $\T_{131}\sim \text{II,\Izerostarns{},II}$ in many contexts. These curves may support $[\g_{\Sigma'}]\cong\su(2).$ This accounts for many of the non-trivial contributions in longer bases. 

\item D-node linking curve contributions: Note that we are not treating the bare quiver cases $41$, $412$ and $413$ here, but rather focusing on longer side-links. Hence, $[\g_{\Sigma'}]$ in $\alpha\sim 41_{\Sigma'}\cdots$ for the remaining cases is limited to one of the following: (i) $[\g_{\Sigma'}]\cong \su(2),$ where we must have $\T\sim\text{\Instarns<1>{},\Inns<3>{},\Izerostarns{}}\cdots$ or 
$\T\sim \text{\Instars<1>{}/\Instarns<1>{},\Izero{},III}$ (relevant in links beginning as $413$ and $412,$ respectively), (ii) $[\g_{\Sigma'}]\cong \su(3)$ for $\T\sim \text{\Izerostars{},\Izero{},III}$ (applicable in $412\cdots$).

\end{itemize}

\end{subsubsection}

\begin{subsubsection}{Outermost curve contributions}
To treat the outer curves of a side-link, it will be helpful to separate our discussion of such curves with $m=1.$ The remaining cases contribute small summands and are easily characterized. 

\begin{paragraph}{Outer curves with $m\geq 2$}
\begin{itemize}
\item $\cdots 23_\Sigma$: Side-links of this form with $\T_{23}\sim\text{III,\Izerostarss{}}$ permit $[\g_{\Sigma}]\cong\su(2).$

\item $\cdots 13_\Sigma$: Here instead $\T_{13}\sim\text{II,\Izerostarns{}}$ permits $[\g_{\Sigma}]\cong\su(2).$ This is supported on many links ending as $\cdots 513.$ 

\item $\cdots 5_\Sigma 1_{\Sigma'}2$: Here $\Sigma'$ must be gaugeless and left attachments require $\g_{\Sigma}\cong \f_4.$ Hence, the $\f_4$ compatible enhancements of $512$ appearing in Table~\ref{t:GS512} determine $[\g_{512}]$ for the side-links of interest. 

\end{itemize}

\end{paragraph}

\begin{paragraph}{Outer curves with $m=1$}
We begin by treating the outermost $-1$ curves bordering a $-5$ curve. With two exceptions we shall discuss shortly for the links 
\begin{align}
  2\overset{1}{3}1 \qquad \text{and} \qquad 2\overset{1}{3}21,
\end{align}
these are the only outer $-1$ curves which can arise off the linear portion of a link, instead bordering a $-5$ curve T-junction. Such curves can give rise to $\su(3)$ or $\g_2$ flavor summands, the latter requiring Kodaira type II along the $-1$ curve. At most two type II curves are permitted to meet a $-5$ curve as discussed in Section~\ref{s:noG2Trifecta}. Links with a $-5$ curve bearing a T-junction often require type II curves on each side of the junction (e.g.\ those with linear portion having truncation of the form $2315132$). Hence, only an $\su(3)$ contribution is typically possible from the outermost $-1$ curves branching from the linear portion of a link away from its ends. When the T-junction is in the outermost (penultimate) position, however, a $\g_2$ summand may appear from (typically at most one of) these outermost $-1$ curves.  All possible configurations such inner and outer T-junctions appear in Table~\ref{t:TJunctionGS}.
     \begin{table}[!h] \small
    \begin{center}
      \begin{tabular} {ccccccccccccccc}
      Interior T-junctions & \\
      \hline
        & & $\overset{[\su(3)]}{\underset{\text{\Izero{}}}{1}}$ &  \\
      &  $\cdots\underset{\text{II}}{1}$ & $\underset{\text{\IVns{}}}{5}$ & $\underset{\text{II}}{1} \cdots$ & \\
    Exterior T-junctions & \\
      \hline  
        & & $\overset{[\su(3)]}{\underset{\text{\Izero{}}}{1}}$ &  \\
      &  $\cdots\underset{\text{II}}{1}$ & $\underset{\text{\IVns{}}}{5}$ & $\underset{\text{II}}{1}\ \ \ [\g_2]$ & \\
      \\
        & & $\overset{[\su(3)]}{\underset{\text{\Izero{}}/\text{\Ione{}}}{1}}$ &  \\
      &  $\cdots\underset{\text{II}}{1}$ & $\underset{\text{\IVns{}}}{5}$ & $\underset{\text{\Izero{}}/\text{\Ione{}}}{1} \ \ \ [\su(3)]$ & \\
      \\
        & & $\overset{[\g_2]}{\underset{\text{II}}{1}}$ &  \\
      &  $\cdots\underset{\text{\Izero{}}}{1}$ & $\underset{\text{\IVns{}}}{5}$ & $\underset{\text{II}}{1} \ \ \ [\g_2]$ & 
   \end{tabular}
        \caption{$-5$ curve T-junction flavor symmetry contributions from links.} 
        \label{t:TJunctionGS}
           \end{center} 
        \end{table}

\end{paragraph}
The remaining outermost $-1$ curve contributions from side-links with a node attachment arise in $\cdots 21$ and $\cdots31.$ 
\begin{paragraph}{Outermost $-1$ curves in $\cdots 31$}
We begin with treating the two special cases noted above in which such a $-1$ curve can appear over a $-3$ curve T-junction. 
\begin{itemize}
\item $2\overset{1}{3}21$: This link $L$ permits $\e_6$ and $\e_7$ node attachments. Both result in the same $\T_L$ with the only non-trivial flavor summand arising from the outer $-1$ curve. The same essential details persist for $\f_4$ attachments. All configurations are among
\begin{align}\label{eq:trivalentConfig}
   \underset{\text{III}}{2} \ \ \  \overset{  \underset{\text{\Izero{}}}{ \overset{[\so(8)]}{1}}    }{\underset{\text{\Izerostarss{}} }{3} }
 \   \ \ \underset{\text{III}}{2} \ \ \ \underset{\text{\Izero{} } }{1} \ \ \   \underset{\text{\IVstars{}/\IIIstar{}}}{5\leq m \leq 8}
\end{align}
\item $2_{\Sigma_L}\overset{1_{\Sigma_U}}{3_{\Sigma_M}}1_{\Sigma_R}$: The only node $\Sigma$ which may attach to this link $\alpha$ has $m_\Sigma=4,$ as required for blow-down consistency. Further attachment to $\Sigma$ is similarly barred making our discussion irrelevant for longer bases. We condense it accordingly while providing enough detail to illustrate a few subtleties. Gauge enhancements of $\alpha$ are highly constrained by the subquiver $23$ allowing only $\su(2),\so(7)$ and $\su(2),\g_2$ enhancements of~\eqref{eq:GS23} from which we read that $T_{\Sigma_L}\in\{\text{III,\IVns{}}\}$, the latter possible only in select $\g_2$ cases. 

Enhancing to $\g_\Sigma\cong\f_4$ requires $\g_{\Sigma_M}\cong\g_2$ and $T_{\Sigma_U}\sim\text{\Izero{}}$ since $T_{\Sigma_R}\sim\text{II}$; this yields $[\g_{\Sigma_U}]\cong\so(8).$ In the remaining cases, $\g_{\Sigma}\cong\so(N\leq 13).$ For $N=8,$ the maximum $[\g_{\Sigma_U}]\cong\so(8)$ requires $T_{\Sigma_R}\sim\text{\Izero{}}.$ When $\g_{\Sigma_M}\cong\g_2,$ the maximum $[\g_{\Sigma_U}]\cong\f_4$ requires $T_{\Sigma_U}\sim\text{II};$ this is reduced to $[\g_{\Sigma_U}]\cong\su(4)$ when $T_{\Sigma_U}\sim\text{\Ione{}}$, in turn demanding $T_{\Sigma_L}\sim\text{\IVns{}}$ and $\T_{14}\sim\text{\Izero{},\Izerostars{}}.$ For $N>8,$ we have $\g_{\Sigma_M}\cong\so(7);$ the above completes all $\g_2$ cases with those remaining having $T_{\Sigma_L}\sim\text{III}.$ This in turn requires $T_{\Sigma_R}\sim\text{\Inns<n\leq3>{}},$ and $T_{\Sigma_U}$ constrained by $n$ (since $d_{\Sigma_U}+d_{\Sigma_R}\leq 3$) with $[\g_{\Sigma_U}]$ then determined from compatible pairs $T_{\Sigma_U},T_{\Sigma_R}$ via Table~\ref{t:outerOneGS}. 

Note that an $\su(2)$ flavor summand can also arise from $\Sigma_M$ in $\so(7)$ cases with $d_{\Sigma_U}+d_{\Sigma_R}\leq 1.$ Non-trivial $[\g_{\Sigma}]$ may also arise when $N\geq 10$ (and $[\g_{\Sigma}]\subset\sp(4)$), as can non-trivial $[\g_{\Sigma_R}]\subset \su(2)$ provided $T_{\Sigma_R}\sim\text{\Inns<3>{}}$ and $\g_{\Sigma_M},\g_{\Sigma}\sim\so(7),\so(9).$ For $N=13,$ the maximum for $[\g_{\Sigma}]$ is $\sp(4).$ For $N=12,$ this becomes $\sp(3)$ for $T_{\Sigma_R}\sim\text{\Itwo{}}$ and $\sp(2)$ for $T_{\Sigma_R}\sim\text{\Inns<3>{}}.$ For $N=11$, these $\T_{\Sigma_R}$ cases instead yield $\su(2)\oplus \sp(2)$ and $\sp(2),$ respectively; these become $\su(2)$ and trivial summand, respectively, for $N=10.$ 
     \begin{table}[!h] \small
    \begin{center}
      \begin{tabular} {ccccccccccccccc}
        Type on $\Sigma$ & Compatible types on $\Sigma'$ & GS summand maxima from $\Sigma$ \\
        III & \Izero{} & $\su(2) \oplus \so(7)$ \\
        & & $\sp(3)$ \\
        \Inns<3>{} & \Izero{} & $\su(3)^{\oplus 3}$ \\
        & & $\so(13)$ \\
        & & $\su(2) \oplus \su(6)$\\
        & & $\su(3) \oplus \su(5)$\\
        \Itwo{} & \Izero{} & $\so(12)$ \\
        & & $\su(2) \oplus \su(3)^{\oplus 2}$\\
        & & $\su(2) \oplus \su(5)$\\  
        II & \Izero{} & $\g_2$ \\
        \Ione{} & \Izero{}/\Itwo{} &$ \so(9)$ \\
         & &  $\su(3)^{\oplus2}$ \\
         & & $ \su(3) \oplus \su(2)^{\oplus 2}$ \\ 
         \Izero{} & \Izero{}/\Itwo{}/\Inns<3>{} & $\so(8)$ 
   \end{tabular}
        \caption{$-1$ curve T-junction flavor symmetry contributions for $2\overset{1_\Sigma}{3}1_{\Sigma'}4$.} 
        \label{t:outerOneGS}
           \end{center} 
        \end{table} 
\end{itemize}

A similar case merits mention with those above, that for $12\overset{2}{3}1,$ a left-attachment permitting link. The structure of enhancements and flavor symmetries here follows at once from our discussion since this is simply a rearrangement of the curves in the first case above obtained by swapping the outer $-2$ and $-1$ curves. Details are thus also captured by~\eqref{eq:trivalentConfig}.

Having concluded discussion of the special cases involving T-junctions at a $-3$ curve, we now turn to the cases involving linear $\cdots 31$ terminations. We begin with a few cases meriting separate discussion as they have highly constrained enhancement structure.
\begin{itemize}
\item $\cdots 5131:$ There are many links of this form permitting left node-attachment. Since the $-5$ curve requires a gaugeless neighbor and the $\e_8$ gauging condition applies to the neighboring curve, the resulting configurations are highly limited. In fact, there are only two possible configurations for the inner curves and we can easily detail the resulting outer $-1$ curve flavor summand possibilities in each case as in Table~\ref{t:outerOneGS5131}. Note that an $\su(2)$ flavor summand may arise from the $-3$ curve in precisely the cases when it has a pair of gaugeless neighbors and Kodaira type \Izerostarns{}.
     \begin{table}[!h] \small
    \begin{center}
      \begin{tabular} {ccccccccccccccc}
        Type on $\Sigma_a$ & Type on $\Sigma_b$ & Type on $\Sigma_c$ & GS summand maxima from $\Sigma_c$ \\
        \Izerostarns{} & II & \IVns{} & $\g_2 \oplus \su(3)$\\ 
        & & & $\sp(2)$ \\
        &  & \Inns<3>{} &  $ \so(13)$ \\
        & & III & $ \so(7) \oplus \su(2) $ \\ 
        & & &  $ \sp(3)$ \\
        & & \Itwo{} & $\so(12)$ \\
        &  & II & $\f_4$\\ 
        &  & \Ione{} & $\su(4)$ \\        
        & & \Izero{} & $\so(8)$  \\
        \IVs{} & \Izero{} & \Izero{} & $\e_6$                 
   \end{tabular}
        \caption{$-1$ curve flavor symmetry contributions for $\cdots 5 1_{\Sigma_a} 3_{\Sigma_b}1_{\Sigma_c}$.} 
        \label{t:outerOneGS5131}
           \end{center} 
        \end{table} 

\item $\cdots 2_{\Sigma_a}3_{\Sigma_b}1_{\Sigma_c}$: Since the subquiver $23$ again severely restricts assignments on $\alpha\sim231$, we can easily list all configurations for links of this form along with $[\g_\Sigma]$ (as shown in Table~\ref{t:outerOneGS231}). These are the only flavor summands from $\alpha$ except for $\su(2)$ contributions from $\Sigma_b$ when $T_{\Sigma_b}\sim\text{\Izerostarss{}}$ while $T_{\Sigma_c}\in\{\text{\Izero{},\Ione{}}\}.$ Note that left attachment of an $\e_{\geq6}$ node to $12231$ allows only those configurations with Kodaira type assignments \IVns{},\Izerostarns{} to the subquiver $23.$
     \begin{table}[!h] \small
    \begin{center}
      \begin{tabular} {ccccccccccccccc}
        Type on $\Sigma_a$ & Type on $\Sigma_b$ & Type on $\Sigma_c$ & GS summand maxima from $\Sigma_c$ \\
        III & \Izerostarss{} & \Inns<3>{} & $\so(13)$ \\
          & & III & $ \so(7) \oplus \su(2) $ \\ 
           & & & $ \sp(3)$ \\
           & & \Itwo{} & $ \so(12)$ \\
           & & II & $ \sp(3)$\\
           & & & $\g_2$ \\
           & & \Ione{} & $\so(9)$ \\
           & & & $\su(3)^{\oplus 2}$ \\
           & & & $\su(3) \oplus \su(2)^{\oplus 2}$ \\
           & & \Izero{} & $\so(8)$ \\
        III/\IVns{}& \Izerostarns{}& II & $\f_4$\\ 
        & & \Ione{} & $\su(4)$ \\
        & & \Izero{} & $\so(8)$ 
   \end{tabular}
        \caption{$-1$ curve flavor symmetry contributions for $\cdots 2_{\Sigma_a} 3_{\Sigma_b}1_{\Sigma_c}$.} 
        \label{t:outerOneGS231}
           \end{center} 
        \end{table}
\item $\g 131:$ One can confirm via the comprehensive link listing appearing in Appendix D of~\cite{atomic} that the only remaining link allowing a left node to attach and $\cdots 31$ right termination is $\alpha\sim131$ (as all remaining links feature terminations $\cdots 5131$ or $\cdots 231$). We now work through the permitted node attachments to $\alpha$ while noting that only nodes carrying $\so(N)$ or $\e_6$ are permitted by the $\e_8$ gauging condition. The unique configuration when carrying $\e_6$ is
\begin{align}
\e_6 \ \ \ \underset{\text{\Izero{} } }{1} \ \ \ \overset{\su(3)}{\underset{\text{\IVs{}} }{3} }
 \ \ \ \underset{\text{\Izero{} } }{1} \ \ [\e_6] \ ,
\end{align}
and the unique flavor symmetry contribution from $n\alpha$ is also $\e_6.$

We abbreviate our discussion for $\so(N)$ attachments since explicit listing for each of the finitely many allowed configurations of this type captures all relevant details and is available from the listing provided for $4131$ via the accompanying workbook.

\end{itemize}

\end{paragraph}

\begin{paragraph}{Outermost $-1$ curves in $\cdots 21$}\label{s:quiversEnding21}
The only curves permitting $\e_7$ and $\e_8$ global symmetry summands are of this form. We proceed through the forms for endings of such links. 
\begin{itemize}
\item $\cdots 3221$: The unique flavor symmetry maximum arising from the outer $-1$ curve of such links is $\e_8.$ The type assignments on these curves must appear as
\begin{align*}
  \cdots\ \  \underset{\text{\Izerostarns{}}}{3} \ \ \ \underset{\text{\IVns{}}}{2} \ \ \ \underset{\text{II}}{2} \ \ \  \underset{\text{\Izero{}}}{1}\ \ [\e_8] \ .
\end{align*} 
\item $\cdots 2321$: The unique flavor symmetry maximum arising from the outer $-1$ curve of any such link with a node attachment is $\e_7.$ Configurations for these links simply appear as
\begin{align*}
  \cdots\ \ \underset{\text{III}}{2} \ \ \   \underset{\text{\Izerostarss{}}}{3} \ \ \ \underset{\text{III}}{2} \ \ \   \underset{\text{\Izero{}}}{1}\ \ [\e_7] \ .
\end{align*} 

\item $1 \overset{2}{3} 2 1_{\Sigma_R}. $ This link permits left attachment to a $-4$ curve $\Sigma$ provided $\g_\Sigma\cong\so(8).$ All curves give trivial flavor summand except $\Sigma_R$ which has $[\g_{\Sigma_R}]\cong\e_7.$

\item $\cdots 51321$: Links of this form permit various maximal global symmetry summands to arise from the outermost curve depending on the Kodaira types realizing the unique gauge assignment. All assignments are of the form
\begin{align*}
  \cdots\ \ \underset{\text{\IVns{}}}{5} \ \ \  \underset{\text{II}}{1} \ \ \  \underset{\text{\Izerostarns{}}}{3} \ \ \ \underset{T_2}{2} \ \ \   \underset{T_1}{1}\ \ [\g] \ , 
\end{align*} 
where $[\g]$ is determined from $T_1,T_2$ via Table~\ref{t:51321GSRules}. 
     \begin{table}[!h] \small
    \begin{center}
      \begin{tabular} {ccccccccccccccc}
        Type on $\Sigma_2$ & Types on $\Sigma_1$ & GS summand maxima from $\Sigma$ \\
        III & \Izero{} & $\e_7$ \\
        \IVns{} & \Izero{} & $\e_6$ \\
        \IVns{} & \Ione{} & $\su(7)$ \\
         & & $\su(N)\oplus\su(M), \ N+M= 8 ,\ N\geq 2$ \\
         & & $\so(11)$ \\
         \IVns{} & II & $\f_4$ \\
         & & $\sp(2)^\oplus 2$
   \end{tabular}
        \caption{$-1$ curve flavor symmetry contributions for $\cdots 513 2_{\Sigma_2} 1_{\Sigma_1}$.} 
        \label{t:51321GSRules}
           \end{center} 
        \end{table} 
        
\item $\cdots 3_{\Sigma_a}2_{\Sigma_b}1_{\Sigma_c}:$ Now that we have treated the special cases above with constrained gauge summands, we move to the general case which essentially merges the others. We again read the three assignments for $32$ from Table~\eqref{eq:GS23}. Permitted $[\g_{\Sigma_c}]$ appearing when $T_{\Sigma_b}\sim\text{\Izerostarns{}}$ follow from Table~\ref{t:51321GSRules}. When instead $T_{\Sigma_b}\sim\text{\Izerostarss{}},$ we follow the above discussion for $\cdots2321.$ Note that we have also now discussed all enhancements of the bare quiver $321$. The remaining links not in the above special case groupings are $(3)1321.$ These are quickly handled by short explicit workbook enabled listings. 
\end{itemize}
\end{paragraph}

\end{subsubsection}

\end{subsection}

\begin{subsection}{Instanton-links with a node attachment}\label{s:instantonLinksWithNodes}

In this section we detail the flavor symmetry contributions which can arise from instanton-links joined to nodes. The attachments of instanton-links to a $-4$ curve are limited in longer bases other than $(2)1414\cdots$ to have at most a single curve. We hence confine our analysis to $\e_{\geq 6}$ nodes as Section~\ref{s:41rules} treats $41$ and $412$ while $4122$ is treated by Sections~\ref{s:41rules},\ref{s:GS22} (in particular, by~\cref{eq:141GSConstraint,eq:mprime,eq:mprime2,eq:GS22Nonclassical,eq:22GSCondition}). Such attachments cannot occur in bases with more than one node. Consequently, their treatment does not affect our discussion of more elaborate bases. Note that instanton-link attachments to a $-5$ curve $\Sigma$ must have length $l\leq4$ with this becoming $l\leq2$ for $\Sigma$ an interior curve. The latter are covered by Table~\ref{t:GS512} and we relegate $l\geq 3$ cases to workbook listings since they are structurally similar to the $\e_6$ attachments treated explicitly here.

\begin{subsubsection}{Gaugeless instanton links}

We begin by discussing the flavor symmetry contributions from a gaugeless instanton-link $\alpha\sim\Sigma_1\Sigma_2\cdots$ attached to an $\e$-type node $\Sigma_0\leftrightarrow\g_0$ with self-intersection $-m_0.$ The permitted $\T_\alpha$ appear in Table~\ref{t:gaugelessInstantonLinks}.
     \begin{table}[!h]
    \begin{center}
      \begin{tabular} {ccccc}
      $m_0$ & $-1$ & $-2$ & $\cdots$ & $-2$ \\
  $\g_0$ & \Izero{} & \Izero{} & $\cdots$ & \Izero{} \\
  & [$\g_1$] & & $\cdots$ & [$\g_k$] \\
  & \\
  $\g_0$ & \Izero{} & \Ione{} & $\cdots$ & \Ione{} \\
  & [$\g_1$] & & $\cdots$ & [$\g_k$] \\  
    & \\
  $\g_0$ & \Izero{} & II & $\cdots$ & II \\
    & [$\g_1$] & & $\cdots$ & [$\g_k$] 
   \end{tabular}
        \caption{All possible gaugeless instanton-link type assignments permitting attachment to an $\e$-type node for links with at least three curves.} 
        \label{t:gaugelessInstantonLinks}
           \end{center} 
        \end{table} 
        
The flavor summands [$\g_i$] which can arise depend on $\T_\alpha$ and $\g_0.$ We will proceed by moving through each $\g_0$ option. We first treat instanton-links having at least three curves before returning to discuss a few exceptions to the following rules for short instanton-links.
\begin{itemize}
\item $\g_0 \cong \e_8:$ In this case, all flavor summands [$\g_i$] are trivial.

\item $\g_0 \cong \e_7:$ Only [$\g_1$] can be non-trivial. Assignments featuring type II curves give all [$\g_i$] trivial, while the others allow $[\g_1]\cong\su(2).$

\item $\g_0 \cong \e_6:$ Again, only [$\g_1$] can be non-trivial. For the last type assignment form above, non-trivial $[\g_1]\subset\su(2)$ can arise. Those without type II curves each can allow $[\g_1]\cong\su(3).$ 
\end{itemize}

To treat the short instanton-links having length $l\leq2$, we can simply extract these configurations from \eqref{eq:GS51} and Table~\ref{t:GS512}, the former directly treating $l=1.$ For $l=2$ these simply obey the following.
\begin{itemize}
\item $\g_0 \cong \e_8:$ In this case, $[\g_2]\cong\su(2)$ may arise for second and third type assignment forms above.

\item $\g_0 \cong \e_7:$ For the second type assignment form above, $[\g_{i>0}]\cong \su(2)$ may are yielding an $\su(2)^{\oplus 2}$ flavor summand from $\alpha.$ The only non-trivial contribution which may arise for the third form above appears as $[\g_2]\cong\su(2).$  

\item $\g_0 \cong \e_6:$ The summands which can arise in the three cases for link type assignment forms are $[\su(3),0], [\su(3),\su(2)],$ and $[\su(2),\su(2)],$ respectively. 
\end{itemize}
Note that for $l=1,$ allowed values of $[\g_1]$ are simply the maximal complementary Lie subalgebras of $\g_0$ in $\e_8.$

\end{subsubsection}

\begin{subsubsection}{Gauged instanton links}
In this section we consider an $\e_q$ node attached to an instanton-link $\Sigma_1\cdots \Sigma_k$ carrying any non-trivial gauge summand. We will break our discussion into two parts. The first concerns ``classical enhancements'' of the form
\begin{align}
      \overset{\text{\Ins<n_0>{}}}{-1} \  \overset{\text{\Ins<n_1>{}}}{-2} \ \overset{\text{\Ins<n_2>{}}}{-2} \  \cdots \ \overset{\text{\Ins<n_k>{}}}{-2} \ .
\end{align}
The second addresses enhancements involving any other Kodaira type, these being confined among types II, III, IV, and \Izerostar{}.

\begin{paragraph}{Classical enhancements of an instanton link with node attachment}
While we have already characterized flavor summands which can arise in subquivers of this form via characterizations for $12$ and $22\cdots2$ in Section~\ref{s:GSRules}, we pause here to make a few comments. First note that $\e$-node attachment requires $T_{\Sigma_0}\sim\text{\Izero{}}$ and this imposes $\g_{\Sigma_1}$ is a subalgebra of $\su(3),$ $\su(2),$ and $n_0$ in the cases for attachment to $\e_6,\e_7,$ and $\e_8,$ respectively. Strong restrictions $\T_\alpha$ follow via propagation of constraints along the link. 

The possible $[\g_\alpha]$ for various classical enhancements are subalgebras of those for a particular enhancement with $n_1\leq n_2 \leq \cdots \leq n_k.$ Convexity conditions for these enhancements as detailed in~\cite{atomic} follow from the rules of Section~\ref{s:GSRules}. These require that for enhancements of such bases having increasing arguments of $\su(n_i)$ algebras with $k>2$ and $i>1,$ we have $n_{i+1}-n_{i}=m_{q}$ with $n_1 = m_q \leq 9-q.$ 

The aforementioned maximal $[\g_\alpha]$ yielding configuration among all enhancements of $\e_q\alpha$ with $q$ fixed has $n_1$ chosen to allow $m_q$ and consequently $n_k$ as large as possible. The resulting $[\g_\alpha]$ appears as
\begin{align}\label{eq:InInstantonChain}
   \e_q \  
   \underset{-}{\overset{\text{\Izero{}}}{-1}} \  
   \underset{-}{\overset{\text{\Ins<n_1>{}}}{-2}} \ 
   \underset{-}{\overset{\text{\Ins<2n_1>{}}}{-2}} \  \cdots \ 
   \underset{[\su((k+1)n_1)]}{\overset{\text{\Ins<kn_1>{}}}{-2}} \ .
\end{align}

The remaining classical enhancements of the instanton-link are required via the convexity conditions for classical enhancements $-2$ curve chains to have $n_i$ increasing to a maximal $n_{i_\text{max}}$ at some $i_\text{max}$ and are non-increasing for $i>i_\text{max}.$ Flavor summands appear only from the curves immediately to the left of any decreases and from the final $-2$ curve with ranks are governed by~\eqref{eq:222convexity}. Explicit workbook enabled listings of the permitted flavor summands for these remaining classical enhancements can be quickly procured.

\end{paragraph}

\begin{paragraph}{Non-classical instanton-link enhancements with node attachment}
In cases with any curve of the instanton-link having a Kodaira type other than \In{}, every curve other than the $-1$ curve is prevented from having type \In{}. These $\T_{\alpha}$ and $\g_i$ obey convexity conditions on $d_i$ and $r_i\equiv\text{rank}(\g_i)$ reading
\begin{align}
d_1\leq d_2 \leq \cdots \leq & d_m\geq d_{m+1}\geq d_{m+2}\geq \cdots\\
r_1\leq r_2 \leq \cdots \leq & r_{m}\geq r_{m+1} \geq r_{m+2} \geq \cdots  \ .
\end{align}

All possible forms for enhancements of instanton-links not of the form~\eqref{eq:InInstantonChain} with $\e$-type node attachment appear in Table~\ref{t:nonclassicalInstantonLinks} with any flavor summands indicated. Intersection contribution tallying eliminates several possibilities which obey the required gauging and convexity conditions, e.g.\ $\T_{\alpha}$ beginning as II,III,IV. Trivial $[\g_{1}]$ results for $q\geq7$ while $q=6$ allows $[\g_{1}]\cong\su(2)$ provided $T_{\Sigma_1}$ is gaugeless.
     \begin{table}[!h] \tiny
    \begin{center}
    \bgroup
    \def\arraystretch{0.9}
      \begin{tabular} {ccccccccccccccc}
      $\g_a\cong \e_{\geq 6}$: & \\
      \hline
      $m_0$ & $-1$ & $-2$ & \multicolumn{7}{c}{$\cdots$} & $-2$ \\
  $\g_a$ & \Izero{} & II & III & III & $\cdots$ & $\cdots$ & & $\cdots$ & III & II \\
  & [$\g_b$]   \\
  \\
  $\g_a$ & \Izero{} & II & \IVns{} & \IVs{} & \multicolumn{4}{c}{$\cdots$} & \IVs{} & \IVns{}/III/II \\
  & [$\g_b$] \\
  \\  
  $\g_a$ & \Izero{} & II & III & & \multicolumn{4}{c}{$\cdots$} & III & III \\
  & [$\g_b$] 
  & & & & & & & & & $[\su(2)]$ \\
  \\
  $\g_a$ & \Izero{} & II & \IVns{} & \IVs{} & \multicolumn{4}{c}{$\cdots$} & \IVs{} & \IVs{} \\
  & [$\g_b$] 
  & & & & & & & & & $[\su(3)]$ \\
  \\
  $\g_a$ & \Izero{} & II & \IVns{} & \Izerostarns{} & \IVns{} & (II) \\
  & [$\g_b$] &  &  & [$\sp(2)$] &  &  \\
  \\
  $\g_a$ & \Izero{} & II & \IVns{} & \Izerostarns{} & III/\IVns{} \\
  & [$\g_b$] &  &  & [$\sp(2)$] &  &  \\
    \\
  $\g_a$ & \Izero{} & II & \IVns{} & \IVs{} & \IVns{} & (II) \\
  & [$\g_b$] &  &  &  & &  \\ \\
  $\g_a$ & \Izero{} & II & \IVns{} & \IVns{} & (II) \\
  & [$\g_b$] &  &  &  &   \\ \\
  $\g_a$ & \Izero{} & II & \IVns{} & \Izerostarns{} \\
  & [$\g_b$] &  &  & [$\sp(3)$]  \\     \\
  $\g_a$ & \Izero{} & II & \IVns{} & II \\
  & [$\g_b$] &  & [$\su(3)$] &  \\  \\   
  $\g_a$ & \Izero{} & II & \IVns{}  \\
  & [$\g_b$] &  & [$\g_2$]  \\  \\    
  $\g_a$ & \Izero{} & II & III  \\
  & [$\g_b$] &  & [$\su(2)$]  \\  \\      
      $\g_a \cong \e_{\leq 7}$: & \\
      \hline  
  $\g_a$ & \Izero{} & III & III & \multicolumn{5}{c}{$\cdots$} & III & II \\
  & & [$\su(2)$] 
  & & & & & & &  \\
  \\      
  $\g_a$ & \Izero{} & III & III & \multicolumn{5}{c}{$\cdots$} & III & III \\
  & & [$\su(2)$] 
  & & & & & & & & $[\su(2)]$ \\
  \\       
  $\g_a$ & \Izero{} & III & \IVs{} & \multicolumn{5}{c}{$\cdots$} & \IVs{} & \IVs{} \\
  & & 
  & & & & & & & & $[\su(3)]$ \\
  \\        
  $\g_a$ & \Izero{} & III & \IVs{} & \multicolumn{5}{c}{$\cdots$} & \IVs{} & III/\IVns{} \\
  & & 
  & & & & & & & & \\
  \\ 
  $\g_a$ & \Izero{} & III & \Izerostarns{} & III/\IVns{} \\
  &  &  &  [$\sp(2)$] &  &  \\
    \\  
  $\g_a$ & \Izero{} & III & \Izerostarss{} & III/\IVns{} \\
  &  &  &  [$\sp(3)$] &  &  \\
    \\ 
  $\g_a$ & \Izero{} & III & \Izerostarns{} \\
  &  &  &  [$\sp(3)$] &  &  \\
    \\           
  $\g_a$ & \Izero{} & III & \Izerostarss{} \\
  &  &  &  [$\sp(4)$] &  &  \\
    \\            
  $\g_a$ & \Izero{} & III & II \\
  &  &  [$\su(2)$] &  &  &  \\
    \\             
   $\g_a$ & \Izero{} & III & \IVns{} \\
   &  &  & &  &  \\
     \\         
      $\g_a \cong \e_6$: & \\
      \hline     
  $\g_a$ & \Izero{} & \IVs{} & \IVs{} &\multicolumn{5}{c}{$\cdots$} & \IVs{} &  \IVs{}  \\
  & &  $[\su(3)]$
  & & & & & & & &  $[\su(3)]$ \\
  \\               
 $\g_a$ & \Izero{} & \IVns{} & \IVs{} &\multicolumn{5}{c}{$\cdots$} & \IVs{} &  \IVs{}  \\
  & & 
  & & & & & & & &  $[\su(3)]$ \\
  \\                 
  $\g_a$ & \Izero{} & \IVs{} & \IVs{} &\multicolumn{5}{c}{$\cdots$} & \IVs{} &  \IVns{} (II) / III \\
  & &  $[\su(3)]$
  & & & & & & & &  \\  
  \\
  $\g_a$ & \Izero{} & \IVns{} & \IVs{} &\multicolumn{5}{c}{$\cdots$} & \IVs{} &  \IVns{} (II) / III  \\
  & & 
  & & & & & & & &  \\    
  $\g_a$ & \Izero{} & \IVns{} & \Izerostarns{} & III/\IVns{} \\ 
  & & 
  & $[\sp(2)]$ & & & & & & &  \\  
  \\   
  $\g_a$ & \Izero{} & \IVns{} & \IVns{} &(II)   \\
   \end{tabular}
   \egroup
        \caption{Non-classical gauged instanton-link type assignments with attachment to an $\e$-type node for links with at least three curves.} 
        \label{t:nonclassicalInstantonLinks}
           \end{center} 
        \end{table} 

\end{paragraph}

\end{subsubsection}

\end{subsection}
\end{section}

\begin{section}{Gauging global symmetries}\label{s:gaugingGS}
We now briefly discuss which flavor symmetries can be consistently gauged to yield a new SCFT, i.e.,\ which 6D SCFT transitions result from promotion of a global symmetry subalgebra to a gauge summand. While room for additional gauging is permitted should we consider the theories coupled to gravity where $B$ is compact (when more strongly, all continuous symmetries must then be gauged as argued for example in~\cite{banks2011symmetries}), the permitted SCFT transitions resulting from global symmetry promotion are highly constrained and have not been fully studied. In this section, we comment briefly on which of these transitions are visible within F-theory using the tools we now have at our disposal.

As a first step, we shall inspect the flavor symmetry maxima for single curve theories to examine when promoting global symmetry subalgebras to gauge summands carried on added neighboring curves results in at most reduced-rank gaugings. Consider an SCFT base with a single compact component of the discriminant locus $\Sigma$ with $\g_\Sigma\equiv\g_M$ and a choice $\ggs$ from among the geometrically realizable flavor symmetry maxima for such a theory. We now ask which other 6D SCFT bases with specified gauge assignment can arise via promotion of a $\g_{\text{GS}}$ subalgebra to neighboring curve gauge summands then yielding a configuration
\begin{align}\label{eq:neighbors}
  \overset{[\ggst]}{\underset{\g_M}{m_M}} \ \ \underset{\g_R}{m_R}\ , \qquad \qquad
 \underset{\g_L}{m_L} \ \  \overset{[\ggst]}{\underset{\g_M}{m_M}} \ \ \underset{\g_R}{m_R}\  ,\qquad \text{or}\qquad  \underset{\g_L}{m_L} \ \ \overset{\underset{\g_U}{m_U}}{ \overset{[\ggst]}{\underset{\g_M}{m_M}}} \ \ \underset{\g_R}{m_R} \ .
\end{align}
This of course requires that $\g_L \oplus \g_R \oplus \g_U\subset \g_{\text{GS}}$ (with terms omitted for absent neighboring curves in the first two cases).
Note that the new base may have a different discrete $U(2)$ subgroup $\Gamma$ associated to it. We can instead study a variant of this approach by requiring $\Gamma$ stays fixed or focus on the nature of permitted $\Gamma$ transitions, but we shall focus on whether $\ggst$ can be made trivial while $\g_L \oplus \g_R \oplus \g_U$ has smaller rank than $\g_{\text{GS}},$ i.e.,\ whether ``sub-maximal gauging'' can occur. We will soon refine this characterization since we often cannot gauge sufficiently many degrees of freedom via an SCFT transition to yield trivial $\ggst,$ as we now show.

\begin{subsection}{Normal crossings constraints}
We now pause to observe that the normal crossings condition barring three gauged neighbors from meeting a $-1$ curve provides a variety of cases where SCFT transitions fully gauging $\ggs$ (or any of its maximal subalgebras) away to neighboring curves is not possible. All single curve enhancements of a $-1$ curve having global symmetry maxima with at least three summands are examples since each neighboring curve can contribute only a single gauge summand. In particular, this restriction applies to the cases shown in Table~\ref{t:incompleteGauging} which we can read off from Table (6.1) of~\cite{BertoliniGlobal1} with the further tightenings for type III and IV curves appearing above in Table~\ref{t:updatedSummary}. 
{ \small\setlength{\tabcolsep}{5pt} \begin{table}[!h] \begin{center}\begin{tabular}{cccccc}
Gauge algebra on $\Sigma$ &  Kodaira type & Flavor symmetry maxima\\
$\su(2)$ & III & $\so(7)\oplus \so(7) \oplus \su(2)$ \\
             & \IVns{} & $ \g_2 \oplus \g_2 \oplus \su(3) $ \\
$\su(3)$ & \IVs{} &  $\su(3)^{\oplus 4}$ \\
                       & & $\su(3)^{\oplus 2} \oplus \sp(2) $ \\
$\so(8)$ & \Izerostars{} & $\sp(3)\oplus\sp(3) \oplus \sp(1)^{\oplus 3}$
\end{tabular}
\caption{Selected single curve theories on a $-1$ curve permitting only sub-maximal gauging of flavor symmetry maxima to yield neighboring curve gauge summands for a valid SCFT base.}
\label{t:incompleteGauging}
\end{center} \end{table} } 

Note that this restriction is irrelevant in determining which theories can be eliminated for lack of a consistent gauging of global symmetries upon considering compact bases and coupling to gravity. In that context, we may relax the positive definite adjacency matrix condition to allow multiple $-1$ curves intersecting $\Sigma.$ Hence, full gauging of the flavor symmetries in the cases of Table~\ref{t:incompleteGauging} again becomes plausible in that context.

Having trimmed the types of configurations where we might find more meaningful sub-maximal gaugings, we move on and discard this form of restriction as a minor curiosity. 

\end{subsection}

\begin{subsection}{Global to gauge symmetry promotion and 6D SCFT transitions}
To motivate a more useful notion of ``sub-maximal gauging,'' we now detail an example illustrating that complete gauging may be possible for a particular choice of resulting base while others can yield a somewhat surprising loss of continuous degrees of freedom. Consider a $-2$ curve with gauge algebra $\so(7).$ The unique flavor symmetry maximum for this single curve theory is $\ggs\cong \sp(1) \oplus \sp(4).$ While we are able to gauge the entire flavor symmetry and can even do so to yield an SCFT base, e.g.,\ $221$ with the configuration 
\begin{align*}
 \underset{(\text{III},\su(2))}{2} \ \ \ \underset{(\text{\Izerostarss{}},\so(7))}{2_\Sigma}  \ \ \  \underset{(\text{\Inns<8>{}},\sp(4))}{1} \ ,
\end{align*}
we cannot do so for the base $\alpha \sim2\overset{2}{2_\Sigma}2.$ Furthermore, can we cannot even gauge the neighboring curves in this base to obtain a maximal subalgebra of $\g_{\text{GS}}.$ Hence, even though the number of neighboring curves we can add allows for the possibility of gauging a maximal Lie subalgebra of $\g_{\text{GS}},$ careful inspection of the available type assignments to these curves reveals only sub-maximal gauging is permitted. 

More surprisingly, the neighboring curve gauge summands together with the remaining global symmetries on $\Sigma$ in the presence of these curves gives a sub-maximal sub-algebra $\ggst\oplus \su(2)^{\oplus 3}$ of $\ggs.$ The required gauging of $\g_{\text{GS}}$ to form this base in fact gives a rank-reducing breaking of $\ggs,$ as we now confirm. Consider that the neighboring $-2$ curves must have Kodaira type III for intersection with $\Sigma.$ The resulting configuration appears as
\begin{align*}
 \underset{(\text{III},\su(2))}{2} \ \ \ \underset{[\ggst]}{\overset{\underset{(\text{III},\su(2))}{2}}{\underset{(\text{\Izerostarss{}},\so(7))}{2_\Sigma}}}  \ \ \  \underset{(\text{III},\su(2))}{2} \ 
\end{align*}
and requires contributions to $(\at,\bt,\dt)_\Sigma\sim(\geq4,\geq6,12)$ for the T-junction given by at least $(3,\geq6,9).$ As a result, the only symmetry bearing curve Kodaira types which can simultaneously intersect $\Sigma$ are \Inns<\leq 3>{} and III. Further gauging of any remaining global symmetry is prevented by adjacency matrix requirements for SCFT bases. The remaining global symmetry for the resulting base (which arises purely from $\Sigma$ and hence matches $\ggst$) together with the neighboring curve gauge summands yields a sub-maximal algebra $\g_{\text{max}}$ of $\g_{\text{GS}}$ since 
\begin{align*}
\widetilde{\g}_{\text{GS}}\oplus \su(2)^{\oplus 3} \subset \su(2)^{\oplus 4} \subset \su(2)^{\oplus 5} \subset \g_{\text{GS}}.
\end{align*}
This phenomenon motivates our working definition of {\it sub-maximal gauging} to be a case in which the sum of neighboring gauge algebras with the residual global symmetry algebras $\ggst$ along $\Sigma$ are not maximal splittings of $\ggs$ in the sense of being among the (relatively) maximal subalgebras of $\ggs$ having the required number of summands to match the count of non-trivially gauged neighboring curves. Observe that the discrete $U(2)$ subgroup, $\Gamma_{2}$ associated to the original base with the single compact curve $\Sigma$ is trivial, while that for the base in the above gauging, namely $\Gamma_{\alpha},$ is not. This suggests that the emergence of non-trivial discrete $U(2)$ gauge fields may be an ingredient in determining permitted global symmetry gauging rules and hence a helpful tool in classifying 6D SCFT RG flows.

Considering this phenomenon more generally defines an interesting structure associating to each single curve flavor symmetry maximum the collection of relatively maximal algebras which can be gauged given fixed neighboring curves defining a base with discrete $U(2)$ subgroup $\Gamma$. Said differently, we have a distinguished class of Lie subalgebras for each of the flavor symmetry maxima with each member in this class of subalgebras associated to a discrete $U(2)$ subgroup. Additional structure emerges since not all $\Gamma$ associated permissible SCFT gauging transitions are shared for the distinct flavor maxima with fixed Kodaira type. Further refinements to the data appear since multiple Kodaira types in some cases can realize a given gauge algebra.

A broad survey of single curve gaugings appears in Table~\ref{t:incompleteGaugings}.
{\setlength{\tabcolsep}{2pt} \begin{table}[!h] \tiny \begin{center}
\bgroup
\def\arraystretch{0.85}
\begin{tabular}{ccccccccc} 
$\gg$ on $\Sigma$ &  Type & GS max. & Gaugable & Base(s) & Types & $\g_{\text{gauged}}\oplus [\widetilde{\g}_{\text{GS}_\Sigma}]$ max(s) & Sub-maximal \\
$m=1$: \\
\hline
$\su(2)$ & \Itwo{} & $\so(20)$ & $\checkmark$ & $14$ & $(-)$\Instars<6>{} \\
\\
               & & & X & $12$ & $\overset{[\text{\Ins<3>{}}/\text{\IVs{}}]}{(-)}\text{\Instarns<3>{}}$ & $\so(13)^*\oplus[\su(3)]^\dagger$ & $\checkmark$ \\
\\            
               & & & & & $\overset{[\text{\Instars<3>{}}]}{(-)}\text{\IVs{}}$ & $\su(3)\oplus[\so(14)]$ & X \\
\\            
               & & & & & $\overset{[\text{\Ins<7>{}}]}{(-)}\text{\IVs{}}$ & $\su(3)\oplus[\su(7)]^\dagger$ & $\checkmark$ \\
\\           
               & & & & & $\overset{[\text{\Instars<4>{}}]}{(-)}\text{III}$ & $\su(2)\oplus[\so(16)]$ & X \\
\\            
               & & & X & $12,13$ & 
               $\underset{N+M=2}{\overset{[\text{\Instars<N>{}}]}{(-)}\text{\Instars<M>{}},}$ & $\so(2M+8)\oplus[\so(2N+8)]$ & X \\
\\            
               & & & & & 
               $\underset{N+M=2}{\overset{[\text{I${}_N^{ns/ss}$}]}{(-)}\text{I${}_M^{ns/ss}$},} $ & $\so(2M+7)\oplus[\so(2N+7)]$ & $\checkmark$ \\
\\                       
               & & & X & $312,313$ & 
               $\underset{N+M=2}{\text{\Instars<N>{}}{(-)}\text{\Instars<M>{}},}$& $(\so(2M+8)\oplus\so(2N+8))^*$ & X \\
\\            
             & III & $\so(7)^{\oplus 2} \oplus \su(2)$ & X${}^{!!}$ & $313,213$ & $\text{\Izerostarss{}}\overset{[\text{III/\Itwo{}}]}{(-)}\text{\Izerostarss{}}$ & $(\so(7)^{\oplus 2 })^\ast\oplus[\su(2)]^\ddagger$ & X \\ 
\\
             & \IVns{} & $ (\g_2^{ \oplus 2})^\ast \oplus \su(3) $ & X${}^{!!}$ & $313,213$ &
             $\text{\Izerostarns{}}\overset{[\text{\IVs{}/\Ins<3>{}}]}{(-)}\text{\Izerostarns{}}$ & $\g_2^{\oplus 2}\oplus[\su(3)]^\ddagger$ & X \\ 
\\ 
$\su(3)$ & \IVs{} &  $\su(3)^{\oplus 4}$ & X${}^{!!}$ & $12$ &  $\overset{[\text{3\IVs{}}]}{(-)}\text{\IVs{}}$ & $\su(3)^*\oplus [\su(3)^{\oplus 3}]^\ddagger$ & X \\
\\
& & $\su(3)^{\oplus 2} \oplus \sp(2) $  & X${}^{!!}$ & $12$ &  $\overset{[\text{\Ins<3>{}}/\text{\IVs{}},\text{\Inns<4>{}}]}{(-)}\text{\IVs{}}$ & $\su(3)^*\oplus [\su(3)\oplus \sp(2)]^\ddagger$ & X \\
\\
$\so(8)$ & \Izerostars{} & $\sp(3)^{\oplus2}\oplus \sp(1)^{\oplus 3}$ & X${}^{!}$ & $12$ & $\overset{[\text{2\Inns<6>{},(III/\Itwo{}/\Inns<3>{}))}]}{(-)}$III & $\sp(1)^\ast \oplus [\sp(3)^{\oplus2}\oplus \sp(1)]^\dagger$ & $\checkmark$ \\
\\
$m=2$: \\
\hline
$\su(3)$ & \IVs{} &  $\su(3)^{\oplus 2}$ & $\checkmark$ & $222$ &  $\text{\IVs{}}{(-)}\text{\IVs{}}$ & \\
\\
& & & $\checkmark$ & $221$ &  $\text{\IVs{}}{(-)}(\text{\Ins<3>{}}/\text{\IVs{}})$ &  \\
\\
& & $\sp(2)$ &$\checkmark$ & $21$ &  ${(-)}\text{\Ins<4>{}}$ &  \\
\\
& & & X & $222$ &  $(\text{III/\IVns{}}){(-)}(\text{III/\IVns{}})$ & $\su(2)\oplus\su(2)$ & X \\
\\
$\su(2)$ & \IVns{} &  $\g_2$ & $\checkmark$ & $23,22,21$ & $(-)$\Izerostarns{} &  \\
\\
& & & &  $123$ & (\Izero{}/\Ione{}/II)$(-)$\Izerostarns{} &  \\
\\
& & & &  $223,222$ & II$(-)$\Izerostarns{} &  \\
\\
& & & X &  $222$ & II$(-)$\IVs{} & $\su(3)$ & $\checkmark$ \\
\\
& III & $\so(7)$ & $\checkmark$ & $23,22,21$ & $(-)$\Izerostarss{} &  \\
\\
& III & $\so(7)$ & $\checkmark$ & $123,122$ & \Izero{}$(-)$\Izerostarss{} &  \\
\\
& & & X & $222$ & II$(-)$\IVs{} & $\su(3)^*$ & $\checkmark$ \\
\\
& & & X & $222$ & III$(-)$III & $(\su(2)^{\oplus 2})^*$ & $\checkmark$ \\
\\
$\so(7)$ & \Izerostarss{} & $\sp(4)\oplus \sp(1)$ & $\checkmark$ & $221$ & 
III$(-)$\Inns<8>{} \\
\\
& & & X & $222$ & III$\overset{[\text{\Inns<6>{}}]}{(-)}$III & ($\sp(1)^{\oplus 2})^*\oplus[\sp(3)]$ & X\\
\\
& & & X & $2\overset{2}{2}2$ & III$\underset{[\text{III/\Inns<3>{}}]}{\overset{\text{III}}{(-)}}$III & $(\sp(1)^{\oplus 3})^*\oplus[\sp(1)]$ & $\checkmark$ \\
\\
\end{tabular}
\egroup
\caption{Selected gaugings of single curve SCFT GS maxima. Here `$\dagger$' indicates that a GS factor is not gaugable onto any additional neighboring compact curve allowed in any SCFT base, `$\ddagger$' the same due to the normal crossings condition, `X${}^!$' that the GS (relative) maximum is not fully gaugable for any SCFT base, and `X${}^{!!}$' the same due to the normal crossings condition. Here `${}^*$' indicates a relatively maximal gauging for the given base and `${}^{**}$' the same among all bases; unique gauging for a given base is indicated with an `${}^{\ast!}$' symbol.}
\label{t:incompleteGaugings}
\end{center} \end{table} } 

\end{subsection}

\end{section}

\begin{section}{Conclusions and outlook}\label{s:conclusions}
We have carried out a systematic investigation global symmetries for each 6D SCFT with a known F-theory realization having no frozen singularities, namely those appearing in the classification of~\cite{atomic}. We have produced a tentative classification of the geometrically realizable global symmetries of these theories. The tools we have provided include an implementation of an algorithm enabling explicit listing of the Kodaira type realizations for each 6D SCFT gauge enhancement, thus helping to complete the geometry to field theory ``dictionary'' for these theories. We have detailed the structure of global symmetries permitted via this algorithm in terms of the ``atomic decomposition'' of 6D SCFT bases from~\cite{atomic} and in terms of certain shorter chains which may occur. This has enabled us to recast our findings via short listings and simple constraint equations for these symmetries in terms of the geometric realizations of each gauge theory, i.e.,\ the Kodaira type assignments compatible with each gauge assignment. We have made the latter manifest, resolving certain ambiguities in the classification appearing in~\cite{atomic}.

In the process, we have eliminated some of the CFTs detailed in the classification of~\cite{atomic} and shown that the refined classification can be recast in purely geometric terms without appeal to anomaly cancellation. We have also investigated the gauging of 6D SCFT global symmetries which yield transitions between SCFTs and found that these gaugings can result in rank reductions. We have derived novel restrictions on Calabi-Yau threefold elliptic fibrations, carried out local analysis of nearly all permitted singular locus collisions, and found many local and global constraints on permitted collections of singular fiber degenerations for such fibrations. This provides steps towards a general analog of~\cite{PerssonsList} by constraining singular fiber degenerations along chains of curves with arbitrary Kodaira types. Our approach strongly constrains the space of non-compact CY threefolds and makes explicit all potentially viable varieties of this type at finite distance in the moduli space meeting a transverse singular fiber collision hypothesis up to specification of non-compact singular fibers not associated to a non-abelian algebras (with this latter caveat easily removed by trivial adjustments in our algorithm).

We hope these tools prove useful in the classification of 6D RG flows and a complete classification of all 6D SCFTs. In particular, whether the multiple global geometrically realizable global symmetry maxima which arise in many cases correspond to distinct theories (i.e.,\ if there terms under which the global symmetries of F-theory models provide SCFT invariants) is a question we leave to future work. While the relations between gauge and global symmetries of 6D SCFTs with those of discrete $U(2)$ gauge fields determining endpoints we have discussed are suggestive, we hope that additional investigation may help clarify the precise interplay between these ingredients determining the field content of these theories.
\end{section}

\acknowledgments
We extend our sincere thanks to Marco Bertolini for providing essential foundational ideas enabling this work and for significant ongoing input throughout its completion.
We also thank David R. Morrison for proposing this project and providing helpful suggestions, Daryl Cooper, Denis Labutin, and Mihai Putinar for support and advice, and Tom Rudelius for useful discussions.

\appendix

\begin{section}{Intersection contributions and forbidden pair intersections}\label{s:theRestrictions}
In this appendix we extend the intersection contributions data collected in~\cite{BertoliniGlobal1,gaugeless} to include that for pairs of curves with $A,B>0$ in which either curve may be compact. The lone exception to the latter concerns cases of non-compact transverse curves which carry no gauge-summand, namely those with Kodaira types \Izero{}, \Ione{}, or II. Such curves cannot contribute global symmetry summands and hence only require consideration in cases where the transverse curve can be compact component of the discriminant locus. 

Before proceeding, we note that studying the contributions to curves with $A>0$ or $B>0$ from transverse gauged fibers was safely ignored in~\cite{BertoliniGlobal1,gaugeless}. For the theories treated in those works, the maximal flavor symmetry inducing configurations arise for a curve with fixed Kodaira type when $A=B=0.$ However, in treating flavor symmetries for more general SCFTs with a base consisting of more than a single compact curve, the minimal values of $A,B$ along a given curve may be non-zero. For example the $-1$ curve in the base $(12)1_\Sigma2231513221(12)$ requires $\Sigma$ here has type \Izero{} with $A_{\Sigma}\geq 4.$ This minimum value of $A_{\Sigma}$ along $\Sigma$ depends significantly on the presence of non-neighboring curves, even those which do not affect the minimum gauging of the neighboring curves as illustrated in Table~\ref{t:nonlocalAB}. Note that the extent of ``non-locality'' is rather significant. For example, the quiver $(12)1_{\Sigma}22315$ requires the same minimum as $(12)1_{\Sigma}22315132$ given by $A_{\Sigma}=3,$ while addition of the gaugeless type II curve of self-intersection $-2$ in the final position to yield $(12)1_{\Sigma}223151322$ raises this minimum to $A_{\Sigma}=4.$ By the same token, the only assignments of orders of vanishing along each curve of the quiver $1223151322$ compatible with a left attachment of a non-compact \IIstar{} fiber inducing the (unique) $\e_8$ global symmetry maximum for this quiver have assignments to the $-1$ curve of the form $(A\geq 4,0,0).$ 

Our approach stems in part from this subtlety that the minimal values of $A,B$ are hence ``global'' in quiver position as illustrated by the above example. An exhaustive computer search approach to determining global symmetries beginning with assignment of orders of vanishing along each compact component of the discriminant locus is hence merited. Similarly, maximality checks of global symmetry subalgebras which may arise from transverse non-compact singular locus components also merits consideration of compact curves with $A,B>0$ in addition to $A=B=0$ cases. Hence, development of local models for all such intersections a natural first step as these do not appear to be available in the literature. 

We shall go somewhat further than required by developing the intersection contribution data from such local models for a somewhat broader class of intersections than actually arise in the minimal order assignments to SCFT bases. This approach has certain benefits. First, it allows us to be cavalier in assigning larger orders of vanishing and allowing automation to ensure we have not missed any global symmetry maxima inducing configurations which may require non-minimal $A,B$ assignments. Furthermore, this allows us to more tightly constrain non-minimal $A,B$ configurations which we hope may find application addressing codimension-two singularities and perhaps development of a framework which may come to bear on issues concerning the gauging of global symmetries beyond the scope addressed in this work.

{ \small\setlength{\tabcolsep}{5pt} \begin{table}[!h] \begin{center}\begin{tabular}{cccccc}
 Minimum value of $A$ along $\Sigma$ & Compatible subquiver(s)   \\
0 & $(12)1$\ \ \ \ \ \ \ \ \ \ \ \ \ \ \ \ \\
1 & $(12)12$\ \ \ \ \ \ \ \ \ \ \ \ \ \ \  \\
2 & $(12)122$\ \ \ \ \ \ \ \ \ \ \ \ \ \  \\
   & $(12)1223$\ \ \ \ \ \ \ \ \ \ \ \ \ \\
   & $(12)12231$\ \ \ \ \ \ \ \ \ \ \ \  \\
3 & $(12)122315$\ \ \ \ \ \ \ \ \ \ \ \\
   & $(12)1223151$\ \ \ \ \ \ \ \ \ \  \\
   & $(12)12231513$\ \ \ \ \ \ \ \ \ \\  
   & $(12)122315132$\ \ \ \ \ \ \ \   \\
4 & $(12)1223151322$\ \ \ \ \ \ \  \\   
  & $(12)12231513221$\ \ \ \ \ \   \\
  & $(12)12231513221(12)$
\end{tabular}
\caption{Minimum orders of vanishing of $f|_{\Sigma}$ for various subquivers of $(12)1_\Sigma2231513221(12)$ with compatible Kodaira type (and hence gauge) assignments given by truncations of the only viable assignment along this quiver, namely \IIstar{},\Izero{},II,\IVns{},\Izerostarns{},II,\IVstarns{},II,\Izerostarns{},\IVns{},II,\Izero{},\IIstar{}.}
\label{t:nonlocalAB}
\end{center} \end{table} } 

Permitted $A,B$ values in longer quivers are often highly constrained. Certain bases allow only a single globally consistent $A,B$ assignment though locally infinitely many choices are permitted. For example, the unique permitted values on the quiver $71231513221(12)$ are
\begin{align}\label{eq:UniqueABExample}
\overset{\e_7}{\underset{A=0}{\underset{(3,5,9)}{\overset{\text{\IIIstar}}{7}}}} \ \ 
\underset{B=1}{\underset{(0,1,0)}{\overset{\text{\Izero{}}}{1}}} \ \  
\overset{\su(2)}{\underset{B=0}{\underset{(1,2,3)}{\overset{\text{III}}{2}}}} \ \ 
\overset{\g_2}{\underset{A=B=0}{\underset{(2,3,6)}{\overset{\text{\Izerostarns{}}}{3}}}} \ \ \ 
\underset{A=0}{\underset{(1,1,2)}{\overset{\text{II}}{1}}} \ \ 
\overset{\f_4}{\underset{B=0}{\underset{(3,4,8)}{\overset{\text{\IVstarns{}}}{5}}}} \ \ 
\underset{A=1}{\underset{(1,1,2)}{\overset{\text{II}}{1}}} \ \ 
\overset{\g_2}{\underset{A=1}{\underset{(3,3,6)}{\overset{\text{\Izerostarns{}}}{3}}}} \ \ 
\overset{\su(2)}{\underset{A=1}{\underset{(3,2,4)}{\overset{\text{\IVns{}}}{2}}}} \ \ 
\underset{A=2}{\underset{(3,1,2)}{\overset{\text{II}}{2}}} \ \ 
\underset{A=3}{\underset{(3,0,0)}{\overset{\text{\Izero{}}}{1}}} \ \ 
\overset{\e_8}{\underset{A=0}{\underset{(4,5,10)}{\overset{\text{\IIstar{}}}{12}}}} \ ,
\end{align}
while the subquiver $13221(12)$ permits infinitely choices of these values consistent with the Kodaira type assignments above.

In some circumstances, the permitted $A,B$ values force particular gauge assignments. Consider for example the base $232.$ We can bypass anomaly cancellation machinery and more involved global analysis of the monodromy cover for \Izerostar{} fibers by noting that the only orders of vanishing consistent with the na\"ive intersection contribution constraints of~\eqref{eq:aprioriContributions} leave the only configuration
\begin{align}\label{eq:minimalABEx}
\underset{(1,2+B_L,3)}{\overset{(\text{III},\su(2))}{2}} \ \ \ 
\underset{B\geq 1}{\underset{(2,3+B,6)}{\overset{(\text{\Izerostarss{}},\so(7))}{3}}} \ \ \ 
\underset{(1,2+B_R,3)}{\overset{(\text{III},\su(2)}{2}} \ ,
\end{align}
where $B\geq 1$ requires that the \Izerostar{} fiber is semi-split (as we can read from Table~\ref{t:IzerostarMonodromyRules} along with the observation of~\ref{s:so8cover} that $\so(8)$ would have required purely even $\at$ contributions), thus yielding an $\so(7)$ gauge summand.

As an example yielding novel constraints, we consider the minimal left $\f_4$ attachment compatible enhancement of the interior link $1321$ with right $\e_7$-node attachement to a curve $\Sigma$ with self-intersection $-m$ having minimal orders of vanishing along each curve as indicated below.
\begin{align}
\underset{(1+A,1,2)}{\underset{(\text{II},-))}{1}} \ \ \ \underset{(2+A_{\Sigma_L},3+B_{\Sigma_L},6)}{\underset{(\text{\Izerostarns{}},\g_2)}{3_{\Sigma_L}}} \ \ \ 
\underset{(1,2+B_{\Sigma_M},3)}{\underset{(\text{III},\su(2))}{2_{\Sigma_M}}} \ \ \ 
\underset{(A=0,B_{\Sigma_R}\geq 1+B_{\Sigma},0)}{\underset{(\text{\Izero{}},-) }{1_{\Sigma_R}}} \ \ \ 
\underset{(3,5+B_\Sigma,10)}{\underset{(\text{\IIIstar{}},\e_7)}{m_\Sigma\overset{?}{=}8}}
\end{align}
Note that $\Sigma_R$ must type \Izero{} since gaugeless curves with type other than \Izero{} lead to non-minimal intersection with a curve having \IIIstar{} fiber; $B_{\Sigma'}\geq1$ follows immediately from the na\"ive residuals tallying requirements of~\eqref{eq:aprioriContributions}. Provided that $5\leq m_\Sigma\leq 7,$ there is no inconsitency with $B_{\Sigma_R}=1.$ However, when $m=8,$ na\"ive intersection contributions of~\eqref{eq:aprioriContributions} from $\Sigma_R$ to $\Sigma$ rule out the configuration. Since additional orders of vanishing of $\gt_{\Sigma_R}$ and $\gt_{\Sigma}$ are coupled, i.e.,\ $B_{\Sigma_R}\geq 1+B_{\Sigma},$ any attempt to permit the required vanishings along $\Sigma$ by raising $B_{\Sigma}$ in turn raises $B_{\Sigma_R}.$ However, this is not compatible with the indicated gauge assignment since it forces $B_{\Sigma_M}>0$ and in turn $B_{\Sigma_L}>0$, thus requiring an $\so(7)$ gauge summand along $\Sigma_M$ as we can read from Table~\ref{t:IzerostarMonodromyRules}. Further note that in these cases with $m_\Sigma=8,$ this further enhancement of $\Sigma_L$ to $\so(7)$ is unacceptable in the presence of a left $\f_4$ attachment (which we can easily confirm via the $\e_8$ gauging condition since $\so(7)\oplus\f_4 \not \subset \e_8$).

We next consider the quiver $2231322$ which minimally supports an assignment with Kodaira types leading to trivial contribution from the $-1$ curve (via ``Persson's list'' restrictions). Now consider raising the Kodaira type on the $-1$ curve to type II which allows for a potentially non-trivial flavor symmetry summand to appear there. Note that such a summand is {\it a priori} permitted while maintaining the $E_8$ gauging condition since $\g_2\oplus \su(2)\subset \f_4$. We have the following assignment with the minimal orders along each curve required by~\eqref{eq:aprioriContributions} indicated below.
\begin{align}\label{eq:ABExample}
\underset{(1,1,2)}{\overset{\text{II}}{2}} \ \ \ 
\underset{(2,2,4)}{\overset{\text{\IVns{}}}{2}} \ \ \ 
\underset{(3,3,6)}{\overset{\text{\Izerostarns{}}}{3}} \ \ \ 
\underset{A\geq1}{\overset{[\sp(1)]?}{\underset{(\geq2,1,2)}{\overset{\text{II}}{1}}}} \ \ \ 
\underset{(3,3,6)}{\overset{\text{\Izerostarns{}}}{3}} \ \ \ 
\underset{(2,2,4)}{\overset{\text{\IVns{}}}{2}} \ \ \ 
\underset{(1,1,2)}{\overset{\text{II}}{2}} \ \ \ 
\end{align}
The only possible global symmetry can only occur from a type \Itwo{} fiber. It hence is relevant in our analysis to determine the local models for intersections of curves with fibers of various Kodaira types having $A,B>0.$ 

Before turning to a systematic analysis of all intersection contributions of potential relevance to constraining F-theory SCFT models, we begin with a simple example to demonstrate the method we follow to derive the intersection contribution data appearing in this section. This data plays a key role in our algorithm to eliminate various gauge enhancements of quivers and possible global symmetry summand inducing non-compact transverse curve configurations. Additional examples for certain $A=B=0$ cases following the same approach can be found in~\cite{BertoliniGlobal1,gaugeless}.

The $A,B$ values which may appear in arbitrary configurations consisting of an SCFT base decorated with non-compact transverse curves are somewhat non-trivial to constrain {\it a priori}. Furthermore, the results of a comprehensive analysis of intersection contributions for all permitted intersections of Kodaira fibers may find alternative uses in constraining elliptically fibered Calabi-Yau threefolds more generally. We shall hence tabulate intersection contributions for nearly all curve pairs (with the exceptions of $-1,-1$ intersections and gaugeless non-compact fibers which clearly do not arise in SCFT bases with non-compact global symmetry inducing fiber configurations). Though many of the indicated intersections do not arise in minimal assignments for SCFT bases, these cases often become relevant when we consider any alternative settings which allow us to relax the positive definite adjacency matrix condition or motivate careful study of all options for codimension two singularities. Though extending the rest of our analysis to these settings is beyond the scope of this work, we hope that the tools developed here facilitate such investigations.

\begin{subsection}{Computing intersection contributions}
We now detail an example illustrating the method underlying our approach to determining intersection contributions. Certain subtleties make some cases significantly more involved than illustrated by our particularly simple choice of example. However, most relevant issues for the types of computations we carry out have been discussed at length in~\cite{BertoliniGlobal1,gaugeless}. A few novel subtleties arise in treating compact pair intersections as we shall discuss in various cases treated in this appendix.

As in the configuration above, we let $\Sigma$ be a type II curve along $\{z=0\}$ with orders of vanishing given by $(1+A,1,2)=(2,1,2)$ and $\Sigma'$ a type \Itwo{} curve along $\{\sigma=0\}$ (with orders of vanishing $(0,0,2)$). Let $P$ denote the point of their intersection at $\sigma =z=0.$ The general form for $f,g,\Delta$ of a type \Itwo{} curve can be read from (A.8) of~\cite{BertoliniGlobal1} as
\begin{align}
\label{eq:I2summary}
f &= -\ff1{48}\phi^2 + f_1\sigma + f_2\sigma^2+O(\sigma^3)~,\nonumber\\
g &= \ff1{864}\phi^3 -\ff1{12}\phi f_1\sigma + (\gt_2 - \ff1{12}\phi f_2)\sigma^2 +  O(\sigma^3)~, \nonumber\\
\Delta &= \ff{1}{16}\left( \phi^3 \gt_2 - \phi^2f_1^2 \right)\sigma^2 + O(\sigma^3)~.
\end{align}
Imposing these divisibility conditions then yields
\begin{align}
\label{eq:I2Example}
f &= -\ff1{48}z^2\phi^2 +z^2 f_1\sigma + z^2f_2\sigma^2+O(z^2\sigma^3)~,\nonumber\\
g &= \ff1{864}z^3\phi^3 -z^3\ff1{12}\phi f_1\sigma + (z\gt_2 - z^3\ff1{12}\phi f_2)\sigma^2 +  O(z\sigma^3)~, \nonumber\\
\Delta &= \ff{1}{16}z^4\left( \phi^3 \gt_2 - \phi^2f_1^2 \right)\sigma^2 + O(z\sigma^3)~,
\end{align}
where we have replaced $\phi/z,f_i/z^2,\gt_2/z$ with $\phi,f_i,\gt_2$ for simplicity of notation.
We read the minimal orders of vanishing giving ``intersection contributions'' to $\Sigma$ via the minimum $\sigma$-degrees of the lowest order terms in $\ft_\Sigma,\gt_\Sigma,\Deltat_\Sigma$ (namely those of $z$-degrees $2,1,2,$ respectively, as required to match the $z$-orders along $\Sigma$) to obtain
\begin{align}
(\at_P,\bt_P,\dt_P)_{\Sigma} \geq (0,2,4).
\end{align}
Thus, the potential $\sp(1)$ flavor symmetry summand appearing in~\eqref{eq:ABExample} is impossible. Note this would have yielded a configuration with transverse curve algebras contained in the global symmetry maxima for a type II curve appearing in~\cite{gaugeless}. The contributions analysis therein does eliminate the configuration, but requires the above augmentation to produce the {\it lowered} $\at_P$ minimum indicated above. 

\end{subsection}

\begin{subsection}{Preliminaries}
In the following sections, we let $\Sigma$ be a curve at $\{z=0\}$ with $-\Sigma\cdot \Sigma = m$, having transverse intersections with curves $\Sigma_{j}'$ located at $\{\sigma_j = 0\}.$ Let the orders of vanishing along $\Sigma$ be $(a,b,d)$ and those along $\Sigma_j'$ be given by $(a_j',b_j',c_j').$
\end{subsection} 

\begin{subsection}{Type II curve intersections}\label{s:II}
We introduce further constraints on collisions involving a Kodaira type II curve in~\ref{s:Instar} for cases involving an \Instar{} curve. Here we focus on the remaining non-trivial cases, those involving an \In{} curve. Such collisions were studied at length in~\cite{gaugeless} when the \In{} curve is non-compact and $A=0$ along the type II curve. The tight restrictions we find in sections~\ref{s:III},\ref{s:IV} for type III and IV curves have only weaker analogues here. Underlying this distinction is that we are farther from non-minimality in II,\In{} collisions. As a consequence, we are required to consider intersections where the most general form for the relevant local models of type \In{} curves is unknown since the cases $7\leq n\leq 9$ are included.

We now proceed to make mild generalizations of the intersection contribution data first appearing in~\cite{gaugeless} that are required in the present work. Part of our work is dispatched by reading from Table~\ref{t:maxn} taken from~\cite{BertoliniGlobal1}. We find that $A\geq3$ is not possible for intersections with \In{} curves having $n\geq 4,$ a bound we revise here since this only holds for non-compact \In{} curves. The $A=0$ results can be read directly from the contribution tables of~\cite{gaugeless}. In one such case, we find a small correction. This leaves us to determine the relevant contributions from compact \In{} curves for all $n,$ and non-compact curves only for $n\leq 4.$ Since type II curves do not carry a non-abelian gauge algebra, we can safely ignore collisions of compact \In{} curves with non-compact type II curves.

We collect intersection contributions for the remaining cases in Table~\ref{t:IIm1_InNoncompactContributions}. The `$!$' symbol there indicates disagreement with~\cite{gaugeless}. Entries marked with `$\dagger$' are not permitted via the inductive form for \In{} curves appearing as (A.25)-(A.28) in~\cite{BertoliniGlobal1}. The `$*$' symbol indicates the contributions are only valid for a non-compact \In{} curve, `X.' that the intersection is valid only for non-compact \In{} curve, and `X..' that the intersection exceeds numbers of vanishings available for a type II curve even with $m=1.$ Entries to the right of those indicated with an `X', `X.', `X..' are similarly forbidden.

 \begin{table}[htbp] \footnotesize
  	\begin{center} \setlength{\tabcolsep}{2pt}
  		\begin{tabular}{c  |  c c c c c c c c c c c c c}
  		\diaghead(-5,6){cccccccccccc}%
  		{$n_{ns/s}$}{$A=a-1$}  &  0 & 1 & 2 & 3 & $\cdots$ & $\cdots$ & $a-1$\\ 
            \hline      
            $2$ & $(1,2,4)$ & \multicolumn{6}{c}{$\cdots(a \mod 2,2,4)(X. \text{ if } A\geq4) \cdots$} \\
            $3_{ns}$ & $(1,2,4)$ & $(0,3,6)$ & $(1,3,6)^*X.$ & \multicolumn{4}{c}{ $\cdots(a \mod 2,3,6)^*X. \cdots$ } \\
            $3_{s}$  & $(2,3,6)$ & $(2,3,6)$& $(2,3,6)$ & \multicolumn{4}{c}{ $\cdots(2(1-\delta_{4,a}),3,6)(X. \text{ if } A\geq4)^* \cdots$ } \\
            $4_{ns}$ & $(2,4,8)$ & $(0,4,8)$ & $(2,4,8)^*$X. & \multicolumn{4}{c}{ $\cdots(2 (a \mod 2),4,8)^*X. \cdots$ } \\
            $4_{s}$  & $(2,4,8)^!$ & $(2,4,8)$ & \multicolumn{4}{c}{ $\cdots(2 (1-\delta_{a,4}),4,8)(X. \text{ if } A\geq 4)^* \cdots$} \\                                    
            $5_{ns}$ & $(2,4,8)$ & $(0,5,10),(2,4,8)^*$ & $(1,4,8)^*$ & \multicolumn{3}{c}{$\cdots (2,4,8)^*$X.$\cdots$} \\
            $(\geq5)_{s}$ & X \\                                    
            $6_{ns}$ & $(2,4,8)$ & $(0,6,12),(2,4,8)^*$ & \multicolumn{4}{c}{$\cdots(2,4,8)^*$X.$\cdots$} \\           
            $7_{ns}$ &  $(\leq3,\leq6,\leq12)$ & X..${}^\dagger$ \\                                              
            $8_{ns}$ & $(\leq 4,\leq 8,\leq 16)^\dagger$X..${}^\dagger$ \\                                              
            $9_{ns}$ & $(\leq 4,\leq 8,\leq 16)^\dagger$X..${}^\dagger$ \\  
            $(n\geq 10)_{ns}$ & $(\lfloor \frac{n}{2}\rfloor,2\lfloor \frac{n}{2}\rfloor,4\lfloor \frac{n}{2}\rfloor)$X.. \\
  		\end{tabular}
  		\caption{Intersection contributions to type II curve from a (non-compact${}^*$) \In{} curve.}
  		\label{t:IIm1_InNoncompactContributions}
  	\end{center}
  \end{table}

\begin{subsubsection}{Intersection with an \Izerostars{} curve}
The analysis for intersection of a type II curve with a transverse curve holding type \Izerostars{} in~\cite{gaugeless} bars such intersections but implicitly uses that $B=0$ along the \Izerostars{} curve, say $\Sigma'$, to obtain non-minimality of the intersection. Provided $B\geq2,$ such intersections are also non-minimal. However, in the special case that $B=1$ factorization of the monodromy is possible. We can read from~\eqref{eq:IzerostarBNonzero} that intersection contributions to $\Sigma$ are then given by $(2+(a \mod 2),4,8)$ and contributions to \Izerostars{} curve are $(2\lceil a/2 \rceil , 1,4\lceil a/2 \rceil  ).$
\end{subsubsection}

\end{subsection} 

\begin{subsection}{Type III curve intersections}\label{s:III}
\noindent Let $\Sigma$ be a curve with Kodaira type III and orders of vanishing $(a,b,3) = (1,2+B,3)$. Suppose that $\Sigma'$ has orders of vanishing $(a',b',d').$ Note $\Sigma$ has odd type, making $\dt$ contributions thrice those of $\at$ contributions. The only technical cases for such intersections concern transverse curves with type \In{} or \Instar{}. The latter are restricted for $n\geq 1$ due to non-minimality. We treat intersections with \Izerostar{} curves in section~\ref{s:Izerostar}, focusing here on treating $\Sigma'$ with type \In{} in each case relevant to global symmetry computation, namely those involving at least one compact curve. \\
 
\subsubsection{III, \In{} intersections}
 
 \paragraph{Type III with $B\geq 0$ intersection with a non-compact type \In{} curve}
 
We now extend the contribution data for cases with $B=0$ analyzed in~\cite{BertoliniGlobal1} to those with $B>0.$  Reading from Tables A.1,A.2 of~\cite{BertoliniGlobal1}, we find values for the local intersection contributions to a III with $B=0$ and that the restriction that $B\geq 2$ requires $n\leq 3$ for non-minimality. We make a correction to this bound; the revisions appear in the appendices as Tables~\ref{t:maxn},\ref{t:sumloccontr}. First, we collect the results of contributions for intersections analyzed~\cite{BertoliniGlobal1} and the remaining $B>0$ cases of contributions to a type III with $m=1,2$ from a non-compact transverse \In{} curve in Table~\ref{t:IIIm1_InNoncompactContributions}. Note that in the cases with $n\geq 7,$ we compute contributions working from the inductive Tate forms for \In{} curves from (A.25)-(A.28) of~\cite{BertoliniGlobal1} that are potentially not the most general when $7\leq n \leq 9.$ Here $\Sigma$ has residuals given by $(5, 8+B, 15)$ for $m=1$ and $(2,4+2B,6)$ when $m=2.$ 

In Table~\ref{t:IIIm1_InNoncompactContributions}, entries indicated with `$*$' are only valid for non-compact \In{} curves and those with `$\dagger$' are permitted only for \In{} non-compact. Those entries with `$!$' correct Table A.2 of~\cite{BertoliniGlobal1}. The symbol `X' indicates a non-minimal intersection and entries to the right of an `X' are also non-minimal. For $m=2,$ those entries with any contribution exceeding those allowed are forbidden; these are indicated with a subscript `1'.

  \begin{table}[htbp]
  	\begin{center} \setlength{\tabcolsep}{1pt}
  		\begin{tabular}{c  |  c c c c c c c c c c}
  		\diaghead(-3,2){cccccccccccc}%
  		{$n_{ns/s}$ }{ $B$ }  &  0 & 1 & 2 & 3 & $\dots$ & $b-2$\\ 
            \hline      
            $2_{ns}$  & $(1,1,3)$  & $(1,0,3)$ & $(1,1,3)$ & $\dots$ & $\dots$ &  $(1,b \mod 3, 3)$ \\
            $3_{ns}$  & $(2,2,6)$  & $(3,0,9)_1,(2,2,6)^*$ & $(2,2,6)^*$ & $(2,2,6)^\dagger$ & $\dots $ & $(2,2,6)^\dagger$ \\
            $3_{s}$  & $(2,2,6)$  & $(2,2,6)$ & $(2,2,6)^\dagger$ & $\dots$ &  $\dots$ & $(2,2,6)^\dagger$ \\
            $4_{ns}$  & $(2,2,6)^!$  & $(3,0,9)_{{}_1},(2,2,6)^*$ & $(2,2,6)^\dagger $ & $\dots$ &  $\dots$ & $(2,2,6)^\dagger$  \\
            $4_{s}$  & $(2,3,6)$  & $(2,3,6)$ & $(2,3,6)^\dagger$ & $\dots$ & $\dots$ & $(2,3,6)^\dagger$  \\                                    
            $5_{ns}$  & $(3,3,9)_{{}_1}$  & $(3,0,9)_{{}_1}$ & X &   \\
            $(\geq5)_{s}$  & X &   &   &  &  & \\                                    
            $6_{ns}$  & $(3,3,9)_{{}_1}$  & $(5,0,15)_{{}_1}$ & X &   \\           
            $(10\geq n\geq7)_{ns}$  & $(\lceil\frac{n}{2}\rceil,\lceil\frac{n}{2}\rceil,3\lceil\frac{n}{2}\rceil)_{{}_1}$  & X & X &  \\                                               
            $(\geq 11)_{ns}$& X \\
  		\end{tabular}
  		\caption{Intersection contributions to III from a (non-compact) \In{} curve.}
  		\label{t:IIIm1_InNoncompactContributions}
  	\end{center}
  \end{table}

 \paragraph{Type III with $B\geq 0$ and $m=2$ intersection with a compact \In{} curve with $m=1$.}
 The only case of interest for III,\In{} intersections with \In{} compact arising in F-theory quivers involve a type III curve with $m=2$ since two $-1$ curves do not intersect in any valid base. Furthermore, when $m'=2$ on $\Sigma'$ with type \In{} has $(\at,\bt)=(0,0),$ intersection with any type III curve is thus prevented. We collect the contributions for valid intersections in Table~\ref{t:III_m2_In_ContributionsToFromBothCompact}. For large $n,$ it will be helpful in making our table succinct to define
 \begin{align}\label{eqn:qfcn}
                 q(n) &=\begin{cases}4 \text{ if $n$ is even}\\ 
                             5 \text{ if $n$ is odd.}
                             \end{cases}
 \end{align}
 The table entries indicate the minimal intersection contributions to/from a $\Sigma$, a (compact unless marked with `X') type III curve having orders $(1,b,3)=(1,2+B,3)$ with $m=2$ transversely intersecting $\Sigma'$, a compact \In{} curve with $m'=1.$ Note the residuals on a (compact) type III curve with $m=2$ are $(2,4+2B,6)$. Here, `X'/`X.' indicate an intersection forbidden by the number of allowed vanishings along $\Sigma/\Sigma',$ respectively, and `$-$' a non-minimal intersection. Entries to the right of an `$-$', `X', or `X.' are similarly forbidden. Intersections corresponding to entries with an `X' are permitted only for non-compact $\Sigma$.
  \begin{table}[htbp]
  	\begin{center} \footnotesize \setlength{\tabcolsep}{2pt}
  		\begin{tabular}{c  |  c c c c c c c c c c}
  		\diaghead(-3,2){cccccccccccc}%
  		{$n_{ns/s}$  }{ $B$ }  &  0 & 1 & 2 & 3 & 4 & $\geq 5$ \\ 
            \hline      
            $1_{}$  & $(1,1,3)/(2,3,5)$  & $(1,0,3)/(2,3,5)$ & \multicolumn{3}{c}{$\dots \ (1,b \mod 3,3)/\left ( 2\lceil \frac{b}{3} \rceil, 3\lceil\frac{b}{3}\rceil,4\lceil \frac{b}{3}\rceil+1\right) \dots \ $} & X. \\            
            $2_{}$  & $(1,1,3)/(2,3,4)$  & $(1,0,3)/(4,6,6),$ & $(2,1,6)/(4,6,8)$ & $(2,1,6)/(4,6,9)$ & $(2,0,6)/(4,6,9)$ & X. \\
             & & $(2,0,6)/(2,3,5)$ & & & & \\
             $3_{ns}$ & $(2,2,6)/(2,3,5)$ & X$/(2,3,5)$ & X/X. & \\
             $3_{s}$ & $(2,2,6)/(4,6,6)$ & $(2,2,6)/(4,6,7)$ & $-$ \\
             $4_{ns}$ &$(2,2,6)/(2,3,4)$& X$/(2,3,5)$ & X/X. & \\
             $4_{s}$ & $(2,3,6)/(4,6,6)$ & $(2,2,6)/(4,6,6)$ & $-$ \\
             $5_{ns}$ & X$/(2,3,5)$ & X$/(2,3,5)$ & X/X. \\
             $(\geq5)_{s}$ & $-$ &  \\
             $6_{ns}$ & X$/(2,3,4)$& X$/(2,3,5)$& X/X. \\
             $(\geq7)_{ns}$ & X$/(2,3,q(n))$& X$/(2,3,5)$& X/X. \\
   		\end{tabular}
  		\caption{Intersection contributions for (compact $m=2$) type III, compact type \In{} intersections.}
  		\label{t:III_m2_In_ContributionsToFromBothCompact}
  	\end{center}
  \end{table}
  
 \paragraph{Intersection contribution to a compact \In{} curve with $m=1$ from a non-compact III with $B\geq 0$.} The remaining intersections of interest between type III and type \In{} curves concern contributions to an \In{} curve from a type III curve giving a global symmetry summand. The relevant contributions to the type \In{} curves here are also given in Table~\ref{t:III_m2_In_ContributionsToFromBothCompact}. Entries are marked with an `X' when we require non-compactness (or allowing $m=1$, which requires we can relax our adjacency matrix requirements) along the type III condition to hold.

\end{subsection}

\begin{subsection}{Type IV curve intersections}\label{s:IV}
Here we collect information about monodromy rules for type IV curves and intersection contributions involving transverse curves. The only subtle cases involve type \In{} and \Izerostar{} curves since \Instar{} intersections with $n>0$ results in non-minimality.

\subsubsection{Preliminaries}
Let $\Sigma$ be a curve with type IV. Orders of vanishing are then given by $(a,b,4) = (2+A,2,4)$. Note that $\Sigma$ has even type. Reading from~\ref{t:monodromyCoverTable}, we see the monodromy along $\Sigma$ is determined by whether $\frac{g}{z^2}|_{\Sigma}$ is a square; the larger gauge algebra, $\su(3),$ occurs if so.

\subsubsection{Intersections of IV with \In{} curves}
 \paragraph{IV with $\mathbf{A\geq 0}$ meeting \In{} curves for global symmetry or quiver intersections}
 This case is detailed in~\cite{BertoliniGlobal1} when the transverse type \In{} curves are non-compact and $A=0$ along the type IV curve. These contributions appear with a minor correction for intersection with an \Inns<3>{} fiber (which also holds for the compact pair case) in Table~\ref{t:sumloccontr}. We shall use these contributions for the compact pair case when $A=0$ as then they remain unchanged for these type pairings. Note that the actual minimal contributions must be modified from these table values (as they depend on monodromy along the IV curve) to give even $\bt$ contributions to the type IV in the \IVs{} case. From Table (6.1) of~\cite{BertoliniGlobal1}, we have that transverse \In{} curves carry at most $\sp(4)$ symmetry. Note that as indicated in Table~\ref{t:maxn}, $n\leq 3$ is required when $A>0$ for non-minimality. We collect the results of contributions to (and from when applicable) a type IV curve intersecting a type \In{} curve for these remaining cases in Tables~\ref{t:IV_InContributionsPosA} and \ref{t:IV_InContributionsZeroA}, separating the $A=0$ case in which $n>3$ is permitted.
\begin{table}[htbp]
  	\begin{center}\small \setlength{\tabcolsep}{2.3pt}
  		\begin{tabular}{c  |  c c c } 
  		\diaghead(-1,1){ccccccc}%
  		{\hspace{-.4in}$A_{(ns/s)}$}{$n_{(ns/s)}$\hspace{-.4in}}  &  $2_{ns}$  & $3_{ns}$ & $3_{s}$  \\  		
            \hline      
            $1_{ns}$  & $(1,2,4)//(4,6,8)$  & -/$(1,3,6)$ & $(1,3,6)/(1,2,4)//(4,6,8)$  \\
            $1_{s}$   & $(1,2,4)//(4,6,8)$  & -/$(1,3,6)^*$  & $(1,3,6)^*/(1,2,4)//(4,6,8)^*$  \\
            $2_{ns}$   & $(0,2,4)//(4,6,8)$ & -/$(0,3,6),(1,2,4)$  & $(0,3,6)/(1,2,4)//(4,6,8)$  \\            
            $2_{s}$   & $(0,2,4)//(4,6,8)$ & -/$(0,3,6)^*,(1,2,4)$   & $(0,3,6)^*/(1,2,4)//(4,6,8)^*$  \\
            $A\geq 3_{ns},$ $2+A\in 4\Z$ & -/ $(0,2,4)$  & -/$(0,3,6)/(1,2,4)$ & -/$(0,3,6),(1,2,4)$  \\
            $A\geq 3_{ns},$ $2+A\in2\Z\setminus 4\Z$ & -/ $(0,2,4)$  & -/$(0,3,6),(1,2,4)$ & -/$(1,3,6),(1,2,4)$  \\
            $A\geq 3_{ns},$ $A\notin 2\Z$ & -/ $(1,2,4)$  & -/$(1,3,6)$ & -/$(1,3,6),(1,2,4)$  \\
            $A\geq 3_{s},$ $2+A\in 4\Z$ & -/ $(0,2,4)$  & -/$(0,3,6)^*/(1,2,4)$ & -/$(0,3,6)^*,(1,2,4)$  \\
            $A\geq 3_{s},$ $2+A\in2\Z\setminus 4\Z$ & -/ $(0,2,4)$  & -/$(0,3,6)^*/(1,2,4)$ & -/$(1,3,6)^*,(1,2,4)$  \\
            $A\geq 3_{s},$ $A\notin 2\Z$ & -/ $(1,2,4)$  & -/$(1,3,6)^*$ & -/$(1,3,6)^*,(1,2,4)$  \\
  		\end{tabular}
  		\caption{Intersection contributions to a (compact) IV${}^{ns/s}$ with $A>0$ from$//$to a transverse compact/non-compact \In{} curve (with `to contributions' only when compact). An `X' indicates the intersection is forbidden by non-minimality considerations and a `-' an intersection forbidden by residuals considerations. Here (${}^*$) indicates an intersection allowed only when $m=1$ on IV via limitations of $\bt$ on IV.}
  		\label{t:IV_InContributionsPosA}
  	\end{center}
  \end{table}

  \begin{table}[htbp]
  	\begin{center}
  		\begin{tabular}{c  |  c c c c c c }
  		\diaghead(-3,2){Ccccccccccc}%
  		{$n_{(ns/s)}$}{  IV${}^{ns}/$IV${}^s$\hspace{-.1in}}  &  $\su(2)$  & $\su(3)$ \\ 
            \hline      
            $2_{ns}$  & $(0,2,4)$ & $(0,2,4)$  \\
            $3_{ns}$    & $(0,3,6),(1,2,4)$ & $(0,3,6)^*,(1,2,4)$  \\
            $3_{s}$    & $(1,3,6),(1,2,4)$ & $(1,3,6)^*,(1,2,4)$  \\
            $ \geq 4_{s}$  & X & X  \\                        
            $4_{ns}$    & $(0,4,8)$ & $(0,4,8)^*$  \\            
            $9_{ns} > n\geq 5_{ns}$    & $(0,n,2n)^*$ & $(0,2\left \lceil \frac{n}{2} \right \rceil ,4\left \lceil \frac{n}{2} \right \rceil)^*$  \\            
            $\geq 9_{ns}$    & - & -  \\            
  		\end{tabular}
  		\caption{Intersection contributions to IV${}^{ns/s}$ with $A=0$ from a transverse non-compact \In{} curve. An X indicates the intersection is forbidden by minimality considerations. Here (${}^*$) indicates a permitted intersection only when IV has $m=1$, and (-) an intersection $\bt$ on IV forbids in all cases.}
  		\label{t:IV_InContributionsZeroA}
  	\end{center}
  \end{table}
\end{subsection}

\begin{subsection}{Type \Izerostar{} curve intersections}\label{s:Izerostar}
Here we compile information about monodromy rules for \Izerostar{} curves and intersection contributions to residuals counts involving transverse curves from an \Izerostar{} curve. 

\subsubsection{Preliminaries}
Let $\Sigma= \{\sigma = 0 \}$ be a curve with type \Izerostar{} and orders of vanishing $(a,b,6) = (2+A,3+B,6)$. Suppose that $\Sigma' = \{z = 0 \}$ is a transverse curve with orders of vanishing $(a',b',d').$  Recall that we refer to the cases with $2b' =3a'$ as `hybrid type', those with $2b'>3a'$ as `odd type' and those with $2b'<3a'$ as `even type'. The contributions to residual vanishings of $\Delta$ along $\Sigma'$ from intersection with the transverse curve $\Sigma$ are given by $3\at$ and $2\bt$ in the odd and even type cases, respectively, where $\at$ and $\bt$ are the residuals contributions to $\Sigma'$ from a curve $\Sigma$ for vanishings of $f,g$, respectively. 

The monodromy along $\Sigma$ is determined by whether $Q(\psi) = \psi^3 + (f/\sigma^2)|_{\{\sigma=0\}} \psi + (g/\sigma^3)|_{\{\sigma=0\}}$ is fully split (giving algebra $\so(8)$), partially split (giving algebra $\so(7)$), or irreducible (giving gauge algebra $\g_2$). 
 
\paragraph{Gauge algebra $\g_2$:} Here we can write the monodromy cover as $Q = \psi^3 + p \psi + q$, with $Q$ irreducible.\\

\noindent{$\mathbf{A>0:}$} In this case, $Q$ becomes $\psi^3 + q$ with irreducibility implying that $q$ is not a cube; otherwise, we could factor $Q$ as $(\psi - \alpha)(\psi^2 + \alpha \psi+ \alpha ^2 )$ with $q = - \alpha^3.$  Since $q = (g/\sigma^3)|_{\sigma=0},$ we conclude that $\Sigma$ cannot have intersections with curves $\Sigma_j'$ with types $(-,3b_j',-)$ using all available $g$ residuals along $\Sigma.$  Said differently, $g = \sigma^3( \nu \phi^3 + g_4 \sigma^4 + O(\sigma^5))$ with $\nu$ cube-free and $\deg(\nu) > 0.$  \\

\noindent {$\mathbf{B>0:}$} Here $Q = \psi (\psi^2 + p),$ and since this is not an irreducible cubic, this case is not possible. 

\paragraph{Gauge algebra $\so(7)$:} Since the cover is split in this case, $Q = (\psi - \alpha) ( \psi^2 + \lambda \psi + \mu).$  Since the $\psi^2$ term vanishes, we have $\alpha = r$ and $$Q = \psi^3 + (\mu - \lambda^2) \psi - \mu \lambda.$$ 

\noindent{$\mathbf{ A>0:}$} This case requires $\mu = \lambda^2,$ and hence $Q = \psi^3 - \lambda^3.$  This means that $Q$ is fully split since $\psi^2 + \lambda \psi + \mu = \psi^2 + \lambda \psi + \lambda^2,$ which has discriminant $\lambda^2 - 4 \lambda^2 = -3 \lambda^2,$ which indeed has a square root. We conclude this case is prohibited. \\

\noindent{$\mathbf{B>0:}$} In this case, $Q = \psi(\psi^2 + \mu)$, and that this is not fully split implies $\mu = (f/\sigma^2)|_{\sigma=0}$ is not a square. Hence, $\Sigma$ cannot have intersections with curves of types $(2a_j', - , -)$ using all $f$ residuals along $\Sigma$ and we conclude that $\Sigma$ cannot receive purely even $f$ intersection contributions using all $f$ residual vanishings. Said another way, $f = \sigma^2 ( \mu \phi^2 + f_3 \sigma + O(\sigma^2))$ with $\mu$ square-free and $\deg(\mu) >0.$  When $m=3,$ $\at =2$, but the form of $f$ clearly bars even contributions to $\at$ since $\deg(\mu)>0.$ Hence, intersection with a type II curve with orders $(2,1,2)$ is prohibited as are intersections with an \Izero{} with orders $(2,0,0).$  

\paragraph{Gauge algebra $\so(8)$:}  The monodromy cover is fully split here and appears as $(\psi - \alpha) (\psi - \beta) (\psi - \gamma).$  To have the $\psi^2$ term vanish we have $$Q = \psi^3 + (-\beta^2 - \alpha^2 - \alpha \beta) \psi - \alpha \beta (\alpha+\beta).$$  Now we note that for $m=4,$ we have that for either $A,B>0$ this is the only possible gauge algebra. For $A>0,m=4$, we have $\bt=0$ and hence the monodromy cover after appropriate rescaling appears as $\psi^3 +1$ and hence factors completely. For $B>0,m=4$ the cover appears as $\psi(\psi^2 +1)$ after rescaling since here $\at=0,$ and hence the cover can be fully split. \\

\noindent{$\mathbf{A>0:}$} In this case, $-\beta^2 -\alpha^2 - \alpha \beta  = 0.$   Substituting for $\beta^2$ using this identity gives $(g/\sigma^3)_{\sigma = 0} = -\alpha \beta (\alpha + \beta) = \alpha^3.$  Thus $g = \sigma^3 ( \alpha^3 + g_4 \sigma + O(\sigma^5)).$  This implies that all contributions to the residuals in $g$ come in multiples of 3. For example, we have larger than expected contributions to $g$ residuals from intersections with curves of type II. Since $A>0,$ inspecting the case of an intersection with type III shows that we have a $(4,6,12)$ point as another consequence. This follows since the remaining terms in $g$ are of total order at least 6. \\

\noindent{$\mathbf{B>0:}$} Here we instead have that $\alpha \beta (\alpha + \beta) = 0,$ giving three cases: $\alpha = 0, \beta = 0,$ and $\alpha = -\beta.$  In each, we have $Q = \psi^3 + (\alpha^2) \psi$ (renaming $\beta$ as $\alpha$ as needed). Thus, 
\begin{align}\label{eq:IzerostarBNonzero}
f = \sigma^2 (-\alpha^2 + f_3 \sigma + O(\sigma^2)).
\end{align} 
We conclude that all $\at$ contributions must be even.

\paragraph{Summary of Restrictions} 
We collect the restricted monodromy assignments in Table~\ref{t:IzerostarMonodromyRules}. 
\begin{table}[!h]
	\begin{center}
		\begin{tabular}{c | c c c }
          ~ & $A>0$  & $B>0$    \\
          \hline
               $\g_2$ & \checkmark  & X \\
              $\so(7)$ & X & \checkmark \\ 
              $\so(8)$ & \checkmark & \checkmark  
		\end{tabular}
		\caption{Forbidden monodromy assignments on \Izerostar{}$\ \sim (2+A, 3+B,6).$}
		\label{t:IzerostarMonodromyRules}
	\end{center}
\end{table}

\subsubsection{Intersections contributions from \Izerostar{}}
Here we study the contributions to the residual vanishings along $\Sigma'$ from a transverse intersection with $\Sigma$ in each of the cases above. 
\paragraph{Gauge algebra $\so(8)$:}\label{s:so8cover}
Using the above preliminaries, we have the following table of intersection contributions from $\Sigma$ with $\so(8)$ gauge algebra to $\Sigma'.$ This depends on the orders of vanishing along $\Sigma',$ whether the order in $g$ there is a multiple of three, and if $\Sigma'$ is of even or odd type. Recall that $\Sigma'$ with orders $(a',b',d')$ is of even type if $3a>2b$, odd type if $3a <2b,$ and hybrid type otherwise. We do not explore the latter case in Table~\ref{t:InstarRules}. Note that \Izero{} can appear in any of the three types as orders of vanishing for \Izero{} are given by $(a',b',0)$ with one of $a'$ or $b'$ necessarily zero. 
 \begin{table}[!h]
 	\begin{center}
 		\begin{tabular}{c | c c c }
           $A>0$ & $b' \equiv 0 \mod 3$  & $b' \not \equiv 0 \mod 3$    \\
           \hline
             ~  & $(2+A, 3, -)$ & $(a, 4,-)$  \\
             Even Type on $\Sigma'$:  & $(a, 3, 6)$ & $(a, 4,8)$  \\
             Odd Type on $\Sigma'$:  & $(a, 3, 3a)$ & $(a,4,3a)$ \\
 		\end{tabular}
 		\caption{Intersection contributions from \Izerostar{}${}^s \ \sim (a =2+A, b=3+B,6).$}
 		\label{t:InstarRules}
 	\end{center}
 \end{table}
\paragraph{Gauge algebra $\mathbf{\so(7)}$:}
Here we only need to study the case $B>0.$  When $\Sigma'$ has even type, we have minimal contributions to residuals along $\Sigma'$ given by $(2, b, 2b).$ In the odd type on $\Sigma'$ case these are instead $(2,b, 6).$  

\paragraph{Gauge algebra $\g_2$:} Only $A>0$ is relevant here.  We have contributions given by $(a,3,6)$ in the even type case and by $(a, 3, 3a)$ in the odd type case. 

\subsubsection{Restricted tuples}  We now discuss some consequences of the above \Izerostar{} restrictions. Residual vanishing counts along $\Sigma$ before any intersections are $(8-2m + 2A, 12-3m + 3B, 24 - 6m),$ where $m = \Sigma \cdot \Sigma.$  We list forbidden collections of transverse curves simultaneously meeting $\Sigma$ for various values of $m$ and a given monodromy assignment in Table~\ref{t:IzerostarBannedTuples}. An $X$ indicates the monodromy assignment for specified values of $A,B,m$ is forbidden. Separate forbidden collections are semicolon-separated. Note that we do not use all available vanishings with some collections. Rather, any collection containing a forbidden collection is also ruled out since the required vanishing conditions cannot be met with any of the indicated transverse subcollections. For example, in the case with data given by $\so(8),\ A>0,\ m=3$, the presence of two transverse curves with types $(.,1,.)$ prevents $\gh=(g/\sigma^2)|_{\sigma =0}$ from being a cube (since these each require 2 additional vanishings of $\gh$ at their intersections with $\Sigma$). However, this contradicts $\bt =3.$  
{  \begin{table}[htbp] \footnotesize \setlength{\tabcolsep}{3pt}
  	\begin{center}
  		\begin{tabular}{c | c c c c }
            $A>0$ & $m = 4$ &  $m=3$ & $m=2$ & $m=1$  \\ 
            \hline 
              $\g_2$  & X & $(.,3,.)$ & $2(.,3,.);(.,6,.)$, & $3(.,3,.);(.,6,.)(.,3,.);(.,9,.)$  \\
              $\so(7)$ & X & X & X & X\\
              $ \so(8)$ & ~ & $2(.,1,.);(.,2,.)(.,1,.)$&$3(.,1,.);2(.,2,.)(.,1,.);3(.,2,.);$&$4(.,1,.);4(.,2,.);2(.,3,.)2(.,1,.);$    \\
                              &     &                                 &$(.,3,.)(.,2,.)(.,1,.)$ &  $(.,4,.)(.,2,.)(.,1,.); $  \\
                              &     &                                  & $(.,3,.)2(.,1,.);(.,4,.)(.,1,.)$ & $4(.,1,.);(.,2,.)3(.,1,.);\dots $ \\
 \hline 
 \hline
                $B>0$ &  \\ 
                \hline 
                $\g_2$ & X & X & X & X\\
                $\so(7)$ & X & $(2,.,.)$ & $2(2,.,.);(4,.,.)$ & $3(2,.,.);(4,.,.)(2,.,.);(6,.,.)$ \\
                $\so(8)$ & ~ & $2(1,.,.)$ & $(2,.,.)2(1,.,.);3(1.,.)$ & $4(1,.,.);(2,.,.)2(1,.,.);$ \\
                               &     &                 &                                    & $(2,.,.)3(1,.,.); (3,.,.)2(1,.,.); \cdots $              
  		\end{tabular}
  		\caption{Forbidden transverse curve collections meeting \Izerostar{} with orders $(a =2+A, b=3+B,6).$}
  		\label{t:IzerostarBannedTuples}
  	\end{center}
  \end{table} }

The form of the relevant restricted polynomials for $\so(8)$ with $A,B>0$ rule out a significant number of intersections that satisfy naive residuals tallying. Notable cases include $\so(8),B>0$ intersection with a type III curve (with orders $(1,\geq 2,3)$) since this hence induces non-minimality and $\so(8),B>1$ intersection with type II curves. Other non-trivial prohibitions include $\so(8),A>0$ intersections with any type IV curves or type III curves. 

\subsubsection{Intersection contributions to \Izerostar{}}
\paragraph{\Izerostar{} with $\mathbf{A>0}$ meets multiple \In{} curves:}

Let $\Sigma$ be a curve with type \Izerostar{} at $z=0$ having orders of vanishing $(2+A,3,6)$. Consider first $A>0.$  Here $\Sigma$ has even type and the contributions to residual vanishings of $\Delta|_{z=0}$ are induced precisely by those of $g|_{z=0}.$ 

In the case that $m = -\Sigma\cdot \Sigma=3,$ the residual vanishings along $\Sigma$ are given by $(-,3,6).$  Working from the general local form of an \Itwo{} curve found in (A.2) of~\cite{BertoliniGlobal1}, we can place further restrictions to give the forms for I${}_n$ with $n>2$. For an \Itwo{} curve at $\sigma=0$ to meet $\Sigma$, we have $z|\phi.$ Since $A>0,$ $\Sigma$ has even type so each vanishing of $\Deltat$ along $\Sigma$ corresponds to a vanishing of $g$ there. Hence, for two \Itwo{} curves at $\sigma, \sigma',$ (using the general form separately for each) restricting to $\sigma =0$ and $\sigma'=0$ and using that there is a $\bt$ contribution at each intersection, we have $z^2| \phi$ in each expansion (rather than only the {\it a priori} requirement that $z| \phi$ needed for intersection with an arbitrary \Izerostar{} curve). Since $g$ goes as $\phi^3 + O(\sigma),$ we thus have a residuals contribution minimum from the \Izerostar{} given by $(-, \geq 1, \geq 2)$ at each \Itwo{} intersection and contributions $(\geq 4, \geq 6, \geq 8)$ to each \Itwo{} curve. When compact, the \Itwo{} curves must have self intersections given by $-1$ and they cannot meet other type \In{} with $n\leq 4$ curves.\footnote{More generally, for $\Sigma'\sim $\In{} rather than \Itwo{} with $\Sigma' \cdot \Sigma' = -1$, we cannot have intersection with curves other than I${}_p$ with $p\leq (n+12)-2n$ when $n$ is even and $(n+12)-3n$ when $n$ is odd.} Since $z^2 | f_1$ and $z^2 | \phi$ we can read from the form of $g$ in (A.2) of~\cite{BertoliniGlobal1} that there are larger residuals contributions to $\Sigma$ given by $(-, \geq 2, \geq 4)$ from each intersection with an \In{} when $n\geq 2$. Hence these triples are disallowed as they exceed the permitted number of vanishings along $\Sigma$. Note that results of this kind follow from intersection contribution tallying as read from the general forms of \In{} unless $A=B=0$ when special treatment is required.

\subsubsection{\Izerostar{} restricted tuples with ${A} = {B}= 0$}
We now briefly discuss certain forbidden configurations for a curve $\Sigma$ of type \Izerostar{} curves with orders of vanishing given precisely by $(2,3,6).$ We will proceed by working through the cases for $m_\Sigma.$ The restrictions we find are helpful in allowing us to eliminate the need for anomaly cancellation machinery while characterizing 6D SCFT gauge enhancement structure.
 \paragraph{$\mathbf{\Sigma \cdot \Sigma} = -3$:}
Here $(\at,\bt,\dt)_\Sigma=(2,3,6).$ We shall inspect intersections of $\Sigma$ lying along $z=0$ with pairs of transverse singular curves $\Sigma_1,\Sigma_2$ having intersections at $\sigma=0$ and $\sigma'=0$, respectively, with these related by $\sigma=1/{\sigma'}.$ Let us denote the restrictions to $\Sigma$ of $\ft,\gt,\Deltat$ to $\Sigma$ by $\fh,\gh,\Deltah.$  

\begin{result} The configuration {\bf \Itwo{} \Izerostar{} \Itwo{}} requires at least a semi-split \Izerostar{} curve $\Sigma$ for $m_\Sigma=3.$
\end{result}
\noindent \underline{Proof:} Let the two \Itwo{} fibers be denoted $\Sigma_1,\Sigma_2.$ Now $\Sigma_i$ intersection with $\Sigma$ requires that the forms of $\fh,\gh$ are those giving the general form for \Itwo{} of~\eqref{eq:I2summary} modified such that the relevant coefficient functions are instead constant.

Proceeding in this fashion while imposing both \Itwo{} intersections, we shall simply obtain a partial splitting of the monodromy cover polynomial $\psi^3+\fh\psi+\gh$ via explicit factorization. The most general $\fh, \gh,\Deltah$ are given in the patch with coordinates $z, \sigma$ by 
\begin{align*}
\fh &= -3 \phi^2 + f_1 \sigma - 3 \Phi^2 \sigma^2\\
\gh &= 2\phi^3 - f_1 \phi \sigma -f_1 \Phi \sigma^2 + 2\Phi^3 \sigma^3
\end{align*}
where $\phi,\Phi,f_1$ are unspecified constants. One can then partially factor the monodromy cover as 
\begin{align*}
\psi^3 + \fh \psi + \gh  = ((\phi + \Phi \sigma) - \psi)((2 \phi^2 - f_1 \sigma - 2 \phi \Phi \sigma + 2 \Phi^2 \sigma^2) - (\phi + \Phi \sigma) \psi - \psi^2)
\end{align*} 
in this coordinate patch. Note that we can now also read off the factorization of the monodromy cover polynomial on the other patch. \qed

This implies the gauge assignment on the base $131$ given by $\su(2), \g_2, \su(2)$ is not viable. Similar methods forbid the triple \Itwo{} \Izerostar${}^{ns}$ III and show that when the III here has orders $(1,3,3),$ we can forbid that $\text{\Itwo{} \Izerostar{} III}$ regardless of monodromy. Likewise, the triples $\text{\Itwo{} \Izerostar{} IV}$ and $\text{\Ifour{} \Izerostar{} \Itwo{}}$ are also forbidden for all of monodromy choices. 

\begin{result}\label{r:semisplit} If $\Sigma$ is semi-split, intersection contributions to $\Sigma'$ having type \Inns<3>{} (even if non-compact) are at least $(2,3,7).$ Contributions to $\Sigma$ from $\Sigma'$ are at least $(0,0,4).$
\end{result}
\noindent \underline{Proof:} Since $\Sigma$ is semi-split with $A=B=0,$ the monodromy cover appears as 
\begin{equation}
\psi^3+(f/z^2)|_{z=0}\cdot\psi+(g/z^3)|_{z=0}
= (\psi - \lambda) (\psi^2 + \lambda \psi + \mu)~.
\label{eq:partially-split}
\end{equation}
and the residual discriminant $\Deltah_\Sigma = (\Delta/z^6)|_{z=0}$ is
\begin{align}
 4(\mu-\lambda^2)^3+27\lambda^2\mu^2 = (4\mu-\lambda^2)(\mu+2\lambda^2)^2~. 
\end{align}
Since $\lambda$ is a section of $-2K_B-\Sigma$ and $\mu$ is a section of $-4K_B-2\Sigma$, $m=3$ gives $\deg(\lambda) = 1$ and $\deg(\mu)=2.$ Define $\varphi\equiv4\mu-\lambda^2$ and $\rho\equiv \mu+2\lambda^2,$ noting $\deg(\varphi) = \deg(\rho)=2.$ Let $\Sigma'$ be an \Inns<3>{} curve at $\{\sigma=0\}.$ Now $\varphi$ and $\rho$ cannot share any roots since this would require $\sigma|\mu,\sigma|\lambda,$ in turn giving $\sigma|\ft_\Sigma$ and forcing splitting of $\Sigma'$ (as it yields $z^3|\ft_{\Sigma'}$ and further consequences necessary in the non-compact case which are obtained by inspection of~\eqref{eq:partially-split} together with (A.11) of~\cite{BertoliniGlobal1}), contrary to hypothesis. Since $\varphi$ and $\rho$ cannot share any roots and we have $\sigma^3|\Deltah_\Sigma,$ we must have $\sigma^3|\rho^2$ and hence $\sigma^4|\rho^2.$ Thus, $\sigma^4|\Deltah_\Sigma$ and the discriminant expanded along $\{\sigma=0\}$ reads
\begin{align}\Delta= z^7 \Delta_3 \sigma^3 + O(\sigma^4,z^6),
\end{align}
from which we conclude intersection contributions to $\Sigma'$ are at least $(2,3,7)$ and those from $\Sigma'$ are at least $(0,0,4).$\qed

This result prevents $\so(13),\sp(1),\so(7)$ gauge assignments to $m13$ for $2\leq m\leq 4$ upon consideration of other residual contribution considerations, thus reproducing a key anomaly cancellation result of~\cite{atomic} via geometric restrictions. 

Since $\varphi$ is the discriminant of the quadratic term in~\eqref{eq:partially-split}, this cannot be a square, as $\Sigma$ would then be a fully split \Izerostars{} curve carrying $\so(8)$ algebra. Using this fact allows extension of the above argument to eliminate several configurations including \begin{align}
\begin{array}{cccc}
& \text{II}\ \Sigma\ \text{\Inns<3\leq l\leq 4>{}}\ , &  \text{\Itwo{}}\ \Sigma\ \text{\Inns<3\leq l\leq 4>{}}\ , & [\text{\Itwo{}}]\ \Sigma\ \text{\Inns<3\leq l\leq 4>{}}\ , \\ \\
& \text{II}\ \overset{[\text{\Itwo{}}]}{\Sigma}\ \text{\Itwo{}} \ , 
& \text{and} &  \text{II}\ \overset{[\text{\Itwo{}}]}{\Sigma}\ \text{II}\ .
\end{array}
\end{align}
Similar argument easily eliminates $\Sigma$,\Inns<5>{}. This is also key in matching anomaly cancellation restrictions.

\paragraph{$\mathbf{\Sigma \cdot \Sigma} = -2$:}

Additional constraints can be derived by extending the argument of Result~\ref{r:semisplit}.
\begin{enumerate}
\item The configuration 
\begin{align}
\text{II}/\text{III}\ \ \ \underset{[\g]}{\underset{\text{\Izerostarss{}}}{2_\Sigma}} \ \ \  \text{III}
\end{align}
has $[\g] \cong [\sp(2)\oplus \sp(1)],$ as we now show. An \Inns<\geq4>{} intersection requires $A=B=0$ along $\Sigma$ and $\deg(\varphi)=4$ for $m=2.$ Let the type III intersections with $\Sigma$ be at $\{\sigma_i=0\}.$ We have $\sigma_i | \rho,$ $\sigma_i|\varphi.$ Any \In{} fiber at $\{\sigma=0\}$ has $n\leq 4$ since we must have $\sigma^n | \varphi^2.$ The latter holds since $\rho$ and $\varphi$ cannot share a root $\sigma$ unless $\sigma | \lambda$ and $\sigma|\mu,$ but this is impossible since $\deg(\lambda)=2$ and $\sigma_1\sigma_2|\lambda.$ An additional $\sp(1)$ summand from an \Itwo{} fiber at $\{\sigma'=0\}$ giving $\sigma'^2|\rho$ exhausts all roots of $\Deltat_\Sigma.$ This applies to restrict global symmetries of $122$ and $222$ with the above assignment. 

\item Similarly,
\begin{align}
\text{II}/\text{III} \ \ \ {\underset{\text{\Izerostarss{}}}{2_\Sigma}}\ \ [\g]
\end{align}
has $[\g] \cong [\sp(3)\oplus \sp(1)].$ Here we have $\sigma_1|\varphi,$ $\sigma_1|\rho.$ This leaves the maximal configuration for the remaining roots of $\varphi$ and $\rho$ occupied by \Inns<6>{} and \Inns<2>{} fibers, respectively, as two \Inns<4>{} fibers can be eliminated. The latter requires three distinct shared roots of $\varphi$ and $\rho$ though $\deg(\lambda)=2.$ In particular, this applies to the bases $12$ and $22$ with type III required for $22.$ 

\end{enumerate}

A similar extension of the argument yields following holding for $\Sigma$ with any $m.$
\begin{result}Let $\Sigma$ be an \Izerostarss{} curve at $\{z=0\}$ and $\Sigma'$ a type II with $(a,b,d)_{\Sigma'}=(1+A,1,2)$ curve at $\{\sigma=0\}.$ Contributions to $\Sigma$ are at least $(1,2,3)$ for $A=0$ and $(1+A,3,4)$ for $A>0.$ Those to $\Sigma'$ are at least $(2,4,8).$
\end{result} 
 
\paragraph{Additional tools to match anomaly cancellation conditions from geometry}
One of the few tools for which we require further geometric insight (beyond those available via tracking contribution counts and single curve global symmetry maxima) in order to match known constraints derived via anomaly cancellation machinery is the following result.

\begin{result}
That the quiver $322$ cannot have algebras $\so(7),\su(2),-$ follows from geometric considerations. 
\end{result}
This condition appears in~\cite{atomic} and is argued on partially on anomaly cancellation grounds. We now show this follows from geometry without field theory considerations. Along with the other geometric constraints derived here, in~\cite{BertoliniGlobal1}, and those in the appendices of~\cite{atomic}, non-minimality and intersection contribution tallying more than suffice to match all 6D SCFT enhancements constraints on all links and hence on all bases to those one can reach while also employing anomaly cancellation considerations. Some enhancements are eliminated via the present considerations, as discussed in Section~\ref{s:newGaugeRestrictions}.

\noindent {\underline{Proof}:} With residuals contribution tracking, we find that the only enhancements of 322 remaining are the options $\so(7),\su(2)$ and $\g_2,\su(2)$ on the 32 subquiver. To see that $\so(7)$ is forbidden on the $-3$ curve $\Sigma,$ we first note that in all the available type assignments on the link $3_\Sigma2_{\Sigma'},$ $\Sigma,\Sigma'$ have orders precisely $(2,3,6),(2,2,4),$ respectively. It thus suffices to show that under the following conditions, the assignment $\so(7)$ to $\Sigma$ is not possible. In fact, this argument now follows directly from Appendix E.3 of~\cite{atomic}. Nonetheless, we shall give a pair of alternative arguments demonstrating the utility of various tools. 

Our setup leaves only one possibility for the residuals on $\Sigma$: they are given by $(0,1,2)$ and hence the form of $\fh_\Sigma,\gh_\Sigma$ are given by $f\sim c_1w^2$ and $g\sim c_2 z w^2$ where $c_1,c_2$ are nonzero constants and the IV lies at $w=0.$  Observe that $z\neq w$ (or we would have a codimension two $(4,6,12)$ point along the \Izerostar{} curve). The resulting monodromy cover is then irreducible. We have the cover given by $P = \psi^3 + c_1 w^2 \psi + c_2 z w^2$ that cannot be semi-split, since as we saw above the $\psi^2$ term vanishing requires that $$Q= (\psi - \alpha)(\psi^2 + \lambda \psi + \mu) = \psi^3 + (\mu - \lambda^2) \psi - \mu \lambda.$$ With self intersection $-3,$ the residuals on are $(2,3,6)$ for type \Izerostar{} with orders $(2,3,6).$ Hence $\deg ( \mu - \lambda^2) = 2$ and $\deg(s \lambda) = 3$ with the degrees of $\mu,\lambda$ being $2,1$ respectively. We can then identify $\mu \sim w^2$ and $\lambda \sim z$ where $z$ gives the other vanishing of $g$ along the \Izerostar{}. We have $\gh = c_2 \lambda \mu = (c w^2) (c' z)$ and $\fh = (\mu - \lambda^2)=  (c w^2 - c'^2 z^2),$ where $c,c'$ are nonzero constants. This means $f$ has two distinct roots along the \Izerostar{}, contradicting that we use both available vanishings at once in meeting the type IV curve. \qed \\

\noindent {\underline{Alternate proof}:} Suppose we have the $\so(7)$ algebra on the $-3$ curve. We know that $\deg \mu =2$ and $\deg \lambda =1$ in the case with orders $(2,3,6)$ along the \Izerostar{}. Expanding each and imposing that $\mu-\lambda^2 =\fh = f_2 w^2$ and $-\lambda \mu = \gh = g_2 w^2 + g_3 w^3$ using the divisibility requirements from meeting a IV, where here $\fh = (f/\sigma^2)|_{\sigma=0}$ and $\gh = (g/\sigma^3)|_{\sigma=0},$ we have
\begin{align*}
\lambda&= \lambda_0 + \lambda_1 \sigma \\
\mu&= \mu_0 + \mu_1 \sigma + \mu_2  \sigma^2 \hfill \\
 \psi^3 + (\mu - \lambda^2) \psi - \mu \lambda &= \psi^3 + \fh \psi + \gh,\\
\ \ \ \ \ \ \implies  \lambda_0 = \mu_0= \mu_1 =0. \hfill
\end{align*}
This gives $\mu\lambda = O(\sigma^3)$, preventing matching the $\sigma^2$ term of $\gh$ unless $g_2 = 0.$ The latter induces non-minimality. \qed

\begin{result} When $\Sigma$ is a curve with $m=3$ and type \Izerostar{}${}^{ss}$ having orders $(a,b,d)=(2,3,6),$ a type III curve $\Sigma'$ with orders $(1,3,3)$ at $\sigma=0$ contributes at least $(2,3,6)$ to the allowed vanishings along $\Sigma.$
\end{result}
\noindent \underline{Proof:} We proceed as above, here imposing the requirements that 
\begin{align*}
\lambda &= \lambda_0 + \lambda_1 \sigma \\
\mu &= \mu_0 + \mu_1 \sigma + \mu_2 \sigma^2,\\
\psi^3 + (\mu - \lambda^2) \psi - \mu \lambda  &= \psi^3 + \fh \psi + \gh \\
\fh &= f_1 \sigma + f_2 \sigma^2 \\
\gh &= g_3 \sigma^3.
\end{align*}  
Proceeding to match the terms of $\fh$ and $\mu - \lambda^2$ order by order and then those of $g,$ we find before completing the matching that 
\begin{align*}
\lambda_0 &= 0, \ \ \ \mu_0 = 0, \\
\mu_1 &= f_1, \ \ \ \mu_2 =f_2 + \lambda_1^2,\\
\implies \mu-\lambda^2 &= \fh, \\
-\mu \lambda &= -\lambda_1 f_1 \sigma^2 - \lambda_1 (\lambda_1^2 + f_2)^2 \sigma^3. 
\end{align*} 
From the latter we see that one of $\lambda_1,f_1$ must be zero since $\gt$ is zero at order $\sigma^2.$  We rule out the first case as it induces infinite intersection contribution, leaving $f_1=0.$ Intersection contributions to the \Izerostar{}${}^{ss}$ are then given by $(2,3,6).$ (Note this forbids any additional intersections along $\Sigma,$ even with an \Izero{} curve having $A\geq 1.$) \qed

The above seemingly mild restriction plays an important role in determining which enhancement configurations are permitted and the degrees of freedom which remain to become global symmetry summands.

\subsubsection{Contributions to \Izerostar{} with $A,B>0$ from \In{}}
We now investigate the details of intersection contributions in the few permitted values for $n$ in for \Izerostar{} intersections. When $A,B>0$, the situation is even more restrictive. The maximal allowed $n$ for \In{} meeting \Izerostar{} in each case of $A,B>0$ is given in Table~\ref{t:sumloccontr}. So that we may refer to the general forms for \In{} type curves, let's suppose our \Izerostar{} here lies at $z=0.$  

\paragraph{$\mathbf{A>0}$, $\so(8)$:}  
We will use the observation that since we have algebra $\so(8)$, intersection contributions to $\bt$ are multiples of 3. Those to $\dt$ are then the doubles of those $\bt$ contributions as $A>0$ is an even type \Izerostar{}. From Table~\ref{t:maxn}, we see that the maximal allowed intersection with \In{} in this case with $A>0$ is for $n=3.$  

\noindent{\bf \In{} compact:}

We will treat the three possible values of $n$ separately when $\Sigma'$ here is a Kodaira type \In{} curve that is compact and has self-intersection $-1.$ Note that we must have $A\leq 2,$ since otherwise we exceed the 4 allowed $f$ residual contributions to $\Sigma'.$ Since $A>0,$ we must be without monodromy along $\Sigma'$ in the one relevant case where $n=3.$ When $n=1,2$ we note that $z^2| \phi$ with $\phi$ as in (A.4,A.8) of~\cite{BertoliniGlobal1}, respectively. Considering this fact together with the restriction that we have contributions in threes yields Table~\ref{t:IzerostarS_InContributionsA}.
 \begin{table}[!h]
 	\begin{center}
 		\begin{tabular}{c | c c c c }
 		\diaghead(-1,1){cccc}%
 		{$A$}{$n$\hspace{-0.05in}} &  1 &  2 & 3   \\ 
           \hline      
           1   & $(1,3,6)/(4,6,10)$ & X & X \\
           2  & $(0,3,6)/(4,6,10)$ & X  & X 
 		\end{tabular} 		
 		\caption{Intersection contributions to \Izerostar{}$^{s}$ with $A>0$ from/to \In{}. An `X' indicates the intersection is forbidden by non-minimality.}
 		\label{t:IzerostarS_InContributionsA}
 	\end{center}
 \end{table}

\noindent Note that in the case when $A=1$ and $n=1$, we must have at least $z^4|g_1$ and $z^4|g_2$ in (A.4) of~\cite{BertoliniGlobal1} via the above consideration that $g$ intersection contributions must come in threes here. This has total order of $f,g$ given by $4,5$ in both cases, falling barely short of non-minimality. The other cases $n>1$ are barred by similar considerations. \\

\noindent{\bf {Non-compact \In{} curves for global symmetry}:}

We carry out a similar study here with the change that $\phi$ is allowed higher degree here, introducing the possibility of intersections with \Itwo{} or \Ithree{}. Since we are concerned with global symmetry, we safely ignore $n=1$ (as \Ione{} curves carry the trivial algebra). However, intersections for $n>1$ are forbidden as a result of non-minimality following from the strong requirement that contributions to $\bt$ are divisible by 3. This result is indicated in Table~\ref{t:IzerostarS_InContributionsA_Noncompact}.
\begin{table}[!h]
 	\begin{center}
 		\begin{tabular}{c | c c c c }
 		\diaghead(-1,1){cccc}%
 		{$A$}{$n$\hspace{-0.05in}} &  2 &  $\geq 3$   \\ 
           \hline      
           $\geq 1$    & X & X 
 		\end{tabular} 		
 		\caption{Intersection Contributions to \Izerostar{}$^{s}$ with $A>0$ from an \In{} non-compact curve. An `X' indicates the intersection is forbidden by  non-minimality.}
 		\label{t:IzerostarS_InContributionsA_Noncompact}
 	\end{center}
 \end{table}
\paragraph{$\mathbf{B>0}$ $\so(8)$:}
This case has similar requirements to those above with the main difference being that the contributions to residuals in $f$ are required to be even here as we saw in the preceding analysis. We collect results for this case in Table~\ref{t:IzerostarS_InContributionsB}.
  \begin{table}[!h]
  	\begin{center}
  		\begin{tabular}{c | c c c c }
  		\diaghead(-1,1){cccc}%
  		{$B$}{$n$\hspace{-0.05in}} &  1 &  2    \\ 
            \hline      
            1   & $( 2, 1,6 )/(4,6,10)$ & X  \\
            2  & X & X   \\
            3  & X & X   
  		\end{tabular} 		
  		\caption{Intersection Contributions to \Izerostar{}$^{s}$ with $B>0$ from/to \In{}. An `X' indicates the intersection is forbidden by non-minimality.}
  		\label{t:IzerostarS_InContributionsB}
  	\end{center}
  \end{table}

\paragraph{$\mathbf{B>0},$ $\so(7)$:}
Again we consider intersections with a transverse type \In{} curve, $\Sigma'$ using restrictions from Table~\ref{t:maxn} which dictate that the maximum value of $n$ here is 2. We will use that we have odd type on the \Izerostar{}. We will not need to use that intersection contributions to the \Izerostar{} in this case are dictated by the lowest order $z$ term in $f$, say $\mu_1 \gamma^2$ with $\mu_1$ square-free and of nonzero degree (since in this case the relevant term cannot be a square, or equivalently, contributions to $\at$ along an \Izerostar{}${}^{ss}$ cannot have purely even $f$ residual contributions).\\ 

\noindent{\bf{\In{} compact}:}\\

\noindent With these constraints, we produce Table~\ref{t:IzerostarSSInContributions} by reading from the general forms for \In{} as they appear in~\cite{BertoliniGlobal1} Appendix A. Note, we have $B\leq 3$, as the degree of $\phi$ (A.4) of~\cite{BertoliniGlobal1} is $2$ in the only relevant compact \In{} intersections, those for a curve with self-intersection $-1.$ 
 \begin{table}[!h]
 	\begin{center}
 		\begin{tabular}{c | c c c c }
 		\diaghead(-1,1){cccc}%
 		{$B$}{$n$\hspace{-0.05in}} &  1 &  2    \\ 
           \hline      
           1   & $(1,1,3)/(4,6,10)$ &  $(1,1,3)/(4,6,8)$   \\
           2  & $(1,1,3)/(4,6,10)$ & X   \\
           3   & $(1,0,3)/(4,6,10)$ &  X                 
 		\end{tabular}
 		\caption{Intersection contributions with \Izerostar{}$^{ss}$ with $B>0$ from/to an \In{} curve. An `X' indicates the intersection is forbidden by minimality considerations.}
 		\label{t:IzerostarSSInContributions}
 	\end{center}
 \end{table}
For a non-compact \In{}, we can raise the degree of $\phi$ to allow large values of $B$ for example when meeting \Itwo{}. Such intersections are barred for $\Sigma'$ compact. We have not used the condition that $\at$ contributions cannot be purely even, instead limitations being induced by minimality considerations. We collect our results in Table~\ref{t:IzerostarSSInContributions}.\\

\noindent{\bf{Non-compact \In{} curves for global symmetry}:}

Our results here only concern $n\geq 2$ since $n=1$ does not yield global symmetry. From Table~\ref{t:maxn}, we cannot exceed $n=2.$ We have the identical result when $B=1.$ When $B\geq 2,$ we can avoid non-minimality by raising the degree of $\phi.$ In this case, we only are interested in the intersection contributions to the \Izerostar{}. We collect the relevant result in Table~\ref{t:IzerostarSSInNoncompactContributions}. 

 \begin{table}[!h]
 	\begin{center}
 		\begin{tabular}{c | c c c c }
 		\diaghead(-1,1){cccc}%
 		{$B$}{$n$\hspace{-0.05in}}  &  2  & $\geq 3$ \\ 
           \hline      
           1   & $(1,1,3)$ & X\\
           2   & $(1,1,3)$ & X\\
           $\geq 3$ & $(1,1,3)$ & X\\
 		\end{tabular}
 		\caption{Intersection Contributions to \Izerostar{}$^{ss}$ with $B>0$ from a transverse non-compact\In{} curve. An `X' indicates the intersection is forbidden by non-minimality considerations.}
 		\label{t:IzerostarSSInNoncompactContributions}
 	\end{center}
 \end{table}
\subsubsection{${A>0}$, $\g_2$}
The {\it a priori} restrictions in this case are similar to those for the $\so(8)$ case with the exception that $g$ contributions are not forced to be multiples of three. On the contrary, we are prevented from having contributions which consist entirely of multiples of three, though this is irrelevant in our analysis here. When $n=3,$ we use that $\phi_0$ rather than $\mu$ in (A.11) of~\cite{BertoliniGlobal1} must carry all divisibility (we have $z|\phi_0$ and have thus used all available roots of $\phi_0$) since we are considering the case of intersection with a compact \In{} curve while $f,g$ residuals are limited by $4,6$, respectively. We find there is non-minimal intersection for $n\geq 3$ as this would require $z^2|\psi_1$ with $\psi_1$ as in (A.11) of~\cite{BertoliniGlobal1}.

 \begin{table}[!h]
 	\begin{center}
 		\begin{tabular}{c | c c c c }
 		\diaghead(-1,1){cccc}%
 		{$A$}{$n$\hspace{-0.05in}} &  1 &  2 & 3   \\ 
           \hline      
           1   & $(1,1,2)/(4,6,9)$ &$(1,2,4)/(4,6,9)$  & X \\
           2  & $(0,1,2)/(4,6,9)$ & $(0,2,4)/(4,6,9)$  & X 
 		\end{tabular}
 		\caption{Intersection contributions to \Izerostar{}$^{ns}$ with $A>0$  from/to \In{}. An `X' indicates the intersection is forbidden by non-minimality considerations.}
 		\label{t:IzerostarNSInContributions}
 	\end{center}
 \end{table}
\end{subsection}

\begin{subsection}{Type \Instar{} curve intersections}\label{s:Instar}
In this section, we collect contributions to residual vanishings from an intersection with an \Instar{} curve. The main focus is the case with transverse curve of type \In{}. 

\subsubsection{Single \Instar{} intersections with a type I${}_m$ curve} 
We begin by finding the minimal simultaneous contributions to the residual vanishings to each curve from an \Instar{},I${}_m$ intersection. The result is that there is a simple general pattern for these contributions which we expect to hold in all cases where intersection should be allowed via the {\it a priori} residual vanishings of each curve. Some of the general forms of such intersections have been constructed for this analysis, but the most general form in the large $n,m$ case for arbitrary $n,m$ has not. Hence a portion of our results here for $7\leq m\leq 9$ is conjectural. We expect these contributions to be a generalization of the cases we have compiled in the table below. The general form of the conjecture is then that we have contributions to the residual vanishings along the \Instar{} and I${}_m$ curves given by $(0,0,m)$ and $(2,3,n)$, respectively, in the case with or without monodromy along the \Instar{} for even $m$; for $m$ odd instead these contributions are given by $(0,0,m)$ and $(2,3,n)$ in the case with monodromy and $(0,0,m+1),$ $(2,3,n+1)$ in the case without monodromy, respectively. Note that the I${}_m$ is non-split (i.e.,\ has monodromy) since we consider an intersection with \Instar{}.

\paragraph{Intersection with type \Ionestar{}}

\noindent{\bf{Meeting \Ione{}:}}

We begin by considering an intersection contributions to an \Ione{} curve from an
\Ionestar{} curve in cases with and without monodromy. This requires working several cases and looking for the minimum. The detailed calculations can be easily carried out with a computer algebra system but are somewhat cumbersome to treat by hand. We find that these minimal contributions to the $I_1$ are given by $(2,3,7)$ and $(2,3,8)$ from \Ionestarns{} and \Ionestars{}, respectively. Note the agreement with~\cite{gaugeless}. \\

\noindent{\bf{Meeting \Itwo{}:}}\\
\indent Here we find that values for the cases with and without monodromy have the same intersection contributions to the \Itwo{}, namely $(2,3,7)$ and those to the I${}_1^*$ are given by $(0,0,2)$ regardless of monodromy. \\

\noindent{\bf{Meeting \Ithree{}:}}

Here we find that the monodromy again matters and the contributions to the \Ithree{} are as in the \Ione{} case given by $(2,3,7)$ and $(2,3,8)$ from \Ionestarns{} and \Ionestars{}, respectively, before we consider the monodromy condition along the \Ithree{}. Investigating monodromy involves detailed inspection, but again can be readily treated using a computer algebra system. We find that for $\su(3)$ along \Ithree{}, the intersection becomes a $(4,6,12)$ point. This same result appears to hold for either monodromy along the \Ionestar{}.

Since $I_n^*$ is obtained by imposing further constraints on an \Ionestar{} and we did not use the monodromy information from the \Ionestar{}, the above also forbids the intersection of I${}_3^{s}$ with higher \Instar{}. Likewise, moving to the cases I${}_{n\geq 3}$ also simply imposes additional constraints on $f,g.$ Meanwhile, the term dictating whether we have monodromy along $\Sigma = $\In{} remains the same. Its form simply becomes more constrained as we increase $n.$ This method is then an alternative demonstration that \Instar{} cannot intersect I${}_m^s$ for ($m\geq 3, n\geq 1$). 

\noindent{\bf{Meeting \Ifour{}:}} 

In this case, the minimal contributions are $(2,3,7)$ with or without monodromy along the \Ionestar{}. 

\paragraph{\Instar{} with $n\geq 4$:}
Here we implement the constraints for \Instar{} beginning with the inductive form of I${}_m$ and vice versa (for large $n$ using the inductive form along the \Instar{} and imposing the constraints for \Ione{}, then \Itwo{}, etc.)\ and concluding by imposing monodromy constraints for the \Instar{}. This allows one to inspect the resulting form to determine the minimal simultaneous intersection contributions in each case. We find the general pattern for low $n$ appears to continue, but proving this in full generality is beyond our reach. Those cases we have explicitly checked are found in Table~\ref{t:InstarLocalContributionsToIn}. 

\subsubsection{Summary}
We find the following minimal simultaneous intersection contributions to \In{} from I${}_m^*$ curves and vice versa, respectively. While it is {\it a priori} possible that the minimal contributions could be realized for each with different intersections giving the minimal ones, this does not appear to be the case. Rather, in all cases we have inspected, simultaneous realization of the minimal contributions to both the \Instar{} and the \In{} curve appears to be possible. Note that only intersections that do not give $(4,6,12)$ points are considered here. Note there is one remaining possible (necessarily simultaneous) singularity of $f,g$ since there are remaining residuals after the \Instar{} intersection given by $(2,3,-)$ (which can be located anywhere we choose other than at the original intersection). Furthermore, there cannot be additional transverse curves over which the fibration is singular not meeting at this point along the \In{} curve that are not of type I${}_p$ for some $p$. The transverse curves also must have $(a,b)\leq (2,3)$. In other words, any other intersection with a curve of which the fibration is singular must give either contribution $(2,3,-)$ (there can be at most one of these) or contributions $(0,0,-)$.
{\setlength{\tabcolsep}{3pt}
\begin{table}[!h] \scriptsize
	\begin{center}
		\begin{tabular}{c | c c c c c c c c c c c c c }
			~ 			& I${}_1^{\ast ns}$	& I${}_1^{\ast s}$ &	I${}_2^{\ast ns}$	& I${}_2^{\ast s} $ &  $\dots $& I${}_4^{\ast ns}$ & I${}_4^{\ast s}$ \\
			\hline
			\Ione{}		 	& $(2,3,7)/(0,0,1)$ &$(2,3,8)/(0,0,2)$ & $(2,3,8)/(0,0,1)$ & $(2,3,9)/(0,0,2)$ & ~ &  $(2,3,10)/(0,0,1)$ & $(2,3,11)/(0,0,2)$ \\ 
			\Itwo{}		    & $(2,3,7)/(0,0,2)$ & $(2,3,7)/(0,0,2)$  & $(2,3,8)/(0,0,2)$ & $(2,3,8)/(0,0,2)$ & ~ & $(2,3,10)/(0,0,2)$ & $(2,3,10)/(0,0,2)$   \\ 
			\Ithree{}		    & $(2,3,7)/(0,0,3)$ & $(2,3,8)/(0,0,4)$  &  $(2,3,8)/(0,0,3)$ & $(2,3,9)/(0,0,4)$ & & &   \\ 	
			\Ifour{}           & $(2,3,7)/(0,0,4)$ & $(2,3,7)/(0,0,4)$  &  & & & &   \\ 	
			\Ifive{}            & $(2,3,7)/(0,0,5)$ & $(2,3,8)/(0,0,6)$ & & & & & \\
			\Isix{}             & $(2,3,7)/(0,0,6)$ & $(2,3,7)/(0,0,6)$ & & & & & \\
			\vdots \\
			\Ieight{} ($\dagger$)      & $(2,3,7)/(0,0,8)$ & $(2,3,7)/(0,0,8)$ & $(2,3,8)/(0,0,8)$ &$(2,3,8)/(0,0,8)$ & $\dots$ &   \\
			\vdots
		\end{tabular}
		\caption{Minimal simultaneous intersection contributions in \In{},I${}_k^\ast$ collisions to \In{} and I${}_k^\ast$, respectively. (The `$\dagger$' symbol indicates the less general inductive form is used for \In{}).}
		\label{t:InstarLocalContributionsToIn}
	\end{center}
\end{table}
}
\subsubsection{Multiple \Instar{} curves meeting a type I${}_m$ curve}
In this case, we must have the self-intersection along the I${}_m$ curve, say $\Sigma,$ with $\Sigma\cdot \Sigma=-1$ since for $-\Sigma \cdot \Sigma>1$ we have no vanishings of $f,g$ possible. The total residuals are $(4,6, 12 + m)$ along $\Sigma,$ so at most two \Instar{} curves can meet $\Sigma,$ and such a pair leaves the no remaining residuals in $f,g.$ Note that intersection with a pair of \Instar{} curves requires that the I${}_m$ curve has monodromy.

\subsubsection{Compact curve intersections for pairs with types II, \Instar{}}
We first observe that the only relevant case here is $n=1,$ as a type II curve cannot meet an \Instar{} curve with $n>1;$ such intersections are non-minimal even when the \Instar{} is non-compact. We consider here the case with both curves compact, thus introducing additional constraints not applicable to the situations considered in~\cite{BertoliniGlobal1}. Along the \Instar{} curve, $(\at,\bt) = (2(4-m),3(4-m))$ and II gives a non-trivial $f,g$ contribution. Hence, $m=4$ does not need to be considered. We summarize the contributions for all remaining cases of such an intersection in Table~\ref{t:InstarIIContributions}. Since I${}_{\geq 1}^{\ast s}$ and I${}_{>2}^{\ast ns}$ intersections with a type II curve are non-minimal, all relevant local contribution data for II,I${}_{n\geq1}^*$ intersections are captured in our table.

 \begin{table}[!h]
  	\begin{center} \small
  		\begin{tabular}{c|cccccccc}
  		\diaghead(-1,1){cccccccc} % \\
  		{$A$}{$m$} & 3	& 2	& 1  \\
            \hline      
            0   & $(2,3,4)/(3,4,8)$ &     \\
            1  &  $(2,3,4)/ (2,4,8)$ &   \\
            2 &  X.. & $(4,6,7) / (3,4,8)$  &  \\
            3 & X..& $(4,6,7) / (2,4,8)$ & \\
            4 & X.. & X.. & $(6,9,10)/(3,4,8) $\\
            5 & X.. & X.. & $(6,9,10)/(2,4,8)$ \\
            $\geq 6$ & X.. & X.. & X.. 
  		\end{tabular} 		
  		\caption{Intersection contributions to/from a compact I${}_1^{\ast ns}$ with $\Sigma\cdot \Sigma=-m$ intersecting a II. An X indicates the intersection is non-minimal and an (X..)X. that the intersection is banned by (naive) residuals considerations. Entries to the right of an allowed entry have the same values.}
  		\label{t:InstarIIContributions}
  	\end{center}
  \end{table} 
  Note that Kodaira types beyond II other than \In{} family types are banned from \Instar{} intersection by non-minimality. Thus we have given here the set of possible intersection contributions for types other than \Izero{}, which we have recorded separately. 

\end{subsection}

\begin{subsection}{\Izero{},\Instar{} curve intersections}
We now find the intersection contributions to a curve $\Sigma$ with type \Izero{} from an \Instar{} curve, $\Sigma',$ which may be non-compact. Our computations rely on the general forms of \Instar{} given in Appendix B of~\cite{BertoliniGlobal1}. We impose divisibility conditions on these local models as required for intersection with a compact \Izero{} curve having specified properties. We then read off the intersection contributions to $\Sigma$ in each case while recording which of these allow $\Sigma'$ compact and tabulating intersection contributions to $\Sigma'$ in such cases.

Let the vanishings along $\Sigma$ be given by $(A,B,0),$ noting that we must have one of $A,B$ being zero. We consider $\Sigma'$ along $\{\sigma = 0\}$ whose form is given by one of the expansions from~\cite{BertoliniGlobal1} (B.1)-(B.5). We compile the total order at the intersection point in \ref{t:InstarLocalContributionsMinOrder}, writing `X' for non-minimal intersections. By inspecting the form of \Instar{}, imposing the required divisibility conditions to have $z^A$ dividing $f$ or $z^B$ dividing $g,$ and then checking the monodromy condition for \Instar{}, we arrive at the indicated minimal total orders at the intersection. We do not record the case $A=B=0$ here since the minimal total order is always $(2,3,n)$ in these cases. Hence, all intersections are allowed except when the $\Sigma'$ has self intersection $-4.$ For such curves, all $A,B>0$ intersections are trivially banned by {\it a priori} residuals tracking. Terminal entries with $A$ or $B$ referenced indicate the general pattern holds for all $A$ or $B$, applying also to the entries below such an entry. Note that we need only explore $n\leq 4$ since further intersections are forbidden via~\cite{PerssonsList}.

\begin{table}[htbp]
	\begin{center}\small \setlength{\tabcolsep}{3pt}
		\begin{tabular}{c | c c c c c c c c c c c c c }
			~ 			& I${}_1^{\ast ns}$	& I${}_1^{\ast s}$ &	I${}_2^{\ast ns}$	& I${}_2^{\ast s}$	&  I${}_3^{\ast ns}$	 & I${}_3^{\ast s}$ & I${}_4^{\ast}$ \\
			\hline
			$A=1$		 	& $ (4,4,8) $ & $ (4,4,8) $ & $ (4,5,10) $ & $ (4,5,10) $ &	$ (3+A,5,10) $ & X  & X  \\
			$A=2$		    & $(4,4,8)$ &  $(4,4,8)$  & $ (4,5,10) $ & $ (5,5,10) $   &  & X   & X  \\
			$A=3$		 	 & $(2+ 2\lceil A/2 \rceil ,4,8)$ &   $(2+ 2\lceil A/2 \rceil ,4,8)$  & $ (6,5,10) $ & $ (6,5,10) $  & & X  & X \\
			$A=4$		  & &   & $ (3+A,5, 10) $ & $ (3+A,5,10) $  & & X   & X  \\
			\hline 
		     $B=1$		  & $ (3,5,9) $ & $ (3,5,9) $ &  $(3,4+B, 9) $ & $ (3,4+B, 9) $ & X & X   & X  \\
		     $B=2$		  & $ (3,6,9) $ & $ (3,6,9) $ &  & & X & X   & X  \\
		     $B=3$		  & $ (3,6,9) $ & $ (3,7,9) $ &  & & X & X   & X  \\
		     $B=4$		  & $ (3,7,9) $ & $ (3,7,9) $ &  &               & X& X   & X \\
		\end{tabular}
		\caption{Minimal total order of \Izero{},\Instar{} intersections}
		\label{t:InstarLocalContributionsMinOrder}
	\end{center}
\end{table}
\end{subsection}

\begin{subsection}{Compact \Izero{},\Instar{} curve intersections}

The residual vanishing counts along an \Instar{} curve with negative self-intersection $m$ are given by 
\begin{align*}
\at &= -4(m-2) +2m=8-2m\\
\bt &= -6(m-2)+3m=12-3m\\
\dt &= -12(m-2)+(n+6)m=24+ (n-6)m.
\end{align*}  
In particular, the constraints on $\at,\bt$ and the general forms of \Instar{} together imply that $u_1$ in the general form has the number of roots indicated in Table~\ref{t:InstarU1Degree} where $f=-\frac{1}{3}u_1^2\sigma^2 +O(\sigma^3)$ and $g=\frac{2}{27}u_1^3\sigma^3 + O(\sigma^4).$
\begin{table}[!h]
	\begin{center}
		\begin{tabular}{c | c c c c c c  }
			~ &  $m=4$ &  $m=3$ &  $m=2$ &  $m=1$ \\
			\hline
		$\deg(u_1)$ & 0 & 1 & 2 & 3 	 
		\end{tabular}
		\caption{Degree of $u_1$ for \Instar{} with $\Sigma\cdot \Sigma = -m.$}
		\label{t:InstarU1Degree}
	\end{center}
\end{table}
This data allows us to find the following rules for intersections with type \Izero$\sim(A,B,0)$ singularities which are not necessarily compact nor transverse curves but simply singularities along the \Instar{} locus. Hence, any remaining residuals in purely $f$ or $g$ yield one of the indicated intersection contributions. Note that in some cases indicated as non-minimal, there may be allowed point singularities in $f,g$ to the corresponding order. Said differently, our tables are intended to track contributions from transverse curves but also give the contributions for point singularities when a transverse curve of the corresponding type is allowed. Only in cases where the intersections are indicated as allowed rather than non-minimal do we track the data since our priority is to treat transverse curves rather than simply singular points (but some information for point singularities does result). The contributions to the \Izero{} are also noted for use in the case that the singularity is an intersection with a {\it compact} transverse \Izero{} curve. Note that for $m=4,$ we cannot have intersection with a curve of type \Izero{}$\sim (A,B,0)$ with either of $A>0$ or $B>0.$  This fact and other restrictions of this form are already accounted for via simple residuals tracking without modification from what follows below. For $A=B=0,$ it is easy to see the intersection contributions in both monodromy cases are simply $(2,3,6+n),$ the na\"ive estimate. We collect the relevant results in Table~\ref{t:InstarLocalContributionsSelfInt3}.

\subsubsection{$m=3$}  
Residuals along the \Instar{} here are given by $(2,3,6+3n).$ We find intersection contributions and non-minimal intersections appearing in Table~\ref{t:InstarLocalContributionsSelfInt3}. 
\begin{table}[htbp]
	\begin{center}    \setlength{\tabcolsep}{3pt} \footnotesize
		\begin{tabular}{c | c c c c c c c c c c c c c }
			~ 			& I${}_1^{\ast ns}$	& I${}_1^{\ast s}$ &	I${}_2^{\ast ns}$	& I${}_2^{\ast s}$	&  I${}_{\geq 3}^{\ast}$ \\
			\hline
			$A=1$		 	& $ (3,4,8)/(2,3,3)$ & $ (3,4,8)/(2,3,3) $ & $ (3,5,10)/(2,3,3) $ & X. & X. \\
			$A=2$		    & $(2,4,8)/(2,3,3)$ &  $(2,4,8)/(2,3,3)$  & $ (2,5,10)/(2,3,3) $ & X. & X. \\
            $A\geq 3$  & - & - & - & - & -  \\
			\hline 
		     $B=1$		  & $ (3,4,9)/(2,3,4) $ & $(3,4,9)/(2,3,5)$ &  $(3,4,9)/(2,3,2)$ & $(3,4,9)/(2,3,2)$ & X    \\
		     $B=2$		  & $ (3,4,9)/(2,3,4) $ & X. & X. & X. & X  \\
		     $B=3$		  & $ (3,3,9)/(2,3,4) $ & X. & X. & X. & X  \\
		     $B\geq 4$  & - & - & - & - & -  \\
		\end{tabular}
		\caption{Intersection Contributions to \Izero{}$\sim(A,B,0)$ and \Instar{} with $m=3$, respectively. `X' indicates non-minimal intersection and `-' indicates exceeding allowed residuals. `X.' indicates that the intersection is non-minimal here while in the non-compact case it is was not yet forbidden via the previous analysis in the non-compact case.}
		\label{t:InstarLocalContributionsSelfInt3}
	\end{center}
\end{table}

\subsubsection{$m=2$}  
Here residual vanishings along the \Instar{} are given by $(4,6,12+2n).$  
Hence, $A,B$ are capped for \Izero{} at $4,6.$  The relevant contribution data is recorded in Table~\ref{t:InstarLocalContributionsSelfInt2}. 

\begin{table}[htbp]
	\begin{center}    \setlength{\tabcolsep}{3pt} \footnotesize
		\begin{tabular}{c | c c c c c c c c c c c c c }
			~ 			& I${}_1^{\ast ns}$	& I${}_1^{\ast s}$ &	I${}_2^{\ast ns}$	& I${}_2^{\ast s}$	&  I${}_{3}^{\ast ns}$	 & I${}_{\geq 3}^{\ast s}$ & I${}_{\geq 4}^{\ast ns}$  \\
			\hline
			$A=1$		 	& $ (3,4,8)/(2,3,3 )$ & $ (3,4,8)/(2,3,3)  $ & $ (3,5,10)/(2,3,3)  $ & $(3,5,10)/(4,6,6) $ & X. & X  & X \\
			$A=2$		    & $(2,4,8)/(2,3,3) $ &  $(2,4,8)/(2,3,3) $  & $(2,5,10)/(2,3,3) $ & $(3,5,10)/(4,6,6) $   & X. & X   & X \\
			$A=3$		    & $(3,4,8)/(4,6,6) $ &  $(3,4,8)/(4,6,6) $  & $ (3,5,10)/(4,6,6)  $ & $ (3,5,10)/(4,6,6)  $   & X. & X   & X  \\
			$A=4$		    & $(2,4,8)/(4,6,6) $ &  $(2,4,8)/(4,6,6) $ & $ (2,5,10)/(4,6,6)  $ & $ (2,5,10)/(4,6,6)  $   & X. & X   & X  \\
            $A\geq 5$  & - & - & - & - & - & - & - \\
			\hline 
		     $B=1$		  & $ (3,4,9)/(2,3,4)  $ & $(3,4,9)/(2,3,5) $ &  $(3,4,9)/(2,3,2) $ & $(3,4,9)/(2,3,2)   $ & X   & X   & X  \\
		     $B=2$		  & $ (3,4,9)/(2,3,4)  $ & $(3,4,9)/(4,6,8)  $ & $(3,4,9)/(4,6,4) $  & $(3,4,9)/(4,6,4) $  & X & X   & X  \\
		     $B=3$		  & $ (3,3,9)/(2,3,4)  $ & $(3,4,9)/(4,6,8)  $ & X.   & X. & X & X   & X \\
		     $B=4$		  & $ (3,4,9)/(4,6,8)  $ & $(3,4,9)/(4,6,8)  $    & X. & X. & X & X   & X  \\
		     $B=5$		  & $ (3,4,9)/(4,6,8)  $ & $(3,4,9)/(4,6,8)  $   & X. & X. & X & X   & X  \\
		     $B=6$		  & $ (3,3,9)/(4,6,8)  $ & $(3,3,9)/(4,6,8)  $     & X. & X. & X & X   & X  \\
		     $B\geq 7$  & - & - & - & - & - & - & -  \\
		\end{tabular}
		\caption{Intersection contributions to \Izero{}$\sim(A,B,0)$ and \Instar{} with $m=2$, respectively. `X' indicates non-minimal intersection or not allowed to have a particular \Instar{} on a curve of this self intersection and `-' indicates exceeding allowed residuals. `X.' indicates that the intersection is non-minimal here while in the non-compact case it is was not yet forbidden via the previous analysis in the non-compact case.}
		\label{t:InstarLocalContributionsSelfInt2}
	\end{center}
\end{table}

\subsubsection{$m=1$}  
In this case, the residual vanishings along the \Instar{} are instead given by $(6,9,18+n).$  Hence $A,B$ for \Izero{} are at most $6,9,$ respectively. Contribution data for these intersections appears in Table~\ref{t:InstarLocalContributionsSelfInt1}. 

\begin{table}[!h]
	\begin{center}  \setlength{\tabcolsep}{3pt} \footnotesize 
		\begin{tabular}{c | c c c c c c c c c c c c c }
			~ 			& I${}_1^{\ast ns}$	& I${}_1^{\ast s}$ &	I${}_2^{\ast ns}$	& I${}_2^{\ast s}$	&  I${}_3^{\ast ns}$	 & I${}_{\geq 3}^{\ast s}$ & I${}_{\geq 4}^{\ast ns}$ \\
			\hline
			$A=1$		 	& $ (3,4,8)/(2,3,3 )$ & $ (3,4,8)/(2,3,3)  $ & $ (3,5,10)/(2,3,3)  $ & $(3,5,10)/(4,6,6) $ & X. & X  & X \\
			$A=2$		    & $(2,4,8)/(2,3,3) $ &  $(2,4,8)/(2,3,3) $  & $(2,5,10)/(2,3,3) $ & $(3,5,10)/(4,6,6) $   & X. & X   & X  \\
			$A=3$		    & $(3,4,8)/(4,6,6) $ &  $(3,4,8)/(4,6,6) $  & $ (3,5,10)/(4,6,6)  $ & $ (3,5,10)/(4,6,6)  $   & X. & X   & X  \\
			$A=4$		    & $(2,4,8)/(4,6,6) $ &  $(2,4,8)/(4,6,6) $ & $ (2,5,10)/(4,6,6)  $ & $ (2,5,10)/(4,6,6)  $   & X. & X   & X \\
			$A=5$		    & $(3,4,8)/(6,9,9) $ &  $(3,4,8)/(6,9,9) $X  & $ (3,5,10)/(6,9,9) $ & X. & X. & X   & X  \\
			$A=6$		    & $(2,4,8)/(6,9,9) $ &  $(2,4,8)/(6,9,9) $X & $ (2,5,10)/(6,9,9) $ & X. & X. & X & X  \\
            $A\geq 7$  & - & - & - & - & - & - & - \\
			\hline 
		     $B=1$		  & $ (3,4,9)/(2,3,4)  $ & $(3,4,9)/(2,3,5) $ &  $(3,4,9)/(2,3,2) $ & $(3,4,9)/(2,3,2)   $ & X   & X   & X \\
		     $B=2$		  & $ (3,4,9)/(2,3,4)  $ & $(3,4,9)/(4,6,8)  $ & $(3,4,9)/(4,6,4) $  & $(3,4,9)/(4,6,4) $  & X & X   & X \\
		     $B=3$		  & $ (3,3,9)/(2,3,4)  $ & $(3,4,9)/(4,6,8)  $ &$(3,4,9)/(6,9,6)  $   & $(3,4,9)/(6,9,6)  $  & X & X   & X  \\
		     $B=4$		  & $ (3,4,9)/(4,6,8) $ & $(3,4,9)/(4,6,8)  $    & X.  & X. & X & X   & X \\
		     $B=5$		  & $ (3,4,9)/(4,6,8)  $ & $(3,4,9)/(4,6,8)  $   & X.   & X. & X & X   & X  \\
		     $B=6$		  & $ (3,3,9)/(4,6,8) $ & $(3,3,9)/(4,6,8)  $     & X.   & X. & X & X   & X  \\
		     $B=7$		  & $ (3,4,9)/(6,9,12) $ & X. & X.   & X. & X & X   & X  \\
		     $B=8$		  & $ (3,4,9)/(6,9,12) $ & X.    & X.    & X. & X & X   & X  \\
		     $B=9$		  & $ (3,3,9)/(6,9,12)$ & X.   & X.   & X. & X & X   & X  \\
		     $B\geq 10$  & - & - & - & - & - & - & -  \\
		\end{tabular}
		\caption{Intersection contributions to \Izero{}$\sim(A,B,0)$ and \Instar{} with $m=1$, respectively. `X' indicates non-minimal intersection or not allowed to have a particular \Instar{} on a curve of this self intersection and `-' indicates exceeding allowed residuals. `X.' indicates that the intersection is non-minimal here while in the non-compact case it is was not yet forbidden via the previous analysis in the non-compact case.}
		\label{t:InstarLocalContributionsSelfInt1}
	\end{center}
\end{table}

\end{subsection}

\begin{subsection}{Type \In{} intersections with $A=B=0$ curves}
In Table~\ref{t:sumloccontr} we show intersection contributions from type \In{} curves to those of every other possible gauged Kodaira type having minimal orders of vanishing on the transverse curve, i.e.,\ curves with $A=B=0.$ Up to a couple of minor corrections indicated here with a superscript `!' symbol, this contribution data is taken from~\cite{BertoliniGlobal1}.  
\begin{table}[!h]
\begin{center}
\begin{tabular}{c | c c c c c c c c c }
~ 			& $n$				&gauge algebra&	$\at$		&$\bt$		&$\dt$	\\
\hline
I${}_0^\ast$ 	& $\geq2$				& $\sp([n/2])$ 		&0 		&0 			& $n^!$ 	 \\
IV		 	& 2					& $\su(2)$			&0 		&2 			&4 	 \\
IV		 	& 3					& $\su(2)$			&0 		&3 			&6 	 \\
IV		 	& 3					& $\su(3)^!$			&1 		&2 			&4 	 \\
IV		 	& 4					& $\sp(2)$			&0 		&4 			&8 	 \\
IV		 	& 5					& $\sp(2)$			&0 		&5 			&10 	 \\
IV		 	& 6					& $\sp(3)$			&0 		&6 			&12 	 \\
IV		 	& $\geq7$				& $\sp([n/2])$		&0 		&$n$ 		&$2n$   \\
III		 	& $2$				& $\su(2)$			&1		&1			&3	 \\
III		 	& $3$				& b.p.			&2		&2			&6	 \\
III		 	& $4$				& $\sp(2)$			&2		& $2^!$			&6	 \\
III		 	& $4$				& $\su(4)$			&2		&3			&6	 \\
III		 	& $5$				& $\sp(2)$			&3		&3			&9	 \\
III		 	& $6$				& $\sp(3)$			&3		&3			&9	 \\
III		 	& $\geq7$				& $\sp([n/2])$		&$\lceil n/2 \rceil$		&$\lceil n/2 \rceil$		&$3\lceil n/2 \rceil$	
\end{tabular}
\caption{Local contributions from non-compact type I${}_n$ curves.}
\label{t:sumloccontr}
\end{center}
\end{table}
In the following table taken from~\cite{BertoliniGlobal1} for ease of reference and to provide a minor correction in one case, we collect the largest values of $n$ permitted for \In{} intersection with a non-compact curve having any of the various other Kodaira types with the exception of \Instar{}, which we have analyzed separately. The maximal values are $A,B$ dependent and hence collected into separate columns.
\begin{table}[!h]
\begin{center}
\begin{tabular}{c | c c c c c c c c c }
~ 			& $A=B=0$			&$A=1$		&$A\geq2$		&$B=1$		&$B\geq2$	\\
\hline
II${}^\ast$ 	& 0					& 0			& 0				& -			& - 	 \\
III${}^\ast$ 	& 0					& -			& -				& 0			& 0	 \\
IV${}^\ast$ 	& 1					& 1			& 1				& -			& -	 \\
I${}_0^\ast$ 	& $\infty$				& 3			& 3				& 2			& 2	 \\
IV		 	& $\infty$				& 3			& 3				& -			& -	 \\
III		 	& $\infty$				& -			& -				& $\infty$		& $5{}^!$	 \\
II			& $\infty$				& $\infty$		& 4				& -			& -
\end{tabular}
\caption{Allowed intersections for non-compact type I${}_n$ curves. Here a `$!$' superscript denotes a correction to the corresponding data in~\cite{BertoliniGlobal1}.}
\label{t:maxn}
\end{center}
\end{table}

\end{subsection}

\begin{subsection}{\Izero{},\Izerostar{} curve intersections}

Here we suppose that a curve $\Sigma$ of type \Izero{} with vanishings $(A,B,0)$ at $\{z= 0 \}$ meets a curve $\Sigma'$ of type I${}_{0}^*$ $= \{ \sigma = 0\} := \Sigma'$ and we investigate the restrictions on monodromy on $\Sigma'$ for various values of $A,B$, the additional orders of vanishing of $f,g$ along $\Sigma,$ as above. We let $A_0,B_0$ give the orders of vanishing along $\Sigma'$ as $(2+A_0, 3+B_0,6).$ Here we have the monodromy cover given by $\psi^3 + (f/\sigma^2)|_{\{\sigma=0\}} + (g/\sigma^3)|_{\{\sigma=0\}}$ and in the case where this splits as $(\psi- \alpha)(\psi - \beta ) (\psi -\gamma),$ it is given by 
\begin{align*}\psi^3 + \psi(-\beta^2 - \alpha^2 - \alpha \beta) -\alpha \beta(\alpha+ \beta).
\end{align*}
In the case that the \Izerostar{} curve $\Sigma'$ has $B_0\geq 4$ so that we have orders of vanishing $(2,\geq 6,6),$ we are near non-minimality. In fact, we find that if the \Izero{} has orders $(1,0,0)$ or beyond in $f,$ non-minimality forbids such an intersection. 

To clarify, in this case $\alpha \beta (\alpha + \beta)$ must vanish. Hence one of $\alpha, \beta,$ or $\alpha + \beta$ must vanish. Suppose $\alpha = -\beta.$  Then 
$$ \alpha^2 + \beta^2 + \alpha\beta = \alpha^2.$$
For $z| f$, we then have $z^2 | (f/\sigma^2)|_{\{\sigma=0\}}$, giving a $(4,6,12)$ point. To see this, we note that our requirements result in
\begin{align*} f = \sigma^2( z g_2 (z)  + z g_3(z)\sigma  + z g_4 (z) \sigma^2 + \dots )
\end{align*}
with $(f/\sigma^2)|_{\sigma = 0}$ a square, making $z g_2(z)$ is a square. Hence, orders $(3,6, 9)$ are boosted, with non-minimality resulting after considering monodromy. 
The other cases are similar. When $B>0$ along the \Izero{}, there is no analogous result. 

In the case that we have $A_0>0,$ the coefficient on $\psi$ vanishes and hence $(g/ \sigma^3) |_{\sigma = 0 } = (\alpha^2 + \beta^2 ) ( \alpha + \beta).$  The utility of this condition in considering $z^k| g$ to study $B>0$ is not immediately obvious and we shall proceed without making use of it.

Given $\so(7)$ algebra on \Izerostar{}, the monodromy cover splits partially. We can write it as 
\begin{align*}
\psi^3 + (\mu - \lambda^2) \psi - \lambda \mu.
\end{align*}
When $A_0>0,$ this becomes $\psi^3 - \lambda^3$. Since $\lambda^3 = (g/\sigma^3)|_{\sigma=0},$ we see that $z$ divisibility of $g$ along the \Izero{} becomes $z^3$ divisibility. In the case that $A_0\geq 2$, the order in $f$ is already 4 before intersection with the \Izero{}. This results in a $(4,6,12)$ point as the $(4,4,8)$ point at the intersection is boosted by the monodromy condition. In terms of expanding $g$, this reads
\begin{align*}
g = \sigma^3( z g_3 (z)  + z g_4(z)\sigma  + z g_5 (z) \sigma^2 + \dots )
\end{align*}
with $z g_3(z)$ a cube. The latter requires that $z^3$ divides $g|_{\sigma = 0}.$ 
Hence, $A_0\geq 2$ and $B\geq 1$ is forbidden for I${}_0^{\ast ss}.$
\begin{table}[!thb]
	\begin{center}
		\begin{tabular}{c | c c c c c c c c c c c c c }
          ~ & $\g_2$  & $\so(7)$  &  $\so(8)$  \\
          \hline
             &   $B_0>0$ & $A_0\geq 2 \ \&\  B>0$  & $B_0\geq 3 \  \&\  A>0$          
		\end{tabular}
		\caption{Forbidden intersections and type restrictions}
		\label{t:InstarRulesSummary}
	\end{center}
\end{table}
When we have $\g_2$ algebra, the monodromy cover is irreducible, and appears as $\psi^3 + q \psi + p$. If this had a root, it would appear as $\psi^3 + \psi(\mu - \lambda^2) - \lambda \mu.$  In the case with $A_0>0,$ irreducibility implies that $p \neq \lambda \mu$ where $\mu = \lambda^2,$ i.e., $(g/\sigma^3)|_{\sigma=0}$ is not a cube. In particular, this prevents $(g/\sigma^3)|_{\sigma=0}$ from being constant, hence barring the case with no residual vanishings along the \Izerostar{} after accounting for any other intersections. Here,
$g = \sigma^3 ( g_3 +\sigma g_4 +\sigma^2 g_5 + \dots),$ and the above is simply to say $g_3$ is not a cube. 
This result applies in all configurations involving an \Izerostar{} with $\g_2$ algebra; there is nothing used about an intersection with \Izero{}. To phrase this differently, if we have an \Izerostar{} with $\g_2$ algebra and $A_0>0,$ the other vanishings of $g$ along this curve cannot all appear as cubes. Among other restrictions this imposes, we see that when $\Sigma' \cdot \Sigma' =-m,$ with $m$ given by $1,2,3,$ the residual vanishings of $g$ before considering other intersections read $9,6,3,$ respectively in these cases. This implies for example that a $\g_2$ gauged \Izerostar{} with $A_0>0$ cannot meet $3,2,1$ other type III curves each having types $(1,3,3)$ in the respective cases. 

Similarly, for $B_0>0$ we have monodromy cover given by $\psi^3 + q \psi.$ This is not irreducible, thus ruling out the case with $\g_2$ algebra.

A few of these results are summarized in Table~\ref{t:InstarRulesSummary}.

\end{subsection}
\end{section}
\begin{section}{Notes for using the computer algebra workbook}\label{s:GSWorkbook}
The arXiv submission of this work is accompanied by `gsFunctions.nb,' computer algebra workbook for Mathematica enabling explicit listing of gauge enhancements and global symmetry maxima for nearly all 6D SCFTs. This notebook can be downloaded via the ``Download:'' link in the arXiv webpage featuring the submission history and abstract. Click on the ``Other format'' link and then on ``Download source.'' After unzipping this file, all contents should be moved to a single folder. The `gsFunctions.nb' workbook contains function definitions and the entire notebook should be evaluated to initialize them. At the bottom of this workbook are a series of examples with comments detailing instructions to initiate computation of gauge and global symmetries for an arbitrary at-most-trivalent base given as user input. Results can be written to a LaTex file in the workbook directory which can be compiled to view results condensed to the format of tables appearing in Appendix~\ref{s:GSTables}. Instructions for in-workbook use of several additional functions appear in comments detailing further examples. 
\end{section}

\begin{section}{Tables of flavor symmetries for miscellaneous quivers}\label{s:GSTables}
In this appendix we provide configuration data for a few miscellaneous bases. For each $\T,$ summands of $\gG$ appear under each curve with $\gG$ totals shown in the rightmost column. Note that in listings for $\alpha m$, permitted $\T$ compatible with $m$ are constrained via the permitted gauging rules discussed at the start of Section~\ref{s:GSRules}. 
\subsection{$321m:$}
Configuration data for the bases $321m_\Sigma$ with $m\in\{3,5,6,7,8\}$ appear in Table~\ref{t:GS3213}; note that only some $\T$ are compatible with the permitted gauging rules for $m>3$.\\

\setlength{\columnsep}{0.35cm}
\noindent\begin{minipage}{\textwidth}
\begin{multicols}{2}

\setbox\ltmcbox\vbox{
\makeatletter\col@number\@ne
{ \fontsize{6.3}{7.3}\selectfont\setlength{\tabcolsep}{1pt} \begin{longtable} {ccccc}3 & 2 & 1 & $m$ & GS Total: \\
  (\Izerostarns{},\gText{}) & (III,\suText<2>{}) & (\Izero{},\nzero{}) & (\IVstars{},\eText<6>{}) &  \\
    $   $  &  $   $  &  $   $  &  $   $  &  $ 0 $ \\
   (\Izerostarns{},\gText{}) & (\IVns{},\suText<2>{}) & (\Izero{},\nzero{}) & (\IVstars{},\eText<6>{}) &  \\
    $   $  &  $   $  &  $   $  &  $   $  &  $ 0 $ \\
   (\Izerostarns{},\gText{}) & (III,\suText<2>{}) & (\Izero{},\nzero{}) & (\IIIstar{},\eText<7>{}) &  \\
    $   $  &  $   $  &  $   $  &  $   $  &  $ 0 $ \\
   (\Izerostarns{},\gText{}) & (III,\suText<2>{}) & (\Izero{},\nzero{}) & (\IVstarns{},\fText{}) &  \\
    $   $  &  $   $  &  $   $  &  $   $  &  $ 0 $ \\
   (\Izerostarns{},\gText{}) & (\IVns{},\suText<2>{}) & (\II{},\nzero{}) & (\IVstarns{},\fText{}) &  \\
    $   $  &  $   $  &  $   $  &  $   $  &  $ 0 $ \\
   (\Izerostarns{},\gText{}) & (\IVns{},\suText<2>{}) & (\Izero{},\nzero{}) & (\IVstarns{},\fText{}) &  \\
    $   $  &  $   $  &  $   $  &  $   $  &  $ 0 $ \\
   (\Izerostarns{},\gText{}) & (\IVns{},\suText<2>{}) & (\II{},\nzero{}) & (\Izerostars{},\soText<8>{}) &  \\
    $   $  &  $   $  &  $   $  &  $   $  &  $ 0 $ \\
   (\Izerostarns{},\gText{}) & (III,\suText<2>{}) & (\Izero{},\nzero{}) & (\Izerostarns{},\gText{}) &  \\
    $ 0 $  &  $ 0 $  &  $ A_{1} $  &  $ A_{1} $  &  $ A_{1}^2 $ \\
   (\Izerostarns{},\gText{}) & (\IVns{},\suText<2>{}) & (\Izero{},\nzero{}) & (\Izerostarns{},\gText{}) &  \\
    $ 0 $  &  $ 0 $  &  $ 0 $  &  $ A_{1} $  &  $ A_{1} $ \\
   (\Izerostarns{},\gText{}) & (\IVns{},\suText<2>{}) & (\II{},\nzero{}) & (\Izerostarns{},\gText{}) &  \\
    $ 0 $  &  $ 0 $  &  $ A_{1} $  &  $ A_{1} $  &  $ A_{1}^2 $ \\
   (\gText{}) & (\suText<2>{}) & (\nzero{}) & (\gText{}) &  \\
    $ 0 $  &  $ 0 $  &  $ A_{1} $  &  $ A_{1} $  &  $ A_{1}^2 $ \\
   (\Izerostarns{},\gText{}) & (III,\suText<2>{}) & (\Izero{},\nzero{}) & (\Instars <1>{},\soText<10>{}) &  \\
    $ 0 $  &  $ 0 $  &  $ A_{1} $  &  $ C_{3} $  &  $ A_{1} \oplus C_{3} $ \\
   (\Izerostarns{},\gText{}) & (\IVns{},\suText<2>{}) & (\Izero{},\nzero{}) & (\Instars <1>{},\soText<10>{}) &  \\
    $ 0 $  &  $ 0 $  &  $ 0 $  &  $ C_{3} $  &  $ C_{3} $ \\
   (\Izerostarns{},\gText{}) & (\IVns{},\suText<2>{}) & (\Ione{},\nzero{}) & (\Instars <1>{},\soText<10>{}) &  \\
    $ 0 $  &  $ 0 $  &  $ 0 $  &  $ C_{3} $  &  $ C_{3} $ \\
   (\gText{}) & (\suText<2>{}) & (\nzero{}) & (\soText<10>{}) &  \\
    $ 0 $  &  $ 0 $  &  $ A_{1} $  &  $ C_{3} $  &  $ A_{1} \oplus C_{3} $ \\
   (\Izerostarns{},\gText{}) & (III,\suText<2>{}) & (\Izero{},\nzero{}) & (\Instarns <2>{},\soText<11>{}) &  \\
    $ 0 $  &  $ 0 $  &  $ 0 $  &  $ C_{4} $  &  $ C_{4} $ \\
   (\Izerostarns{},\gText{}) & (\IVns{},\suText<2>{}) & (\Ione{},\nzero{}) & (\Instarns <2>{},\soText<11>{}) &  \\
    $ 0 $  &  $ 0 $  &  $ 0 $  &  $ C_{4} $  &  $ C_{4} $ \\
   (\gText{}) & (\suText<2>{}) & (\nzero{}) & (\soText<11>{}) &  \\
    $ 0 $  &  $ 0 $  &  $ 0 $  &  $ C_{4} $  &  $ C_{4} $ \\
   (\Izerostarns{},\gText{}) & (III,\suText<2>{}) & (\Izero{},\nzero{}) & (\Instars <2>{},\soText<12>{}) &  \\
    $ 0 $  &  $ 0 $  &  $ 0 $  &  $ C_{5} $  &  $ C_{5} $ \\
   (\Izerostarns{},\gText{}) & (III,\suText<2>{}) & (\Izero{},\nzero{}) & (\Izerostarss{},\soText<7>{}) &  \\
    $ 0 $  &  $ 0 $  &  $ A_{1} $  &  $ C_{2} $  &  $ A_{1} \oplus C_{2} $ \\
   (\Izerostarns{},\gText{}) & (\IVns{},\suText<2>{}) & (\Izero{},\nzero{}) & (\Izerostarss{},\soText<7>{}) &  \\
    $ 0 $  &  $ 0 $  &  $ 0 $  &  $ C_{2} $  &  $ C_{2} $ \\
   (\Izerostarns{},\gText{}) & (\IVns{},\suText<2>{}) & (\II{},\nzero{}) & (\Izerostarss{},\soText<7>{}) &  \\
    $ 0 $  &  $ 0 $  &  $ 0 $  &  $ C_{2} $  &  $ C_{2} $ \\
   (\Izerostarns{},\gText{}) & (\IVns{},\suText<2>{}) & (\Ione{},\nzero{}) & (\Izerostarss{},\soText<7>{}) &  \\
    $ 0 $  &  $ 0 $  &  $ A_{1} $  &  $ C_{2} $  &  $ A_{1} \oplus C_{2} $ \\
   (\gText{}) & (\suText<2>{}) & (\nzero{}) & (\soText<7>{}) &  \\
    $ 0 $  &  $ 0 $  &  $ A_{1} $  &  $ C_{2} $  &  $ A_{1} \oplus C_{2} $ \\
    $\vdots$ & $\vdots$&$\vdots$&$\vdots$&$\vdots$\\
    & & & & \\
    & & & & \\    
    $\vdots$ & $\vdots$&$\vdots$&$\vdots$&$\vdots$\\
   (\Izerostarns{},\gText{}) & (III,\suText<2>{}) & (\Izero{},\nzero{}) & (\Izerostars{},\soText<8>{}) &  \\
    $ 0 $  &  $ 0 $  &  $ A_{1} $  &  $ A_{1}^3 $  &  $ A_{1}^4 $ \\
   (\Izerostarns{},\gText{}) & (\IVns{},\suText<2>{}) & (\Izero{},\nzero{}) & (\Izerostars{},\soText<8>{}) &  \\
    $ 0 $  &  $ 0 $  &  $ 0 $  &  $ A_{1}^3 $  &  $ A_{1}^3 $ \\
   (\Izerostarns{},\gText{}) & (\IVns{},\suText<2>{}) & (\Ione{},\nzero{}) & (\Izerostars{},\soText<8>{}) &  \\
    $ 0 $  &  $ 0 $  &  $ 0 $  &  $ A_{1}^2 $  &  $ A_{1}^2 $ \\
   (\gText{}) & (\suText<2>{}) & (\nzero{}) & (\soText<8>{}) &  \\
    $ 0 $  &  $ 0 $  &  $ A_{1} $  &  $ A_{1}^3 $  &  $ A_{1}^4 $ \\
   (\Izerostarns{},\gText{}) & (III,\suText<2>{}) & (\Izero{},\nzero{}) & (\Instarns <1>{},\soText<9>{}) &  \\
    $ 0 $  &  $ 0 $  &  $ A_{1} $  &  $ C_{2} $  &  $ A_{1} \oplus C_{2} $ \\
   (\Izerostarns{},\gText{}) & (\IVns{},\suText<2>{}) & (\Izero{},\nzero{}) & (\Instarns <1>{},\soText<9>{}) &  \\
    $ 0 $  &  $ 0 $  &  $ 0 $  &  $ C_{2} $  &  $ C_{2} $ \\
   (\Izerostarns{},\gText{}) & (\IVns{},\suText<2>{}) & (\II{},\nzero{}) & (\Instarns <1>{},\soText<9>{}) &  \\
    $ 0 $  &  $ 0 $  &  $ 0 $  &  $ C_{2} $  &  $ C_{2} $ \\
   (\Izerostarns{},\gText{}) & (\IVns{},\suText<2>{}) & (\Ione{},\nzero{}) & (\Instarns <1>{},\soText<9>{}) &  \\
    $ 0 $  &  $ 0 $  &  $ 0 $  &  $ C_{2} $  &  $ C_{2} $ \\
   (\gText{}) & (\suText<2>{}) & (\nzero{}) & (\soText<9>{}) &  \\
    $ 0 $  &  $ 0 $  &  $ A_{1} $  &  $ C_{2} $  &  $ A_{1} \oplus C_{2} $ \\
   (\Izerostarns{},\gText{}) & (III,\suText<2>{}) & (\Izero{},\nzero{}) & (\IVs{},\suText<3>{}) &  \\
    $ 0 $  &  $ 0 $  &  $ A_{3} $  &  $ 0 $  &  $ A_{3} $ \\
    $ 0 $  &  $ 0 $  &  $ A_{1} \oplus A_{2} $  &  $ 0 $  &  $ A_{1} \oplus A_{2} $ \\
   (\Izerostarns{},\gText{}) & (\IVns{},\suText<2>{}) & (\Izero{},\nzero{}) & (\IVs{},\suText<3>{}) &  \\
    $ 0 $  &  $ 0 $  &  $ A_{1}^2 $  &  $ 0 $  &  $ A_{1}^2 $ \\
   (\gText{}) & (\suText<2>{}) & (\nzero{}) & (\suText<3>{}) &  \\
    $ 0 $  &  $ 0 $  &  $ A_{3} $  &  $ 0 $  &  $ A_{3} $ \\
    $ 0 $  &  $ 0 $  &  $ A_{1} \oplus A_{2} $  &  $ 0 $  &  $ A_{1} \oplus A_{2} $ \\
   (\Izerostarss{},\soText<7>{}) & (III,\suText<2>{}) & (\Izero{},\nzero{}) & (\IVstars{},\eText<6>{}) &  \\
    $ A_{1} $  &  $ 0 $  &  $ 0 $  &  $ 0 $  &  $ A_{1} $ \\
   (\Izerostarss{},\soText<7>{}) & (III,\suText<2>{}) & (\Izero{},\nzero{}) & (\IIIstar{},\eText<7>{}) &  \\
    $ A_{1} $  &  $ 0 $  &  $ 0 $  &  $ 0 $  &  $ A_{1} $ \\
   (\Izerostarss{},\soText<7>{}) & (III,\suText<2>{}) & (\Izero{},\nzero{}) & (\IVstarns{},\fText{}) &  \\
    $ A_{1} $  &  $ 0 $  &  $ 0 $  &  $ 0 $  &  $ A_{1} $ \\
   (\Izerostarss{},\soText<7>{}) & (III,\suText<2>{}) & (\Izero{},\nzero{}) & (\Izerostarns{},\gText{}) &  \\
    $ A_{1} $  &  $ 0 $  &  $ A_{1} $  &  $ A_{1} $  &  $ A_{1}^3 $ \\
   (\Izerostarss{},\soText<7>{}) & (III,\suText<2>{}) & (\Izero{},\nzero{}) & (\Instars <1>{},\soText<10>{}) &  \\
    $ A_{1} $  &  $ 0 $  &  $ A_{1} $  &  $ C_{3} $  &  $ A_{1}^2 \oplus C_{3} $ \\
   (\Izerostarss{},\soText<7>{}) & (III,\suText<2>{}) & (\Izero{},\nzero{}) & (\Instarns <2>{},\soText<11>{}) &  \\
    $ A_{1} $  &  $ 0 $  &  $ 0 $  &  $ C_{4} $  &  $ A_{1} \oplus C_{4} $ \\
   (\Izerostarss{},\soText<7>{}) & (III,\suText<2>{}) & (\Izero{},\nzero{}) & (\Instars <2>{},\soText<12>{}) &  \\
    $ A_{1} $  &  $ 0 $  &  $ 0 $  &  $ C_{5} $  &  $ A_{1} \oplus C_{5} $ \\
   (\Izerostarss{},\soText<7>{}) & (III,\suText<2>{}) & (\Izero{},\nzero{}) & (\Izerostarss{},\soText<7>{}) &  \\
    $ A_{1} $  &  $ 0 $  &  $ A_{1} $  &  $ C_{2} $  &  $ A_{1}^2 \oplus C_{2} $ \\
   (\Izerostarss{},\soText<7>{}) & (III,\suText<2>{}) & (\Izero{},\nzero{}) & (\Izerostars{},\soText<8>{}) &  \\
    $ A_{1} $  &  $ 0 $  &  $ A_{1} $  &  $ A_{1}^3 $  &  $ A_{1}^5 $ \\
   (\Izerostarss{},\soText<7>{}) & (III,\suText<2>{}) & (\Izero{},\nzero{}) & (\Instarns <1>{},\soText<9>{}) &  \\
    $ A_{1} $  &  $ 0 $  &  $ A_{1} $  &  $ C_{2} $  &  $ A_{1}^2 \oplus C_{2} $ \\
   (\Izerostarss{},\soText<7>{}) & (III,\suText<2>{}) & (\Izero{},\nzero{}) & (\IVs{},\suText<3>{}) &  \\
    $ A_{1} $  &  $ 0 $  &  $ A_{3} $  &  $ 0 $  &  $ A_{1} \oplus A_{3} $ \\
    $ A_{1} $  &  $ 0 $  &  $ A_{1} \oplus A_{2} $  &  $ 0 $  &  $ A_{1}^2 \oplus A_{2} $ \\
    &&&&\\
      \caption{All configurations for $321m,$ $m\in\{3,5,6,7,8\}.$}
         \label{t:GS3213}
  \end{longtable} } 

\unskip
\unpenalty
\unpenalty}
\unvbox\ltmcbox
\end{multicols}
\end{minipage}
\vfill
\setlength{\columnsep}{1cm}
\subsection{$31:$}
\ \\
\setlength{\columnsep}{0.35cm}
\begin{minipage}{\textwidth}
\begin{multicols}{3}

\setbox\ltmcbox\vbox{
\makeatletter\col@number\@ne
{ \fontsize{5.8}{6.4}\selectfont\setlength{\tabcolsep}{0.2pt}
\renewcommand*{\arraystretch}{1.2}
 \begin{longtable} {ccc}3 & 1 & GS Total: \\
   (\IVstars{},\eText<6>{}) & (\Izero{},\nzero{}) &  \\
    $ 0 $  &  $ A_{2} $  &  $ A_{2} $ \\
   (\IIIstar{},\eText<7>{}) & (\Izero{},\nzero{}) &  \\
    $ 0 $  &  $ A_{1} $  &  $ A_{1} $ \\
   (\IVstarns{},\fText{}) & (\Izero{},\nzero{}) &  \\
    $ 0 $  &  $ A_{2} $  &  $ A_{2} $ \\
   (\IVstarns{},\fText{}) & (\II{},\nzero{}) &  \\
    $ 0 $  &  $ \g_{2} $  &  $ \g_{2} $ \\
   (\IVstarns{},\fText{}) & (\Ione{},\nzero{}) &  \\
    $ 0 $  &  $ A_{2} $  &  $ A_{2} $ \\
   (\fText{}) & (\nzero{}) &  \\
    $ 0 $  &  $ \g_{2} $  &  $ \g_{2} $ \\
   (\Izerostarns{},\gText{}) & (\Izero{},\nzero{}) &  \\
    $ A_{1} $  &  $ D_{4} $  &  $ A_{1} \oplus D_{4} $ \\
   (\Izerostarns{},\gText{}) & (\II{},\nzero{}) &  \\
    $ A_{1} $  &  $ \f_{4} $  &  $ A_{1} \oplus \f_{4} $ \\
   (\Izerostarns{},\gText{}) & (\Ione{},\nzero{}) &  \\
    $ A_{1} $  &  $ A_{3} $  &  $ A_{1} \oplus A_{3} $ \\
   (\gText{}) & (\nzero{}) &  \\
    $ A_{1} $  &  $ \f_{4} $  &  $ A_{1} \oplus \f_{4} $ \\
   (\Izerostarns{},\gText{}) & (III,\suText<2>{}) &  \\
    $ 0 $  &  $ A_{1} \oplus B_{3} $  &  $ A_{1} \oplus B_{3} $ \\
    $ 0 $  &  $ C_{3} $  &  $ C_{3} $ \\
   (\Izerostarns{},\gText{}) & (\IVns{},\suText<2>{}) &  \\
    $ 0 $  &  $ A_{2} \oplus \g_{2} $  &  $ A_{2} \oplus \g_{2} $ \\
    $ 0 $  &  $ C_{2} $  &  $ C_{2} $ \\
   (\Izerostarns{},\gText{}) & (\Itwo{},\suText<2>{}) &  \\
    $ 0 $  &  $ D_{6} $  &  $ D_{6} $ \\
   (\Izerostarns{},\gText{}) & (\Inns<3>{},\suText<2>{}) &  \\
    $ 0 $  &  $ B_{6} $  &  $ B_{6} $ \\
   (\gText{}) & (\suText<2>{}) &  \\
    $ 0 $  &  $ B_{6} $  &  $ B_{6} $ \\
   (\Instars <1>{},\soText<10>{}) & (\Izero{},\nzero{}) &  \\
    $ C_{3} $  &  $ A_{3} $  &  $ A_{3} \oplus C_{3} $ \\
    $ C_{3} $  &  $ A_{1} \oplus A_{2} $  &  $ A_{1} \oplus A_{2} \oplus C_{3} $ \\
   (\Instars <1>{},\soText<10>{}) & (\Ione{},\nzero{}) &  \\
    $ C_{3} $  &  $ A_{1}^2 $  &  $ A_{1}^2 \oplus C_{3} $ \\
   (\soText<10>{}) & (\nzero{}) &  \\
    $ C_{3} $  &  $ A_{3} $  &  $ A_{3} \oplus C_{3} $ \\
    $ C_{3} $  &  $ A_{1} \oplus A_{2} $  &  $ A_{1} \oplus A_{2} \oplus C_{3} $ \\
   (\Instars <1>{},\soText<10>{}) & (\Inns<4>{},\spText<2>{}) &  \\
    $ A_{1} $  &  $ D_{7} $  &  $ A_{1} \oplus D_{7} $ \\
   (\Instars <1>{},\soText<10>{}) & (\Inns<5>{},\spText<2>{}) &  \\
    $ A_{1} $  &  $ B_{6} $  &  $ A_{1} \oplus B_{6} $ \\
   (\soText<10>{}) & (\spText<2>{}) &  \\
    $ A_{1} $  &  $ D_{7} $  &  $ A_{1} \oplus D_{7} $ \\
   (\Instars <1>{},\soText<10>{}) & (\Inns<6>{},\spText<3>{}) &  \\
    $ 0 $  &  $ D_{9} $  &  $ D_{9} $ \\
   (\Instars <1>{},\soText<10>{}) & (\Itwo{},\suText<2>{}) &  \\
    $ C_{2} $  &  $ D_{5} $  &  $ C_{2} \oplus D_{5} $ \\
   (\Instars <1>{},\soText<10>{}) & (\Inns<3>{},\suText<2>{}) &  \\
    $ C_{2} $  &  $ B_{4} $  &  $ B_{4} \oplus C_{2} $ \\
   (\soText<10>{}) & (\suText<2>{}) &  \\
    $ C_{2} $  &  $ D_{5} $  &  $ C_{2} \oplus D_{5} $ \\
   (\Instarns <2>{},\soText<11>{}) & (\Izero{},\nzero{}) &  \\
    $ C_{4} $  &  $ A_{1}^2 $  &  $ A_{1}^2 \oplus C_{4} $ \\
   (\Instarns <2>{},\soText<11>{}) & (\Ione{},\nzero{}) &  \\
    $ C_{4} $  &  $ A_{1} $  &  $ A_{1} \oplus C_{4} $ \\  
   (\soText<11>{}) & (\nzero{}) &  \\
    $ C_{4} $  &  $ A_{1}^2 $  &  $ A_{1}^2 \oplus C_{4} $ \\
    $\vdots$ & $\vdots$ & $\vdots$ \\
    & & \\ 
    & & \\
    $\vdots$ & $\vdots$ & $\vdots$ \\          
   (\Instarns <2>{},\soText<11>{}) & (\Inns<4>{},\spText<2>{}) &  \\
    $ C_{2} $  &  $ D_{6} $  &  $ C_{2} \oplus D_{6} $ \\
   (\Instarns <2>{},\soText<11>{}) & (\Inns<5>{},\spText<2>{}) &  \\
    $ C_{2} $  &  $ B_{6} $  &  $ B_{6} \oplus C_{2} $ \\
   (\soText<11>{}) & (\spText<2>{}) &  \\
    $ C_{2} $  &  $ B_{6} $  &  $ B_{6} \oplus C_{2} $ \\
   (\Instarns <2>{},\soText<11>{}) & (\Inns<6>{},\spText<3>{}) &  \\
    $ A_{1} $  &  $ D_{8} $  &  $ A_{1} \oplus D_{8} $ \\
   (\Instarns <2>{},\soText<11>{}) & (\Inns<7>{},\spText<3>{}) &  \\
    $ A_{1} $  &  $ B_{8} $  &  $ A_{1} \oplus B_{8} $ \\
   (\soText<11>{}) & (\spText<3>{}) &  \\
    $ A_{1} $  &  $ B_{8} $  &  $ A_{1} \oplus B_{8} $ \\
   (\Instarns <2>{},\soText<11>{}) & (\Inns<8>{},\spText<4>{}) &  \\
    $ 0 $  &  $ D_{10} $  &  $ D_{10} $ \\
   (\Instarns <2>{},\soText<11>{}) & (\Inns<9>{},\spText<4>{}) &  \\
    $ 0 $  &  $ B_{10} $  &  $ B_{10} $ \\
   (\soText<11>{}) & (\spText<4>{}) &  \\
    $ 0 $  &  $ B_{10} $  &  $ B_{10} $ \\
   (\Instarns <2>{},\soText<11>{}) & (\Itwo{},\suText<2>{}) &  \\
    $ C_{3} $  &  $ D_{4} $  &  $ C_{3} \oplus D_{4} $ \\
   (\Instarns <2>{},\soText<11>{}) & (\Inns<3>{},\suText<2>{}) &  \\
    $ C_{3} $  &  $ B_{4} $  &  $ B_{4} \oplus C_{3} $ \\
   (\soText<11>{}) & (\suText<2>{}) &  \\
    $ C_{3} $  &  $ B_{4} $  &  $ B_{4} \oplus C_{3} $ \\
   (\Instars <2>{},\soText<12>{}) & (\Izero{},\nzero{}) &  \\
    $ C_{5} $  &  $ A_{1}^2 $  &  $ A_{1}^2 \oplus C_{5} $ \\
   (\Instars <2>{},\soText<12>{}) & (\Ione{},\nzero{}) &  \\
    $ C_{5} $  &  $ 0 $  &  $ C_{5} $ \\
   (\soText<12>{}) & (\nzero{}) &  \\
    $ C_{5} $  &  $ A_{1}^2 $  &  $ A_{1}^2 \oplus C_{5} $ \\
   (\Instars <2>{},\soText<12>{}) & (\Inns<4>{},\spText<2>{}) &  \\
    $ C_{3} $  &  $ D_{6} $  &  $ C_{3} \oplus D_{6} $ \\
   (\Instars <2>{},\soText<12>{}) & (\Inns<5>{},\spText<2>{}) &  \\
    $ C_{3} $  &  $ B_{5} $  &  $ B_{5} \oplus C_{3} $ \\
   (\soText<12>{}) & (\spText<2>{}) &  \\
    $ C_{3} $  &  $ D_{6} $  &  $ C_{3} \oplus D_{6} $ \\
   (\Instars <2>{},\soText<12>{}) & (\Inns<6>{},\spText<3>{}) &  \\
    $ C_{2} $  &  $ D_{8} $  &  $ C_{2} \oplus D_{8} $ \\
   (\Instars <2>{},\soText<12>{}) & (\Inns<7>{},\spText<3>{}) &  \\
    $ C_{2} $  &  $ B_{7} $  &  $ B_{7} \oplus C_{2} $ \\
   (\soText<12>{}) & (\spText<3>{}) &  \\
    $ C_{2} $  &  $ D_{8} $  &  $ C_{2} \oplus D_{8} $ \\
   (\Instars <2>{},\soText<12>{}) & (\Inns<8>{},\spText<4>{}) &  \\
    $ A_{1} $  &  $ D_{10} $  &  $ A_{1} \oplus D_{10} $ \\
   (\Instars <2>{},\soText<12>{}) & (\Inns<9>{},\spText<4>{}) &  \\
    $ A_{1} $  &  $ B_{9} $  &  $ A_{1} \oplus B_{9} $ \\
   (\soText<12>{}) & (\spText<4>{}) &  \\
    $ A_{1} $  &  $ D_{10} $  &  $ A_{1} \oplus D_{10} $ \\
   (\Instars <2>{},\soText<12>{}) & (\Inns<10>{},\spText<5>{}) &  \\
    $ 0 $  &  $ D_{12} $  &  $ D_{12} $ \\
   (\Instars <2>{},\soText<12>{}) & (\Itwo{},\suText<2>{}) &  \\
    $ C_{4} $  &  $ D_{4} $  &  $ C_{4} \oplus D_{4} $ \\
   (\Instars <2>{},\soText<12>{}) & (\Inns<3>{},\suText<2>{}) &  \\
    $ C_{4} $  &  $ B_{3} $  &  $ B_{3} \oplus C_{4} $ \\
   (\soText<12>{}) & (\suText<2>{}) &  \\
    $ C_{4} $  &  $ D_{4} $  &  $ C_{4} \oplus D_{4} $ \\
   (\Izerostarss{},\soText<7>{}) & (\Izero{},\nzero{}) &  \\
    $ C_{2} $  &  $ D_{4} $  &  $ C_{2} \oplus D_{4} $ \\
   (\Izerostarss{},\soText<7>{}) & (\II{},\nzero{}) &  \\
    $ C_{2} $  &  $ \g_{2} $  &  $ C_{2} \oplus \g_{2} $ \\
    $\vdots$ & $\vdots$ & $\vdots$ \\
    & & \\ 
    $\vdots$ & $\vdots$ & $\vdots$ \\               
   (\Izerostarss{},\soText<7>{}) & (\Ione{},\nzero{}) &  \\
    $ C_{2} $  &  $ A_{2}^2 $  &  $ A_{2}^2 \oplus C_{2} $ \\
    $ C_{2} $  &  $ B_{4} $  &  $ B_{4} \oplus C_{2} $ \\
    $ C_{2} $  &  $ A_{1}^2 \oplus A_{2} $  &  $ A_{1}^2 \oplus A_{2} \oplus C_{2} $ \\
   (\soText<7>{}) & (\nzero{}) &  \\
    $ C_{2} $  &  $ A_{2}^2 $  &  $ A_{2}^2 \oplus C_{2} $ \\
    $ C_{2} $  &  $ B_{4} $  &  $ B_{4} \oplus C_{2} $ \\
    $ C_{2} $  &  $ A_{1}^2 \oplus A_{2} $  &  $ A_{1}^2 \oplus A_{2} \oplus C_{2} $ \\
   (\Izerostarss{},\soText<7>{}) & (\Inns<4>{},\spText<2>{}) &  \\
    $ 0 $  &  $ D_{8} $  &  $ D_{8} $ \\
   (\Izerostarss{},\soText<7>{}) & (III,\suText<2>{}) &  \\
    $ A_{1} $  &  $ A_{1} \oplus B_{3} $  &  $ A_{1}^2 \oplus B_{3} $ \\
    $ A_{1} $  &  $ C_{3} $  &  $ A_{1} \oplus C_{3} $ \\
   (\Izerostarss{},\soText<7>{}) & (\Itwo{},\suText<2>{}) &  \\
    $ A_{1} $  &  $ D_{6} $  &  $ A_{1} \oplus D_{6} $ \\
   (\Izerostarss{},\soText<7>{}) & (\Inns<3>{},\suText<2>{}) &  \\
    $ A_{1} $  &  $ B_{6} $  &  $ A_{1} \oplus B_{6} $ \\
   (\soText<7>{}) & (\suText<2>{}) &  \\
    $ A_{1} $  &  $ B_{6} $  &  $ A_{1} \oplus B_{6} $ \\
   (\Izerostars{},\soText<8>{}) & (\Izero{},\nzero{}) &  \\
    $ A_{1}^3 $  &  $ D_{4} $  &  $ A_{1}^3 \oplus D_{4} $ \\
   (\Izerostars{},\soText<8>{}) & (\Ione{},\nzero{}) &  \\
    $ A_{1}^2 $  &  $ A_{2}^2 $  &  $ A_{1}^2 \oplus A_{2}^2 $ \\
    $ A_{1}^2 $  &  $ B_{3} $  &  $ A_{1}^2 \oplus B_{3} $ \\
   (\Izerostars{},\soText<8>{}) & (\II{},\nzero{}) &  \\
    $ 0 $  &  $ \g_{2} $  &  $ \g_{2} $ \\
   (\soText<8>{}) & (\nzero{}) &  \\
    $ A_{1}^3 $  &  $ D_{4} $  &  $ A_{1}^3 \oplus D_{4} $ \\
    $ A_{1}^2 $  &  $ A_{2}^2 $  &  $ A_{1}^2 \oplus A_{2}^2 $ \\
   (\Izerostars{},\soText<8>{}) & (\Itwo{},\suText<2>{}) &  \\
    $ A_{1}^2 $  &  $ D_{6} $  &  $ A_{1}^2 \oplus D_{6} $ \\
   (\Izerostars{},\soText<8>{}) & (\Inns<3>{},\suText<2>{}) &  \\
    $ A_{1} $  &  $ B_{5} $  &  $ A_{1} \oplus B_{5} $ \\
   (\soText<8>{}) & (\suText<2>{}) &  \\
    $ A_{1}^2 $  &  $ D_{6} $  &  $ A_{1}^2 \oplus D_{6} $ \\
   (\Instarns <1>{},\soText<9>{}) & (\Izero{},\nzero{}) &  \\
    $ C_{2} $  &  $ A_{3} $  &  $ A_{3} \oplus C_{2} $ \\
    $ C_{2} $  &  $ A_{1} \oplus A_{2} $  &  $ A_{1} \oplus A_{2} \oplus C_{2} $ \\
   (\Instarns <1>{},\soText<9>{}) & (\II{},\nzero{}) &  \\
    $ C_{2} $  &  $ \g_{2} $  &  $ C_{2} \oplus \g_{2} $ \\
   (\Instarns <1>{},\soText<9>{}) & (\Ione{},\nzero{}) &  \\
    $ C_{2} $  &  $ B_{3} $  &  $ B_{3} \oplus C_{2} $ \\
   (\soText<9>{}) & (\nzero{}) &  \\
    $ C_{2} $  &  $ B_{3} $  &  $ B_{3} \oplus C_{2} $ \\
    $ C_{2} $  &  $ A_{1} \oplus A_{2} $  &  $ A_{1} \oplus A_{2} \oplus C_{2} $ \\
   (\Instarns <1>{},\soText<9>{}) & (\Inns<4>{},\spText<2>{}) &  \\
    $ 0 $  &  $ D_{7} $  &  $ D_{7} $ \\
   (\Instarns <1>{},\soText<9>{}) & (\Inns<5>{},\spText<2>{}) &  \\
    $ 0 $  &  $ B_{7} $  &  $ B_{7} $ \\
   (\soText<9>{}) & (\spText<2>{}) &  \\
    $ 0 $  &  $ B_{7} $  &  $ B_{7} $ \\
   (\Instarns <1>{},\soText<9>{}) & (\Itwo{},\suText<2>{}) &  \\
    $ A_{1} $  &  $ D_{5} $  &  $ A_{1} \oplus D_{5} $ \\
   (\Instarns <1>{},\soText<9>{}) & (\Inns<3>{},\suText<2>{}) &  \\
    $ A_{1} $  &  $ B_{5} $  &  $ A_{1} \oplus B_{5} $ \\
   (\soText<9>{}) & (\suText<2>{}) &  \\
    $ A_{1} $  &  $ B_{5} $  &  $ A_{1} \oplus B_{5} $ \\
   (\IVs{},\suText<3>{}) & (\Izero{},\nzero{}) &  \\
    $ 0 $  &  $ \e_{6} $  &  $ \e_{6} $ \\
    & & \\
    & & \\    
   \caption{All configurations for the base $31.$}
   \label{t:GS31}
  \end{longtable} } 
\unskip
\unpenalty
\unpenalty}

\unvbox\ltmcbox

\end{multicols}
\end{minipage}
\vfill
\newpage

\subsection{$m12:$}
\begin{minipage}{\textwidth}
\begin{multicols}{2}
\medskip

\setbox\ltmcbox\vbox{
\makeatletter\col@number\@ne
{ \fontsize{6.3}{7.3}\selectfont\setlength{\tabcolsep}{3pt} \begin{longtable} {cccc}5 & 1 & 2 & GS Total: \\
  (\IVstars{},\eText<6>{}) & (\Izero{},\nzero{}) & (\Izero{},\nzero{}) &  \\
   $ 0 $  &  $ A_{2} $  &  $ 0 $  &  $ A_{2} $ \\
  (\IVstars{},\eText<6>{}) & (\Izero{},\nzero{}) & (\II{},\nzero{}) &  \\
   $ 0 $  &  $ A_{1} $  &  $ A_{1} $  &  $ A_{1}^2 $ \\
  (\IVstars{},\eText<6>{}) & (\Izero{},\nzero{}) & (\Ione{},\nzero{}) &  \\
   $ 0 $  &  $ A_{2} $  &  $ A_{1} $  &  $ A_{1} \oplus A_{2} $ \\
  (\eText<6>{}) & (\nzero{}) & (\nzero{}) &  \\
   $ 0 $  &  $ A_{2} $  &  $ A_{1} $  &  $ A_{1} \oplus A_{2} $ \\
  (\IVstars{},\eText<6>{}) & (\Izero{},\nzero{}) & (III,\suText<2>{}) &  \\
   $ 0 $  &  $ 0 $  &  $ B_{3} $  &  $ B_{3} $ \\
  (\IVstars{},\eText<6>{}) & (\Izero{},\nzero{}) & (\Itwo{},\suText<2>{}) &  \\
   $ 0 $  &  $ 0 $  &  $ A_{3} $  &  $ A_{3} $ \\
  (\IVstars{},\eText<6>{}) & (\Izero{},\nzero{}) & (\IVns{},\suText<2>{}) &  \\
   $ 0 $  &  $ 0 $  &  $ A_{1} \oplus A_{2} $  &  $ A_{1} \oplus A_{2} $ \\
   $ 0 $  &  $ 0 $  &  $ \g_{2} $  &  $ \g_{2} $ \\
  (\eText<6>{}) & (\nzero{}) & (\suText<2>{}) &  \\
   $ 0 $  &  $ 0 $  &  $ B_{3} $  &  $ B_{3} $ \\
   $ 0 $  &  $ 0 $  &  $ A_{1} \oplus A_{2} $  &  $ A_{1} \oplus A_{2} $ \\
  (\IVstars{},\eText<6>{}) & (\Izero{},\nzero{}) & (\Ins<3>{},\suText<3>{}) &  \\
   $ 0 $  &  $ 0 $  &  $ A_{5} $  &  $ A_{5} $ \\
  (\IVstars{},\eText<6>{}) & (\Izero{},\nzero{}) & (\IVs{},\suText<3>{}) &  \\
   $ 0 $  &  $ 0 $  &  $ A_{2}^2 $  &  $ A_{2}^2 $ \\
   $ 0 $  &  $ 0 $  &  $ C_{2} $  &  $ C_{2} $ \\
  (\eText<6>{}) & (\nzero{}) & (\suText<3>{}) &  \\
   $ 0 $  &  $ 0 $  &  $ A_{5} $  &  $ A_{5} $ \\
  (\IIIstar{},\eText<7>{}) & (\Izero{},\nzero{}) & (\Izero{},\nzero{}) &  \\
   $ 0 $  &  $ A_{1} $  &  $ 0 $  &  $ A_{1} $ \\
  (\IIIstar{},\eText<7>{}) & (\Izero{},\nzero{}) & (\II{},\nzero{}) &  \\
   $ 0 $  &  $ 0 $  &  $ A_{1} $  &  $ A_{1} $ \\
  (\IIIstar{},\eText<7>{}) & (\Izero{},\nzero{}) & (\Ione{},\nzero{}) &  \\
   $ 0 $  &  $ A_{1} $  &  $ A_{1} $  &  $ A_{1}^2 $ \\
  (\eText<7>{}) & (\nzero{}) & (\nzero{}) &  \\
   $ 0 $  &  $ A_{1} $  &  $ A_{1} $  &  $ A_{1}^2 $ \\
  (\IIIstar{},\eText<7>{}) & (\Izero{},\nzero{}) & (\Itwo{},\suText<2>{}) &  \\
   $ 0 $  &  $ 0 $  &  $ A_{3} $  &  $ A_{3} $ \\
  (\IIIstar{},\eText<7>{}) & (\Izero{},\nzero{}) & (III,\suText<2>{}) &  \\
   $ 0 $  &  $ 0 $  &  $ B_{3} $  &  $ B_{3} $ \\
  (\eText<7>{}) & (\nzero{}) & (\suText<2>{}) &  \\
   $ 0 $  &  $ 0 $  &  $ B_{3} $  &  $ B_{3} $ \\
   $\vdots$ & $\vdots$ & $\vdots$ & $\vdots$ \\
   & & & \\
   $\vdots$ & $\vdots$ & $\vdots$ & $\vdots$ \\
  (\IVstarns{},\fText{}) & (\II{},\nzero{}) & (\Izerostarns{},\gText{}) &  \\
   $ 0 $  &  $ 0 $  &  $ C_{4} $  &  $ C_{4} $ \\
  (\IVstarns{},\fText{}) & (\Izero{},\nzero{}) & (\Izero{},\nzero{}) &  \\
   $ 0 $  &  $ A_{2} $  &  $ 0 $  &  $ A_{2} $ \\
  (\IVstarns{},\fText{}) & (\Izero{},\nzero{}) & (\II{},\nzero{}) &  \\
   $ 0 $  &  $ A_{1} $  &  $ A_{1} $  &  $ A_{1}^2 $ \\
  (\IVstarns{},\fText{}) & (\Izero{},\nzero{}) & (\Ione{},\nzero{}) &  \\
   $ 0 $  &  $ A_{2} $  &  $ A_{1} $  &  $ A_{1} \oplus A_{2} $ \\
  (\IVstarns{},\fText{}) & (\II{},\nzero{}) & (\II{},\nzero{}) &  \\
   $ 0 $  &  $ A_{1} $  &  $ 0 $  &  $ A_{1} $ \\
  (\IVstarns{},\fText{}) & (\Ione{},\nzero{}) & (\Ione{},\nzero{}) &  \\
   $ 0 $  &  $ A_{1} $  &  $ 0 $  &  $ A_{1} $ \\
  (\fText{}) & (\nzero{}) & (\nzero{}) &  \\
   $ 0 $  &  $ A_{2} $  &  $ A_{1} $  &  $ A_{1} \oplus A_{2} $ \\
  (\IVstarns{},\fText{}) & (\Izero{},\nzero{}) & (III,\suText<2>{}) &  \\
   $ 0 $  &  $ 0 $  &  $ B_{3} $  &  $ B_{3} $ \\
  (\IVstarns{},\fText{}) & (\Izero{},\nzero{}) & (\Itwo{},\suText<2>{}) &  \\
   $ 0 $  &  $ 0 $  &  $ A_{3} $  &  $ A_{3} $ \\
  (\IVstarns{},\fText{}) & (\II{},\nzero{}) & (III,\suText<2>{}) &  \\
   $ 0 $  &  $ 0 $  &  $ A_{1} $  &  $ A_{1} $ \\
  (\IVstarns{},\fText{}) & (\II{},\nzero{}) & (\IVns{},\suText<2>{}) &  \\
   $ 0 $  &  $ 0 $  &  $ \g_{2} $  &  $ \g_{2} $ \\
  (\IVstarns{},\fText{}) & (\Ione{},\nzero{}) & (\Itwo{},\suText<2>{}) &  \\
   $ 0 $  &  $ 0 $  &  $ A_{2} $  &  $ A_{2} $ \\
  (\IVstarns{},\fText{}) & (\Izero{},\nzero{}) & (\IVns{},\suText<2>{}) &  \\
   $ 0 $  &  $ 0 $  &  $ A_{1} \oplus A_{2} $  &  $ A_{1} \oplus A_{2} $ \\
   $ 0 $  &  $ 0 $  &  $ \g_{2} $  &  $ \g_{2} $ \\
  (\fText{}) & (\nzero{}) & (\suText<2>{}) &  \\
   $ 0 $  &  $ 0 $  &  $ B_{3} $  &  $ B_{3} $ \\
   $ 0 $  &  $ 0 $  &  $ A_{1} \oplus A_{2} $  &  $ A_{1} \oplus A_{2} $ \\
  (\IVstarns{},\fText{}) & (\Izero{},\nzero{}) & (\Ins<3>{},\suText<3>{}) &  \\
   $ 0 $  &  $ 0 $  &  $ A_{5} $  &  $ A_{5} $ \\
  (\IVstarns{},\fText{}) & (\Ione{},\nzero{}) & (\Ins<3>{},\suText<3>{}) &  \\
   $ 0 $  &  $ 0 $  &  $ A_{4} $  &  $ A_{4} $ \\
  (\IVstarns{},\fText{}) & (\Izero{},\nzero{}) & (\IVs{},\suText<3>{}) &  \\
   $ 0 $  &  $ 0 $  &  $ A_{2}^2 $  &  $ A_{2}^2 $ \\
   $ 0 $  &  $ 0 $  &  $ C_{2} $  &  $ C_{2} $ \\
  (\fText{}) & (\nzero{}) & (\suText<3>{}) &  \\
   $ 0 $  &  $ 0 $  &  $ A_{5} $  &  $ A_{5} $ \\
      & & & \\
      \caption{All configurations for the base $512.$}
      \label{t:GS512}
  \end{longtable} } 
\unskip
\unpenalty
\unpenalty}

\unvbox\ltmcbox

\medskip
\end{multicols}
\end{minipage}

\end{section}

%\bibliographystyle{unsrt}
%\bibliography{noteBib}

\end{document}